\pdfoutput=1
\documentclass[preprint]{aastex631}
\usepackage[T1]{fontenc}
\usepackage[frenchb, english]{babel}
\usepackage{graphicx}
\usepackage{dcolumn}
\usepackage{bm}
\usepackage{subfigure}
\usepackage{wasysym}
\usepackage{verbatim}
\usepackage{color}
\usepackage{todonotes}
\usepackage{float}
\usepackage{placeins}
\usepackage{xspace}
\usepackage{tabularx} 
\DeclareUnicodeCharacter{2060}{\nolinebreak}
\def\pcam{Prime-Cam\xspace}

\newcommand{\blos}{B$_{\text{LOS}}$}
\newcommand{\bperp}{B$_{\text{POS}}$}

\begin{document}

\title{CCAT-prime Collaboration: Science Goals and Forecasts with Prime-Cam \\
on the Fred Young Submillimeter Telescope}

\begin{abstract}

We present a detailed overview of the science goals and predictions for the \pcam direct detection camera/spectrometer being constructed by the CCAT-prime collaboration for dedicated use on
the Fred Young Submillimeter Telescope (FYST). The FYST
is a wide-field, 6-m aperture submillimeter telescope being built (first light in mid-2024) by an international consortium of institutions led by Cornell University and sited at more than 5600 meters on Cerro Chajnantor in northern Chile. \pcam is one of two instruments planned for FYST
and will provide unprecedented spectroscopic and broadband measurement capabilities to address important astrophysical questions ranging from 
Big Bang cosmology through reionization and the formation of the first galaxies to star formation within our own Milky Way galaxy. 
\pcam on the FYST will have a mapping speed that is over ten times greater than existing and near-term facilities for high-redshift science and broadband polarimetric imaging at frequencies above 300\,GHz.
We describe details of the science program enabled by this system and our preliminary survey strategies.

\end{abstract}


\author{CCAT-Prime collaboration}
\affiliation{CCAT Observatory Inc.}
\author{Manuel~Aravena}
\affiliation{N\'{u}cleo de Astronom\'{i}a, Facultad de Ingenier\'{i}a y Ciencias, Universidad Diego Portales, Av. Ej\'{e}rcito 441, Santiago, Chile}
\author{Jason~E.~Austermann}
\affiliation{Quantum Sensors Group, National Institute of Standards and Technology, Boulder, CO 80305, USA}
\author[0000-0001-5276-8730]{Kaustuv~Basu}
\affiliation{Argelander Institute for Astronomy, University of Bonn, Auf dem H\"{u}gel 71, D-53121 Bonn, Germany}
\author{Nicholas~Battaglia}
\affiliation{Department of Astronomy, Cornell University, Ithaca, NY 14853, USA}
\author{Benjamin~Beringue}
\affiliation{DAMTP, Centre for Mathematical Sciences, Wilberforce Road, Cambridge, UK, CB3 0WA}
\author{Frank~Bertoldi}
\affiliation{Argelander Institute for Astronomy, University of Bonn, Auf dem H\"{u}gel 71, D-53121 Bonn, Germany}
\author{Frank~Bigiel}
\affiliation{Argelander Institute for Astronomy, University of Bonn, Auf dem H\"{u}gel 71, D-53121 Bonn, Germany}
\author[0000-0003-2618-6504]{J.~Richard~Bond}
\affiliation{Canadian Institute for Theoretical Astrophysics, University of Toronto, 60 St. George Street, Toronto, ON M5S 3H8, Canada}
\author[0000-0001-8382-5275]{Patrick~C.~Breysse}
\affiliation{Center for Cosmology and Particle Physics, Department of Physics, New York University, 726 Broadway, New York, NY 1003, USA}
\author{Colton~Broughton}
\affiliation{Department of Physics and Astronomy, University of Victoria, 3800 Finnerty Road, Elliot Building, Victoria, BC, V8P 5C2, Canada}
\author[0000-0001-8468-9391]{Ricardo~Bustos}
\affiliation{Departamento de Ingenier\'{i}a El\'{e}ctrica, Universidad Cat\'{o}lica de la Sant\'{i}sima Concepci\'{o}n, Alonso de Ribera 2850, Concepci\'{o}n, Chile}
\author[0000-0002-8487-3153]{Scott~C.~Chapman}
\affiliation{Dalhousie University, Halifax, NS B3H4R2, Canada}
\affiliation{University of British Columbia, Vancouver, BC V6T1Z1, Canada}
\affiliation{Herzberg Astronomy and Astrophysics Research Centre, National Research Council Canada, 5071 West Saanich Road, Victoria, BC, V9E 2E7, Canada}
\author{Maude~Charmetant}
\affiliation{Argelander Institute for Astronomy, University of Bonn, Auf dem H\"{u}gel 71, D-53121 Bonn, Germany}
\author[0000-0002-9113-7058]{Steve~K.~Choi}
\affiliation{Department of Physics, Cornell University, Ithaca, NY 14853, USA}
\affiliation{Department of Astronomy, Cornell University, Ithaca, NY 14853, USA}
\author[0000-0003-2618-6504]{Dongwoo~T.~Chung}
\affiliation{Canadian Institute for Theoretical Astrophysics, University of Toronto, 60 St. George Street, Toronto, ON M5S 3H8, Canada}
\affiliation{Dunlap Institute for Astronomy and Astrophysics, University of Toronto, 50 St. George Street, Toronto, ON M5S 3H4, Canada}
\author[0000-0002-7633-3376]{Susan~E.~Clark}
\affiliation{Department of Physics, Stanford University, Stanford, California 94305, USA}
\affiliation{Kavli Institute for Particle Astrophysics \& Cosmology, P. O. Box 2450, Stanford University, Stanford, CA 94305, USA}
\author{Nicholas~F.~Cothard}
\affiliation{Department of Applied and Engineering Physics, Cornell University, Ithaca, NY 14853, USA}
\author{Abigail~T.~Crites}
\affiliation{Department of Physics, Cornell University, Ithaca, NY 14853, USA}
\author{Ankur~Dev}
\affiliation{Argelander Institute for Astronomy, University of Bonn, Auf dem H\"{u}gel 71, D-53121 Bonn, Germany}
\author{Kaela~Douglas}
\affiliation{Department of Physics and Astronomy, University of Victoria, 3800 Finnerty Road, Elliot Building, Victoria, BC, V8P 5C2, Canada}
\author{Cody~J.~Duell}
\affiliation{Department of Physics, Cornell University, Ithaca, NY 14853, USA}
\author{Rolando~D\"unner}
\affiliation{Instituto de Astrof\'isica and Centro de Astro-Ingenier\'ia, Facultad de F\'isica, Pontificia Universidad Cat\'olica de Chile, Av. Vicu\~na Mackenna 4860, 7820436, Macul, Santiago, Chile}
\author{Haruki~Ebina}
\affiliation{Department of Physics, Cornell University, Ithaca, NY 14853, USA}
\author[0000-0002-5486-6745]{Jens~Erler}
\affiliation{Argelander Institute for Astronomy, University of Bonn, Auf dem H\"{u}gel 71, D-53121 Bonn, Germany}
\affiliation{Deutsches Zentrum f\"{u}r Luft- und Raumfahrt e.V. (DLR) Projekttr\"{a}ger, Joseph-Beuys-Allee 4, D-53113 Bonn, Germany }
\author[0000-0002-5903-1886]{Michel~Fich}
\affiliation{Waterloo Centre for Astrophysics, Department of Physics and Astrophysics, University of Waterloo, Waterloo, ON N2L 3G1, Canada}
\author[0000-0002-4666-609X]{Laura~M.~Fissel}
\affiliation{Department of Physics, Engineering Physics, and Astronomy, Queen's University, Kingston, ON K7L 3N6, Canada}
\author[0000-0002-0190-2271]{Simon~Foreman}
\affiliation{Dominion Radio Astrophysical Observatory, Herzberg Astronomy and Astrophysics Research Centre, National Research Council Canada, P. O. Box 248, Penticton, BC V2A 6J9 Canada}
\affiliation{Perimeter Institute for Theoretical Physics, 31 Caroline Street North, Waterloo, ON N2L 2Y5, Canada}
\author{Patricio~A.~Gallardo}
\affiliation{Department of Physics, Cornell University, Ithaca, NY 14853, USA}
\author{Jiansong~Gao}
\affiliation{Quantum Sensors Group, National Institute of Standards and Technology, Boulder, CO 80305, USA}
\author[0000-0002-8586-6721]{Pablo~Garc\'ia}
\affiliation{Chinese Academy of Sciences South America Center for Astronomy, National Astronomical Observatories, CAS, Beijing 100101, China.}
\affiliation{Instituto de Astronom\'ia, Universidad Cat\'olica del Norte, Av. Angamos 0610, Antofagasta 1270709, Chile.}
\author{Riccardo~Giovanelli}
\affiliation{Department of Astronomy, Cornell University, Ithaca, NY 14853, USA}
\author{Joseph~E.~Golec}
\affiliation{Department of Astronomy and Astrophysics, University of Chicago, Chicago, IL 60637, USA}
\author{Christopher~E.~Groppi}
\affiliation{School of Earth and Space Exploration, Arizona State University, Tempe, AZ 85287, USA}
\author[0000-0001-5334-5166]{Martha~P.~Haynes}
\affiliation{Department of Astronomy, Cornell University, Ithaca, NY 14853, USA}
\author{Douglas~Henke}
\affiliation{Herzberg Astronomy and Astrophysics Research Centre, National Research Council Canada, 5071 West Saanich Road, Victoria, BC, V9E 2E7, Canada}
\author[0000-0001-7449-4638]{Brandon~Hensley}
\affiliation{Department of Astrophysical Sciences, Princeton University, Princeton, NJ 08544, USA}
\author[0000-0002-3856-8385]{Terry~Herter}
\affiliation{Department of Astronomy, Cornell University, Ithaca, NY 14853, USA}
\author[0000-0001-8195-3900]{Ronan~Higgins}
\affiliation{I.~Physikalisches Institut, Universit\"at zu K\"oln, Z\"ulpicher Str. 77, D-50937 K\"oln, Germany}
\author{Ren\'ee~Hlo\v{z}ek}
\affiliation{David A. Dunlap Department of Astronomy and Astrophysics, University of Toronto, 50 St. George Street, Toronto ON M5S3H4}
\affiliation{Dunlap Institute for Astronomy and Astrophysics, University of Toronto, 50 St. George Street, Toronto ON M5S3H4}
\author{Anthony~Huber}
\affiliation{Department of Physics and Astronomy, University of Lethbridge, Alberta, Canada}
\author{Zachary~Huber}
\affiliation{Department of Physics, Cornell University, Ithaca, NY 14853, USA}
\author{Johannes~Hubmayr}
\affiliation{Quantum Sensors Group, National Institute of Standards and Technology, Boulder, CO 80305, USA}
\author{Rebecca~Jackson}
\affiliation{Department of Astronomy and Astrophysics, University of Chicago, Chicago, IL 60637, USA}
\author[0000-0002-6773-459X]{Douglas~Johnstone}
\affiliation{Herzberg Astronomy and Astrophysics Research Centre, National Research Council Canada, 5071 West Saanich Road, Victoria, BC, V9E 2E7, Canada}
\affiliation{Department of Physics and Astronomy, University of Victoria, 3800 Finnerty Road, Elliot Building, Victoria, BC, V8P 5C2, Canada}
\author[0000-0003-3259-7457]{Christos~Karoumpis}
\affiliation{Argelander Institute for Astronomy, University of Bonn, Auf dem H\"{u}gel 71, D-53121 Bonn, Germany}
\author{Laura~C.~Keating}
\affiliation{Leibniz Institute for Astrophysics Potsdam (AIP), An der Sternwarte 16, 14482 Potsdam, Germany}
\author[0000-0002-0136-2404]{Eiichiro~Komatsu}
\affiliation{Max-Planck-Institut f\"{u}r Astrophysik, Karl-Schwarzschild Str. 1, 85741 Garching, Germany}
\affiliation{Kavli Institute for the Physics and Mathematics of the Universe (Kavli IPMU, WPI), University of Tokyo, Chiba 277-8582, Japan}
\author{Yaqiong~Li}
\affiliation{Department of Physics, Cornell University, Ithaca, NY 14853, USA}
\affiliation{Kavli Institute at Cornell for Nanoscale Science, Cornell University, Ithaca, NY 14853, USA}
\author[0000-0002-6777-6490]{Benjamin~Magnelli}
\affiliation{Argelander Institute for Astronomy, University of Bonn, Auf dem H\"{u}gel 71, D-53121 Bonn, Germany}
\affiliation{AIM, CEA, CNRS, Universit\'{e} Paris-Saclay, Universit\'{e} Paris Diderot, Sorbonne Paris Cit\'{e}, 91191 Gif-sur-Yvette, France}
\author[0000-0003-3017-9577]{Brenda~C.~Matthews}
\affiliation{Herzberg Astronomy and Astrophysics Research Centre, National Research Council Canada, 5071 West Saanich Road, Victoria, BC, V9E 2E7, Canada}
\affiliation{Department of Physics and Astronomy, University of Victoria, 3800 Finnerty Road, Elliot Building, Victoria, BC, V8P 5C2, Canada}
\author{Philip~D.~Mauskopf}
\affiliation{School of Earth and Space Exploration, Arizona State University, Tempe, AZ 85287, USA}
\affiliation{Department of Physics, Arizona State University, Tempe, AZ 85287, USA}
\author{Jeffrey~J.~McMahon}
\affiliation{Department of Astronomy and Astrophysics, University of Chicago, Chicago, IL 60637, USA}
\affiliation{Kavli Institute for Cosmological Physics, University of Chicago, Chicago, IL 60637, USA}
\affiliation{Department of Physics, University of Chicago, Chicago, IL 60637, USA}
\affiliation{Enrico Fermi Institute, University of Chicago, Chicago, IL 60637, USA}
\author{P.~Daniel~Meerburg}
\affiliation{Van Swinderen Institute for Particle Physics and Gravity, University of Groningen, Nijenborgh 4, 9747 AG Groningen, The Netherlands}
\author[0000-0001-8510-2812]{Joel~Meyers}
\affiliation{Department of Physics, Southern Methodist University, 3215 Daniel Ave, Dallas, TX 75275, USA}
\author{Vyoma~Muralidhara}
\affiliation{Max-Planck-Institut f\"{u}r Astrophysik, Karl-Schwarzschild Str. 1, 85741 Garching, Germany}
\author{Norman~W.~Murray}
\affiliation{Canadian Institute for Theoretical Astrophysics, University of Toronto, 60 St. George Street, Toronto, ON M5S 3H8, Canada}
\author[0000-0001-7125-3580]{Michael~D.~Niemack}
\affiliation{Department of Physics, Cornell University, Ithaca, NY 14853, USA}
\affiliation{Department of Astronomy, Cornell University, Ithaca, NY 14853, USA}
\affiliation{Kavli Institute at Cornell for Nanoscale Science, Cornell University, Ithaca, NY 
14853, USA}
\author{Thomas~Nikola}
\affiliation{Cornell Center for Astrophysics and Planetary Sciences, 210 Space Sciences Building, Cornell University, Ithaca, NY 14853, USA}
\author[0000-0002-6838-6435]{Yoko~Okada}
\affiliation{I.~Physikalisches Institut, Universit\"at zu K\"oln, Z\"ulpicher Str. 77, D-50937 K\"oln, Germany}
\author{Roberto~Puddu}
\affiliation{Instituto de Astrof\'isica and Centro de Astro-Ingenier\'ia, Facultad de F\'isica, Pontificia Universidad Cat\'olica de Chile, Av. Vicu\~na Mackenna 4860, 7820436, Macul, Santiago, Chile}
\author{Dominik~A.~Riechers}
\affiliation{I.~Physikalisches Institut, Universit\"at zu K\"oln, Z\"ulpicher Str. 77, D-50937 K\"oln, Germany}
\author[0000-0002-5204-2259]{Erik~Rosolowsky}
\affiliation{Department of Physics, University of Alberta, CCIS 4-181, Edmonton, AB T6G 2E1, Canada}
\author{Kayla~Rossi}
\affiliation{Cornell Center for Astrophysics and Planetary Sciences, 210 Space Sciences Building, Cornell University, Ithaca, NY 14853, USA}
\author{Kaja~Rotermund}
\affiliation{Dalhousie University, Halifax, NS B3H4R2, Canada}
\author[0000-0001-5729-0246]{Anirban~Roy}
\affiliation{Department of Astronomy, Cornell University, Ithaca, NY 14853, USA}
\author[0000-0001-7474-6874]{Sarah~I.~Sadavoy}
\affiliation{Department of Physics, Engineering Physics, and Astronomy, Queen's University, Kingston, ON K7L 3N6, Canada}
\author{Reinhold~Schaaf}
\affiliation{Argelander Institute for Astronomy, University of Bonn, Auf dem H\"{u}gel 71, D-53121 Bonn, Germany}
\author[0000-0003-2141-5689]{Peter~Schilke}
\affiliation{I.~Physikalisches Institut, Universit\"at zu K\"oln, Z\"ulpicher Str. 77, D-50937 K\"oln, Germany}
\author[0000-0002-6878-9840]{Douglas~Scott}
\affiliation{University of British Columbia, Vancouver, BC V6T1Z1, Canada}
\author[0000-0003-2555-4408]{Robert~Simon}
\affiliation{I.~Physikalisches Institut, Universit\"at zu K\"oln, Z\"ulpicher Str. 77, D-50937 K\"oln, Germany}
\author[0000-0003-0064-6867]{Adrian~K.~Sinclair}
\affiliation{School of Earth and Space Exploration, Arizona State University, Tempe, AZ 85287, USA}
\author[0000-0001-6682-916X]{Gregory~R.~Sivakoff}
\affiliation{Department of Physics, University of Alberta, CCIS 4-181, Edmonton, AB T6G 2E1, Canada}
\author{Gordon~J.~Stacey}
\affiliation{Department of Astronomy, Cornell University, Ithaca, NY 14853, USA}
\author[0000-0003-2300-8200]{Amelia~M.~Stutz}
\affiliation{Departamento de Astronom\'{i}a, Universidad de Concepci\'{o}n,Casilla 160-C, Concepci\'{o}n, Chile}
\affiliation{Max-Planck-Institut f\"ur Astronomie, K\"{o}nigstuhl 17, 69117 Heidelberg, Germany}
\author[0000-0001-7658-4397]{Juergen~Stutzki}
\affiliation{I.~Physikalisches Institut, Universit\"at zu K\"oln, Z\"ulpicher Str. 77, D-50937 K\"oln, Germany}
\author[0000-0001-8749-1436]{Mehrnoosh~Tahani}
\affiliation{Dominion Radio Astrophysical Observatory, Herzberg Astronomy and Astrophysics Research Centre, National Research Council Canada, P. O. Box 248, Penticton, BC V2A 6J9 Canada}
\author[0000-0003-1187-2544]{Karun~Thanjavur}
\affiliation{Department of Physics and Astronomy, University of Victoria, 3800 Finnerty Road, Elliot Building, Victoria, BC, V8P 5C2, Canada}
\author{Ralf~A.~Timmermann}
\affiliation{Argelander Institute for Astronomy, University of Bonn, Auf dem H\"{u}gel 71, D-53121 Bonn, Germany}
\author{Joel~N.~Ullom}
\affiliation{Quantum Sensors Group, National Institute of Standards and Technology, Boulder, CO 80305, USA}
\author{Alexander~van~Engelen}
\affiliation{School of Earth and Space Exploration, Arizona State University, Tempe, AZ 85287, USA}
\author{Eve~M.~Vavagiakis}
\affiliation{Department of Physics, Cornell University, Ithaca, NY 14853, USA}
\author{Michael~R.~Vissers}
\affiliation{Quantum Sensors Group, National Institute of Standards and Technology, Boulder, CO 80305, USA}
\author{Jordan~D.~Wheeler}
\affiliation{Quantum Sensors Group, National Institute of Standards and Technology, Boulder, CO 80305, USA}
\author[0000-0002-1061-6154]{Simon~D.~M. White}
\affiliation{Max-Planck-Institut f\"{u}r Astrophysik, Karl-Schwarzschild Str. 1, 85741 Garching, Germany}
\author{Yijie~Zhu}
\affiliation{Department of Astronomy, Cornell University, Ithaca, NY 14853, USA}
\author{Bugao~Zou}
\affiliation{Department of Applied and Engineering Physics, Cornell University, Ithaca, NY 14853, USA}

\shortauthors{CCAT-prime Collaboration}

\correspondingauthor{G.J. Stacey}
\email{stacey@astro.cornell.edu}

\section{Introduction}
\label{sec:intro}
The CCAT-prime collaboration is an international group of institutions led by Cornell University that is brought together by a common desire to address pressing astrophysical questions ranging from Big Bang cosmology and the large scale structure of the Universe, through the formation of the first stars and galaxies and on to the formation of stars and planetary systems in the Galaxy.  These types of studies involve large-scale, high-sensitivity polarimetric, photometric, and spectroscopic mapping at frequencies spanning the electromagnetic spectrum.  Indeed, a wide variety of such programs are ongoing or planned from the optical to the cm-wave bands (e.g. Rubin Observatory (\emph{Rubin}, formerly LSST), Euclid, Roman Space Telescope (\emph{Roman}, formerly WFIRST), ASKAP, SKA etc.).  We have undertaken the construction of a new high-efficiency, wide-field telescope, the Fred Young Submillimeter Telescope (FYST, pronounced ``feest''), which together with its \pcam instrument will enable wide-field and very deep mapping through the telluric windows from 100 to 900 GHz (3 to 0.33 millimeter wavelength). This spectral regime has been explored through a variety of surveys including those using ground-based telescopes (e.g. CSO, JCMT, APEX, SPT and ACT) and space telescopes (e.g. Planck and Herschel) with many striking discoveries. With its 6\,m aperture at a superb site, FYST and its associated instrumentation moves beyond these surveys in terms of confusion limited depth, areal coverage, and/or frequency coverage, enabling new science as well as greatly expanding the science returns from these surveys and those at other frequencies.
Within this paper we outline the science plans for FYST focussing on the following science programs:
\begin{enumerate}

    \item \textbf{Epoch of Reionization:} The first stars and galaxies dominated reionization processes during the epoch of re-ionization (EoR) at $6 \lesssim z \lesssim 20$.  Stars form at local over-densities of matter so that, if we can trace the star formation (SF) process, we both trace the history of star and galaxy formation and the growth of large scale structure, which is governed by fundamental physics in the evolving Universe.  The [\ion{C}{2}] 1.901 THz (158\,$\mu$m) and [\ion{O}{3}] 3.393 THz (88\,$\mu$m) fine-structure lines are exceptionally luminous and provide uniquely powerful probes of star formation in the early Universe. These lines are redshifted into the FYST windows by the cosmic expansion since the EoR, and 
    the aggregate line emission from faint galaxies tomographically reveals the large scale structure of reionization and the galaxy assembly process (see Section \ref{sec:IM}).
    \item \textbf{Tracing galaxy evolution over cosmic time:} Half the star formation in the Universe over cosmic time is obscured to optical/UV observers by the presence of dust.  The dust is heated by stellar photons and re-radiates its energy in the far-infrared, with its spectral energy distribution typically peaking at
    frequencies from 2 to 6 THz.  Tracing this power peak reveals the star forming luminosity of these dusty star forming galaxies (DSFGs).  This power peak is shifted into the FYST windows at redshifts at and beyond the peak of cosmic star formation activity, so called ``Cosmic Noon'' at $1 \leq z \leq 3$.   Therefore, by mapping the 3-d distribution and energetics of DSFGs, we track the process of galaxy assembly and the evolving structure of the Universe from times between the EoR through Cosmic Noon, when dark energy begins to have important impact on the expansion (see Section \ref{sec:highz}).
    \item  \textbf{Measuring CMB foregrounds:} The scale of structures in the cosmic microwave background (CMB) is controlled by the fundamental physics of the Universe. For instance, wide-field mapping of the CMB has revealed that the Universe is flat, it is 13.8 Gyrs old, and its energy content is about 5\% matter, 27\% dark matter, and 68\% dark energy.   Encoded within the polarization patterns and scales of the CMB are signals that reflect the earliest moments of the expansion.  In particular, B-mode polarization patterns promise to constrain models of cosmic inflation. Unfortunately, strong B-mode patterns are present in foreground emission as well (e.g. the Milky Way galaxy's dust emission) so the measurement of primordial B-modes is very challenging.  FYST is uniquely suited to unravelling this foreground emission from the CMB B-mode emission with its access to the high-frequency submillimeter-wavelength bands (see Section \ref{sec:dust}).  
    \item \textbf{Galactic Polarization:} Magnetic fields help govern the accumulation and collapse of neutral gas within galaxies and therefore play an important role in regulating the star formation process.  Magnetic fields are probed in several ways: spectroscopically through, for example Zeeman splitting of line emission or absorption; selective extinction of starlight due to aligned dust grains in the visible/IR; synchrotron emission; or the rotational measures of radio emission from background pulsars. These probes have their limits - extinction for visible light, and limited lines of sight and appropriate excitation conditions for Zeeman splitting or rotation measure observations. Magnetic fields are also traced through linearly polarized thermal dust emission  in the far-infrared to millimeter-wave bands from dust grains aligned with respect to the local magnetic field. The degree of polarization and its morphology reflect intrinsic grain properties and the underlying magnetic field direction, strength, and energetic importance relative to turbulence and gravity. Dust emission arises from all phases of the interstellar medium, and FYST provides the platform to trace this radiation continuously from kpc to the sub-pc scales at which individual stars form (see Section \ref{sec:bfield}).
    \item \textbf{Galaxy cluster evolution:} The CMB is also a backlight for studies of large scale structure and galaxy assembly through the inverse Compton scattering of CMB photons off of hot electrons deep in the gravitational wells of galaxy clusters.  This energy-shifting of CMB photons, called the Sunyaev-Zel'dovich (SZ) effect, is used to trace the mass, spatial distribution, and peculiar motions of these clusters, which are the most massive self-gravitating structures in the Universe. The SZ effect centers at 220 GHz in the mm-wave bands, and its spectrum is very effectively mapped through the FYST windows (see Section \ref{sec:sz}).    
    \item \textbf{Rayleigh Scattering:} The process of recombination was not  instantaneous.  Neutral hydrogen and helium atoms began to appear early in the recombination epoch and could then Rayleigh scatter the CMB radiation, producing a frequency-dependent signal that traces cosmic structure during the epoch of recombination. The detection of this signal is quite challenging but made more manageable with the broad frequency coverage of the \pcam on FYST (see Section \ref{sec:rs}). 
    \item \textbf{Time domain phenomena:} Time domain astrophysics is an exciting area of research that is being reinvigorated by the new generation of large-field survey facilities spanning the electromagnetic spectrum. While typically discovered at other frequencies, the understanding of very energetic transients such as supernovae, gamma-ray bursts, X-ray binaries, merging neutron stars, and tidal disruption events will benefit greatly from the FYST spectral coverage, which, for example, can trace shocks and energy deposition in the enveloping circumstellar or interstellar medium.  The utility of the submillimeter bands are especially evident for dust-enshrouded sources such as protostars, where changes in the submillimeter brightness likely reflect the heating of the enshrouding protostellar core by the accretion luminosity of matter flowing onto the protostar itself (see Section \ref{sec:YSO}). 

\end{enumerate}  

FYST, with its 6-meter aperture, wide-field of view design, as well as its location at more than 5600\,meters elevation on Cerro Chajnantor in northern Chile \citep{parshley/2018a,parshley/2018b}, and together with \pcam \citep{vavagiakis/etal:2018}, is uniquely well suited to pursue these science programs. The aperture is appropriate for diffraction-limited imaging with a beam-size appropriate for our studies, especially the CMB B-mode polarization and sub-Mpc scales for line intensity mapping (LIM) experiments.  \pcam imaging arrays take advantage of the wide-field-of view of FYST to yield mapping speeds that are over ten times greater than existing and near-term facilities for the various experiments. 

The paper is organized as follows:
First we describe the \pcam instrument, the camera optics, the spectrometer design, the arrays, and expected performance (Section \ref{sec:methods}). Next, the surveys to address the science goals are presented in Section \ref{sec:surveys}. Then, each science program and its goal that we expect to accomplish with these surveys in the context of a five-year observing plan is presented in detail in Sections \ref{sec:IM} to  \ref{sec:YSO}. We summarize the paper in Section \ref{sec:summary}.

\section{Instrument Overview} 
\label{sec:methods}

\begin{figure}
\begin{center}
\includegraphics[width=3in]{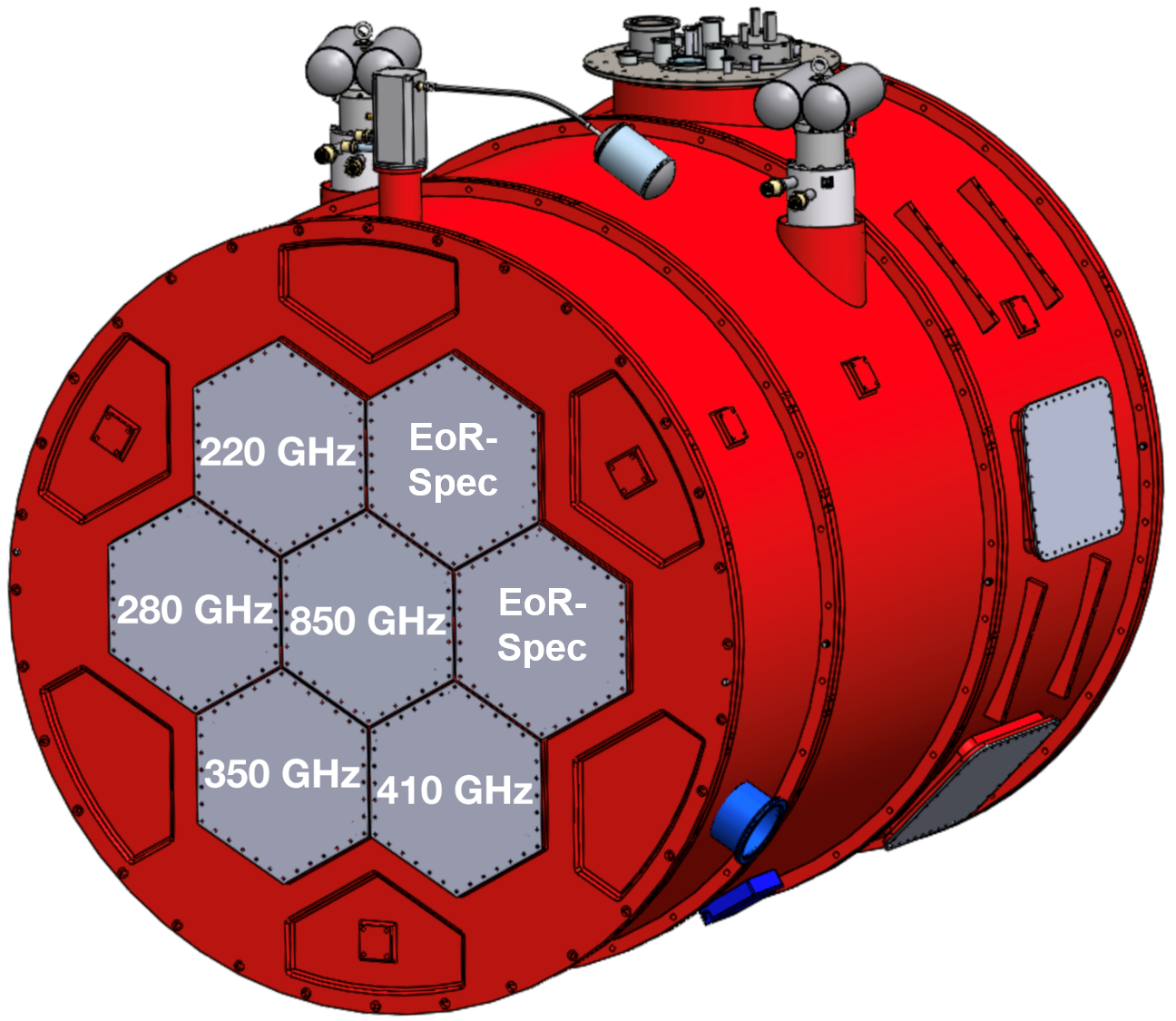}
\end{center}
\caption{\pcam instrument design showing the placement of the seven instrument module entrance windows including the five polarization-sensitive modules at nominal band center frequencies from 220 to 850 GHz, and the two spectrometer (EoR-Spec) modules.  \citep{vavagiakis/etal:2018}. \label{fig:pcam}}
\end{figure}

The \pcam instrument is designed to fill the central 4.9$^{\circ}$ of the 8$^{\circ}$ diameter field of view delivered by the FYST \citep{parshley/2018a,niemack/2016}.  To do so, the field is split up into seven independent instrument modules, with six modules arranged in a compact configuration around a central module (Fig.~\ref{fig:pcam}). Each module fills up to a 1.3$^{\circ}$  diameter field of view, and is separated from its neighbors by 1.8$^{\circ}$ within a 1.8\,m diameter cryostat.  Externally, the instrument modules are close to identical, but the interiors are independently optimized with regard to optical elements and detectors for 
specific science programs.
The baseline plan for the \pcam modules includes: two imaging spectrometer modules that use 
Fabry-Perot Interferometers (FPIs) for line intensity mapping from 210 to 420\,GHz, and five broadband polarization-sensitive modules for observations at five frequencies: 220\,GHz, 280\,GHz, 350\,GHz, 410\,GHz, and 850\,GHz \citep{choi20}. Each module is contained within a cylindrical casing $\sim$45\,cm in diameter and  $\sim$1.6\,m long \citep{vavagiakis/etal:2018}.  

\subsection{Optical Design}
The seven entrance window, silicon optics design of the instrument modules is similar to ideas put forth for the proposed 25\,m CCAT telescope \citep{Stacey/etal:2014}.  These ideas evolved through  
collaboration with the Simons Observatory (SO) project \citep{dicker/2018,Ade19} into the nearly interchangeable instrument module  design for \pcam \citep{vavagiakis/etal:2018}. Each module has a roughly 40\,cm diameter entrance window made of ultra-high-molecular-weight polyethylene (UHMWPE) that has been anti-reflection (AR) coated with a layer of expanded Teflon. The focus of FYST is about 20\,cm inside of the window of each module.  Within the module are lenses and the optical filters that are required to block unwanted radiation.  Each of the lenses and filters are thermally attached to the temperature bus (80, 40, 4, or 0.1 K) appropriate for their function.  The powered optics include a field lens near the FYST focus followed by a pair of plano -convex lenses.  The first of these lenses forms a pupil image at about 20\,cm diameter, and the second reimages the beam onto the detector array at \emph{f}/2.0. The FPI optics incorporate an additional powered lens to provide the smaller (14\,cm) diameter pupil and more collimated illumination required by the FPI relative to the broadband cameras. Lenses are made from high-purity silicon with metamaterial anti-reflection coatings \citep{Datta_2013}.  High resistivity silicon has a high index of refraction and, at low temperatures, has both very low optical loss to mm-wave radiation and high thermal conductivity. These properties combined enable the manufacture of the large diameter ($\sim$ 40 cm) optical components that are required for our instrument modules that can be kept uniformly cool, resulting in both high transmission and very low thermal emission.

\subsection{EoR-Spec}
The Epoch of Reionization Spectrometer (EoR-Spec) line intensity mapping modules are optimized to measure the 158 $\mu$m [\ion{C}{2}] fine-structure line emission from z = 8.05 (which is within the epoch of reionization) to z = 3.5 (which is nearly at the peak of star formation per unit comoving volume in the Universe). The spectrometer is based on a cryogenic imaging FPI that spectrally scans this redshift interval detecting the [\ion{C}{2}] line over wide fields \citep{cothard/2020}.  Since the [\ion{C}{2}] line tracks star formation \citep{stacey/1991,stacey/2010}, it reveals the reionization process that is driven by the first stars and galaxies $\it{and}$ the subsequent growth and evolution of these early galaxies. 

The FPI utilizes silicon-substrate based (SSB) mirrors, a new technology that enables tuning of FPI cavity finesse so as to deliver near-uniform spectral resolving power and high efficiency over the octave bandwidth of the spectrometer (210 to 420 GHz)  \citep{cothard/2018,cothard/2020}. The goal is a resolving power of $\sim$100. Each EoR-Spec module will have three focal-plane arrays.  Two arrays will have architectures and filtration centered at 260\,GHz for line intensity mapping from 210 to 315\,GHz, while the third will be centered at 370\,GHz for work at 315 to 425\,GHz.  The high frequency cut-offs for each array will be set by low-pass filters directly in front of the arrays, while the low-frequency cut-offs are set by the feedhorn design for each pixel in an array. The FPI will be set in second order to address the 260\,GHz array so that the third order fringe will be detected by the 370\,GHz array.   

\subsection{Focal Plane Arrays}
The primary technologies for large-scale arrays suitable for broadband polarimetry and photometry, or low resolving power spectroscopy in the submillimeter to millimeter wave-bands are transition-edge-sensor (TES) bolometers and kinetic inductance detectors (KIDs). While comparatively less mature, the KID technology is significantly easier to manufacture and read out than TES arrays, so that the costs per detector can be substantially smaller. Recent laboratory and on-sky measurements indicate that the KID technology can reach the fundamental background limits required for our instruments in large-format array architectures \citep{hubmayr/2015,calvo/2016}. Therefore, KID arrays have been adopted as the baseline technology for \pcam \citep{duell/2020}.

The focal planes of each camera or EoR-Spec module are well populated with three KID arrays constructed on 150\,mm silicon wafers in a tightly packed configuration \citep{vavagiakis/etal:2018}. For each camera module, the three arrays will typically be identical, while, for the EoR-Spec modules, we will have two low-frequency arrays and one high-frequency array, as described above. The reason for this choice in EoR-Spec is that the low-frequency line intensity mapping signal is much more difficult to detect than the high-frequency (low-redshift) signal. 

The arrays that we baseline for both the 280\,GHz and 350\,GHz cameras and the corresponding bands in EoR-Spec are quite similar. Each element of the focal plane array consists of a feedhorn, which
defines the received radiation pattern (beam), coupled to a waveguide
terminated by the TiN/Ti/TiN inductor of a KID. The inductor is designed
to efficiently absorb the waveguide EM wave.
 \citep{austermann/2018,duell/2020}.  The  absorber is fabricated over a backshort etched into a silicon-on-insulator (SOI) wafer for optimal detector quantum efficiency.  The major differences between the camera arrays and EoR-Spec arrays are that the camera arrays will be sensed in two polarizations by distinct KIDs while the two polarizations are tied together in the EoR-Spec arrays to detect total power through just one KID per feedhorn. For all arrays, we maximize the numbers of beams on the sky, consistent with efficient coupling to the astrophysical source and array readout limitations.  Table  \ref{tab:sens} lists the detector counts for the five camera bands and the two EoR-Spec bands.   More information about the \pcam instrument is available in \cite{vavagiakis/etal:2018} and \cite{choi20}.

\subsection{Noise Model}

The expected sensitivities for each type of module are summarized in Table~\ref{tab:sens}. We estimate the expected system sensitivity by first calculating the fundamental limits imposed by the presence of thermal backgrounds under the parameters imposed by the system as described within \cite{Stacey2011}.  Important parameters include: the sky, telescope and cryostat window emissivities; the sky, telescope, and instrumental transmission efficiencies; the thermal backgrounds that arise from within the instrument (expected to be small); the detector quantum efficiency; the numbers of polarizations accepted; and the spectral resolving power.  The result is then degraded by mixing in the expected detector noise in quadrature. The system performance is further degraded by the noise angular power spectrum 
in the atmospheric transmission commonly referred to as ``sky noise." The integrated angular noise power spectrum is given by $$N_\ell = N_{\rm{red}}\left(\frac{\ell}{\ell_{\rm{knee}}}\right)^{\alpha_{\rm{knee}}}+N_{\rm{white}},$$ 
with $N_{\rm{red}}$ and $N_{\rm{white}}$ given in Table~\ref{tab:sens} for $\ell_{\rm{knee}}=1000$ and $\alpha_{\rm{knee}}=-3.5$ for the temperature noise power spectrum \citep{Ade19, choi20}. For the polarization noise power spectrum, we use $N_{\rm{red}}=N_{\rm{white}}$ (multiplied by 2 to account for polarization noise) and $\ell_{\rm{knee}}=700$ and $\alpha_{\rm{knee}}=-1.4$.
Mapping speed is calculated based on the number of beams per instrument module on the sky taking into account expected pixel yields.  

For convenience, we include sensitivities calculated both in units of Noise Equivalent Temperature (NET) and noise equivalent intensity (NEI).  The NET unit ($\mu$K$\sqrt{\rm{s}}$) is appropriate for CMB-based studies, while the NEI unit (Jy/sr$\sqrt{\rm{s}}$) is appropriate for studies based on dust or spectral line emission.  The sensitivities for the wide survey are given in CMB temperature units.
\begin{table*}[ht!]
\begin{center}
\caption{Sensitivity estimates for three surveys. \label{tab:sens}}
\setlength{\tabcolsep}{8pt}
Broadband channels wide survey (20,000 deg$^{2}$; 4,000 hours)$^a$
\begin{tabularx}{0.95\textwidth}{c c c c c c c c c}
$\nu$ & $\Delta\nu$ & Beam & $N_{\rm{det}}$ &  NEI & Sensitivity & NET & $N_\mathrm{white}$ & $N_\mathrm{red}$ \\

[GHz] & [GHz] & [arcsec] &  & [Jy sr$^{-1}\sqrt{\rm{s}}$] & [$\mu$K-arcmin] & [$\mu$K$\sqrt{\rm{s}}$] & [$\mu$K$^2$~sr] & [$\mu$K$^2$~sr] \\
 \hline
 \hline
220 & 56 & 59 & 7938 & 3300 & 15 & 6.8 & 2.0$\times 10^{-5}$ & 0.021 \\
280 & 60 & 47 & 10368 & 5500 & 28 & 13 & 6.8$\times 10^{-5}$ & 0.14 \\
350 & 35 & 37 & 20808 & 14600 & 107 & 48 & 9.4$\times 10^{-4}$ & 3.6 \\
410 & 30 & 32 & 20808 & 37300 & 407 & 182 & 1.4$\times 10^{-2}$ & 22 \\
850 & 97 & 15 & 20808 & 479000
& 680000 & 310000 & 3.9$\times 10^{4}$ & 8.1$\times 10^{6}$ \\
\hline
\vspace{0.05in}
\end{tabularx}

\setlength{\tabcolsep}{8pt}
Broadband channels for star-formation survey in first quartile PWV (100 deg$^2$; 500 hours)$^b$
 \begin{tabularx}{0.95\textwidth}{c c c c c c c}
$\nu$ & $\Delta\nu$ & Beam & $N_{\rm{det}}$ & NEFD/beam & $\sigma_\mathrm{survey}$/beam  & $\sigma_\mathrm{survey}$/beam\\

[GHz] & [GHz] & [arcsec] & & [mJy$\sqrt{\rm{s}}$/beam] & Q1: [$\mu$Jy/beam] & Q1-Q3: [$\mu$Jy/beam] \\
 \hline
 \hline
220 & 56 & 59 & 7938  & 14/17/23 & 110 & 130 \\
280 & 60 & 47 & 10368 & 16/21/30 & 140 & 170 \\
350 & 35 & 37 & 20808 & 37/52/94 & 270 & 370 \\
410 & 30 & 32 & 20808 & 66/120/230 & 560 & 810 \\
850 & 97 & 15 & 20808 & 170/400/1700 & 3100 & 4900 \\ 
\hline
\vspace{0.05in}
\end{tabularx}
\setlength{\tabcolsep}{8pt}
DSS:  Selected spectrometer channel parameters$^{c}$
\begin{tabularx}{0.95\textwidth}{c c c c c c c c c}
$\nu$ & [\ion{C}{2}] &$\Delta\nu$ & Beam & $N_\mathrm{beams}$ & NEFD/beam & NEI & $N_\mathrm{white}$ \\

[GHz] & redshift & [GHz] & [arcsec] & & [mJy$\sqrt{\rm{s}}$/beam] & [Jy sr$^{-1}\sqrt{\rm{s}}$] & [Mpc$^3$ Jy$^2$ sr$^{-2}$]  \\
 \hline
 \hline
220 & 7.64 & 2.2 & 58 & 6912 & 68/81/110 & 13,000 & 2.6$\times 10^{9}$ \\
280 & 5.79 & 2.8 & 48 & 6912 & 64/81/120 & 18,000 & 4.9$\times 10^{9}$ \\
350 & 4.43 & 3.5 & 37 & 6144 & 100/140/260 & 53,000 & 3.9$\times 10^{10}$ \\
410 & 3.64 & 4.1 & 33 & 6144 & 140/230/470 & 98,000 & 1.2$\times 10^{11}$ \\
 \hline
\end{tabularx}
\vspace{0.05in}

\end{center}
\small{
$^a$ {The top table gives the properties, instantaneous sensitivities, and noise power spectrum parameters of broadband channels for the wide area survey (20,000 deg$^2$). All temperatures are given in CMB temperature units. $N_{det}$ is the total number of KID detectors. For each beam, there is one detector per polarization, so the total number of beams on the sky $N_{\rm{beams}} = N_{\rm{det}}/2$.  In all calculations, we assume a conservative $80\%$ yield for the detectors so that the actual number of active beams on the sky is $0.8 \times N_{\rm{det}}/2$. The detector quantum efficiency is taken as $80\%$.  The quoted sensitivities are the weighted average of the top three quartile (Q1, Q2, Q3) weather expectations. We use the zenith precipitable water vapor (PWV) quartile estimates derived from $350$ $\mu{\rm m}$ radiometer measurements made over a 5 year period at the FYST site (\citep{Radford2016} as recalibrated by \citep{Cortes2020}). The values are 0.36, 0.67, and 1.28 mm zenith PWV in Q1, Q2 and Q3, respectively, averaged over the year. Sensitivity estimates refer to source elevations of 45$^{\circ}$ for all three tables.}\\
$^b$ {The middle table broadband sensitivities refer to the smaller field (100 deg$^2$) pilot DSFG survey (\ref{sec:DSFG}), where the relevant units are Jy/beam.  In the fifth column, the noise equivalent flux density (NEFD) is quoted for weather quartiles Q1/Q2/Q3.  Columns 6 and 7 list the expected noise per beam in the star-formation survey if the observations restricted to Q1 or all three weather quartiles (Q1-Q3) respectively.}\\
$^c$ {The bottom table gives properties and sensitivities at four representative EoR-Spec frequencies. Here we feed both polarizations into a single KID so that $N_{det}$ = $N_{beams}$. As for the middle table, the NEFD per beam in a given resolution element is quoted for Q1/Q2/Q3.  The NEI is the average over the field of view and is taken as a weighted average over the top three weather quartiles. Calculation of $N_\mathrm{white}$ from NEI accounts for scanning of the full spectral range, so the 4000 hour survey time is divided up across the 42 spectral elements for each order of the EoR-Spec FPI. Note that the resolution bandwidth of EoR-Spec is equal to $\Delta\nu = \nu/R$ but the noise bandwidth is $\pi/2$ larger, since the FPI spectral profile is Lorentzian in form.}\\
}

\end{table*}

\section{Surveys Mapping into Science}
\label{sec:surveys}
Multiple complementary surveys will be pursued with the seven instrument modules motivated by the variety of science programs outlined below. Different science programs call for different science survey strategies and instrument modules. It is critical to merge these surveys with an eye towards synergies into a complete plan covering relevant sidereal times during the year with high observing efficiency. An overview of the planned surveys is presented in Table \ref{tab:surveys}.  
\begin{table*}[ht!]
{\small
\hfill{}
\small
\begin{center}
\vspace{-.1in}
\caption{Overview of planned survey regions (see Fig.~\ref{fig:surveys}) and observing parameters.
\label{tab:surveys}}
\begin{tabularx}{0.95\textwidth}{l l c c l l c}

Survey     & Field(s) or & Area        & Time   & Science case & Supporting   & Section \\
           & targets         &  [deg$^2$]  & [$10^2$\,hr]   &       &  surveys$^a$ & number \\
\hline
\hline
DSS       & E-COSMOS       & 4 &  20 & Reionization LIM    & 1,2 & $\phn$\ref{sec:IM} \\
          & $\&$ E-CDFS    & 4 &  20 & Galaxy Evolution &     & $\phn$\ref{sec:highz} \\
          &                &                 &       & Time Domain Astrophysics &  & \ref{sec:YSO} \\
\hline
WFS       & AdvACT/SO      & 20,000 & 40 & Primordial Gravitational Waves & 3 & $\phn$\ref{sec:dust}\\
 & & & &                                  Galaxy Evolution &  & $\phn$\ref{sec:highz}\\
 & & & &                                  Galactic Polarization Science &  & $\phn$\ref{sec:bfield}\\
  & & & &                                 Galaxy Cluster Evolution via SZ  &  & $\phn$\ref{sec:sz}\\
   & & & &                                Rayleigh Scattering &  & $\phn$\ref{sec:rs}\\
   & & & &                                Time Domain Astrophysics &  & \ref{sec:YSO}\\ 
 \hline
 CIB-Mid  & GAMA & 100 & $\phn$5 & CIB Galaxy Evolution &  &$\phn$\ref{sec:highz}\\

\hline
GalPol & MW $\&$ LMC & 125 & 5.25 & Galactic Polarization Science &  &$\phn$\ref{sec:bfield}\\
 & ISM $\&$ clouds& &  &  & \\
\hline
PS &  MW clouds & 10$\times$4  & 5 & Episodic Protostellar Accretion &  &\ref{sec:YSO}\\
Monitor   & $\&$ protostars &  & & &  &\\ 
\hline
\end{tabularx}
\end{center}}
{\small {$^a$(1) Deep Subaru HSC+PFS spectroscopy (current) and
COSMOS X-Ray-to-meter-wave multiwavelength survey; (2) deep Euclid
grism spectroscopy (upcoming), HERA HI 21\,cm (upcoming),
andH-UDF/CDF-S multiwavelength surveys (including JWST GTO);
(3) \citet{planck_dust}, SDSS, DES, ACT (current), SO, DESI, \emph{Rubin}, and
eROSITA (upcoming). }}
\end{table*}
 FYST will support two distinct science instruments during normal operations.  The first, described here, is \pcam with its camera and broadband, low resolving power direct-detection spectrometer modules. The second is the CCAT Heterodyne Array Instrument (CHAI, \citealt{Graf2019}) which is designed for high-efficiency velocity-resolved mapping of important diagnostic lines from the Galaxy and nearby galaxies in the high-frequency submillimeter windows.  Approximately 3/4 and 1/4 of the FYST time on the sky will be devoted to \pcam and CHAI observations, respectively. We plan block campaigns for \pcam and CHAI instrumentation, the fundamental time unit of a block is likely to be 2 weeks. The two instruments cannot be operated simultaneously. 
   \subsection{Operations}
  \label{sec:survey_operations}
 Based on long term Chajnantor weather statistics we estimate about 2500 hours/year will be available for \pcam science observations. Within our 5-year plan, the total available time will be about 12,500 hours.  We expect to devote up to 2500 hours to commissioning and Early Science activities and the remaining 10,000 hours will be devoted to our Baseline Science, described below.  Most of this time (8000 hours) will be spent on two large surveys, while an additional 2000 hours will be spent on focused and unique observational niches (e.g., location on sky, depth, or cadence) surveys.
\\
\\
  \subsection{Survey Overview}
  \label{sec:survey_details}
 Our two large surveys (4000 hours each) are the \textbf{Wide Field Survey (WFS)}, and the \textbf{Deep Spectroscopic Survey (DSS)}. There are also smaller-scale surveys that we outline below.
 
 The WFS is a 4000-hour wide-field survey, covering all declinations from $-61^{\circ}$ to $+18^{\circ}$ over the entire range of right ascension.  This survey will uniformly cover over 20,000 deg$^{2}$ and have high synergy with lower-frequency cosmological surveys envisioned for the Simons Observatory and CMB-S4.  It is optimized to address the science programs that we outline below: ``Measuring CMB Foregrounds to Aid the Search for Primordial Gravitational Waves'' (Section \ref{sec:dust}); ``Probing Galaxy and Galaxy Cluster Evolution with Sunyaev-Zeldovich Effects'' (Section \ref{sec:sz}); and ``Rayleigh Scattering: a New, Blue Surface of Last Scattering'' (Section \ref{sec:rs}).  Its broad spatial coverage includes a large fraction of the Milky Way Galaxy, enabling much of the ``Galactic Polarization Science:  Magnetic Fields and Dust Properties'' (Section \ref{sec:bfield}) and the Time Domain Astrophysics (Section \ref{sec:YSO}) science programs as well.   
The DSS is a 4000-hour survey that will yield very deep spectroscopic images of two fields (2000 hours each). It is designed to measure the growth of galaxies and large-scale structure as well as the process of reionization in the Universe through [\ion{C}{2}] and [\ion{O}{3}] line intensity mapping (LIM), outlined in Section \ref{sec:IM} ``Reionization Intensity Mapping''.  This LIM will focus on two $\sim$4 deg$^{2}$ fields: the Extended-COSMOS (E-COSMOS; \citealt{aihara18}); and Extended-CDFS (E-CDFS) field (\citealt{lehmer05}, which includes the Hubble Ultra Deep Field, or H-UDF; \citealt{beckwith06}).

These two major surveys deliver most of the science outlined in Section \ref{sec:highz} ``Tracing Galaxy Evolution from the First Billion Years to Cosmic Noon with the Cosmic Infrared Background'', Section \ref{sec:bfield} ``Galactic Polarization Science: Magnetic Fields and Dust Properties'', and Section \ref{sec:YSO} ``A New Submillimeter Window into Time Domain Astrophysics''.  Galaxy evolution studies have both shallow/large and deep/small area science requirements that are very well addressed by piggybacking its 850 GHz imaging onto both Surveys above.  The remaining time ($\sim$2000 hours) will be spent commissioning instrument modules and on first-light and/or demonstration science.  
 
 Each of the science programs outlined below will have components that cover either smaller fields or fields with shallower integration time requirements oriented towards first light and/or demonstration science. The Galaxy Evolution program requires the addition of the mid-scale, mid-depth Galaxy and Mass Assembly (GAMA, www.gama-survey.org) fields (CIB-Mid). The Galactic Polarization Science requires eight small-field observations of Galactic star-formation regions and the Large Magellanic cloud, with deep integrations totalling about 525 hours (Gal/LMC SF). The Time Domain Astrophysics program requires repeated short (1/2 hour) observations of 10 fields about 20 times per year, totalling about 500 hours (PS Monitor). An overview of the planned surveys is presented in Table \ref{tab:surveys}.  
 
 \subsection{Scan Patterns}
 \label{sec:scanpatterns}
Prime-Cam will be mounted on the Nasmyth focus of the altitude-azimuth mounted FYST telescope.  As such the field of view rotates with both the parallactic angle and the elevation of the telescope. Furthermore, all instrument modules except the 850 GHz camera module are off the optical axis by 1.8$^{\circ}$ so that their 1.3$^{\circ}$ fields will circulate at a radius of 1.8$^{\circ}$ about the optical axis. As such, it is not trivial 
to optimize scan strategies or to calculate scan efficiencies. Optimization involves understanding the complex interplay between instantaneous field of view, the science field rotation during an observation, cross-linking of polarization knowledge (for polarimetry), cross-linking of spectral sampling (for spectral scanning) and pointing knowledge and stability under both the uniform motion and acceleration of the telescope.  Coupled into these science requirements and telescope performance parameters are the noise spectrum of the detectors and that of the sky which fold into the scan speed required for minimizing sky noise. Sky noise is a function of wavelength and weather conditions and the required spatial scales needed to be recovered for the science case. The final scan pattern must be repeated until the desired sensitivity levels are reached. Here we define scan efficiency as the fraction of time that is spent "on-source and pointed" relative to the wall-clock time, that includes the time "off-source and/or not well pointed" which is presumed mostly to occur in the turn around events.  Clearly these are complex optimizations. Much thought has already been invested in WFS-like surveys with constant elevation scans by Advanced ACTPol, SO, and CMB-S4 \citep{stevens/etal:2018}, and the WFS scan strategy will most likely employ a similar strategy. For these surveys, we estimated scan efficiencies in excess of 90\%. The smaller-scale surveys (e.g., DSS) involve fields not much bigger than the instrument module field of view. For these fields we will most likely adopt ``daisy" or ``pong" patterns widely used at the JCMT \citep{Holland2013}.  Preliminary estimates of scan efficiences for these modes are roughly 70\%.

 We have described the instrument and its capabilities and our survey strategies. The following sections discuss the seven primary areas of scientific focus that are enabled by this instrumentation and these surveys.

\begin{figure*}[ht!]
\begin{center}
\includegraphics[width=6.0in]{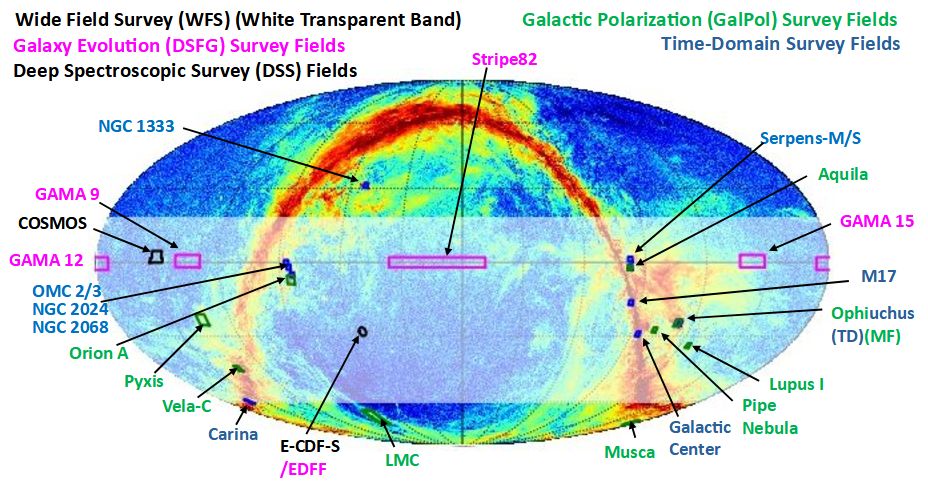}
\end{center}
\caption{ Overview of \pcam survey fields in equatorial coordinates overlaid on the $\emph{Planck}$ dust polarization map \citep{planck_dust}. The planned wide-field survey field for the CMB foregrounds, galaxy cluster evolution and Rayleigh scattering science cases is approximated by the white transparent band (overlapping with Advanced ACTPol and SO fields \citep{debernardis/etal:2016, stevens/etal:2018}). The deep intensity mapping/reionization surveys, the galaxy evolution survey fields, the Galactic polarization science targets, and the specific time-domain science fields are outlined in black, magenta, green and blue frames respectively. See Tables \ref{tab:surveys},  \ref{tab:poln_surveys} and \ref{tab:PS-var} for survey coordinates, areas, and planned integration times.}
\label{fig:surveys}
\end{figure*}

\section{Reionization, structure growth, and galaxy emergence through line intensity mapping}
\label{sec:IM}
Intensity mapping is a technique that measures the spatial
fluctuations of signals due to large-scale structure at low spatial
resolution. The perhaps best-known example of this application are the
temperature and polarization fluctuation maps of the cosmic microwave
background (CMB) as measured by $\emph{COBE}$, $\emph{WMAP}$, and $\emph{Planck}$. While CMB maps are integrated along the line of sight and thus do not contain
three-dimensional information without other assumptions, intensity
mapping studies in spectral lines can recover redshift structure (and
therefore measure signals at early epochs as a function of cosmic
time). Compared to other intensity mapping techniques, spectral line intensity mapping therefore has the advantage of providing three-dimensional spatial information about the sources of emission (or absorption) that can be used to probe the processes of structure formation.  Intensity maps can also be used as cosmological probes, since the fluctuations in the emission/absorption lines are correlated with the underlying dark matter density fluctuations. 

In Section \ref{sec:CII IM}, we describe the background and general concept of the intensity mapping survey planned with Prime-Cam. In Section \ref{sec:CO IM}, we provide additional details on science to be extracted from foregrounds in the survey. In Section \ref{sec:SS IM}, we describe the detailed survey strategy, and provide information on the ancillary data available in the survey fields. In Section \ref{sec:PS IM}, we give a detailed description on the methods used to extract the main survey signal. In Section \ref{sec:FR IM}, we give a detailed description on how to separate the main signal from foreground emission. 

\subsection{[\ion{C}{2}] and [\ion{O}{3}] Intensity Mapping in the Epoch of Reionization}
\label{sec:CII IM}
The Epoch of Reionization (EoR), approximately $6 \leq z \leq 8$ to $20$, is the last largely unexplored cosmic epoch in which stars and galaxies govern and shape the overall properties of the Universe \citep{fan06}. 
With telescopes like {\em HST} and ALMA (and {\em JWST} in the near future), it has become possible to probe small samples of
star-forming galaxies in the EoR. These studies provide compelling
evidence that the first galaxies, which formed within the first billion
years, are likely to be mostly responsible for cosmic reionization \citep[e.g.,][]{robertson15}.  However, the
overall properties of the sources of reionization and the connection between large-scale structure and the topology of reionization remains poorly understood \citep{finlator09} because the main sources of reionization are numerous but intrinsically extremely faint \citep[e.g.,][]{robertson13}. It is therefore challenging to detect them individually. To make significant progress, large-scale three-dimensional surveys of the universe during the EoR are of fundamental importance. Such surveys become possible by employing new observational strategies, in particular spectral line intensity mapping \citep{kovetz17}.  
Since intensity mapping detects the {\em aggregate} signal from faint but numerous sources, the sensitivity requirements are less demanding than for individual source detection studies such as with ALMA or {\em JWST}, providing the potential for a shortcut toward detection with a carefully designed experiment (see model predictions by, e.g., \citealt{silva15,yue15,2019MNRAS.488.3014P}).

21\,cm \ion{H}{1} EoR mapping was proposed as such a pathway a long time ago.
Indeed, as shown by the claimed recent detection of the unexpectedly strong \ion{H}{1} 21\,cm absorption signal at $z\sim 17$ due to the first stars near the onset of reionization \citep{bowman18}, our current understanding is still incomplete, making direct measurements of all phases of reionization critical. These experiments are very challenging and easily contaminated by strong foregrounds. Also, while a firm detection of the cosmological \ion{H}{1} signal should be within reach with pathfinder experiments like HERA \citep{deboer17}, a very large facility like the full Square Kilometer Array \citep[SKA;][]{santos15} is needed to reach the required mapping capabilities. 

Promising complementary and alternative probes of the EoR are [\ion{C}{2}] 158\,$\mu$m and [\ion{O}{3}] 88\,$\mu$m line-intensity mapping. The clear advantage of using these lines are that they much stronger than the \ion{H}{1} 21\,cm line, and it takes only a small single-dish telescope with wide field of view and modest resolution spectroscopic capabilities to map the line intensities. 
Additionally, these observations will not be hampered by radio frequency interference (RFI), a major concern for \ion{H}{1} surveys. 
In contrast to traditional UV/optical diagnostic tracers (e.g., Ly$\alpha$ and H$\alpha$) that will be accessible to {\em SPHEREx}, the [\ion{C}{2}] and [\ion{O}{3}] signal will not be subject to absorption due to the increasingly neutral IGM when probing deeper into the EoR or due to dust extinction in galaxies and along the line of sight \citep[e.g.,][]{silva15,yue15}. As described in more detail in Section \ref{sec:CO IM}, CO will be a powerful probe of structure at low to intermediate redshifts, up to the peak epoch of galaxy formation at $z$=1--4. One concern for CO at higher redshifts is that the observable CO luminosity per
unit H$_2$ mass is strongly affected by metallicity (e.g., \citealt{Bolatto2013}), which is expected to
steeply drop in the EoR when galaxies first form. Thus, fine structure lines like [\ion{C}{2}] and [\ion{O}{3}], which are significantly less affected, are perhaps the most promising pathway towards obtaining tomographic maps of the EoR in the near-term future. While [\ion{C}{2}] typically is the strongest ISM emission line in star-forming galaxies \citep{stacey/1991}, [\ion{O}{3}] 88\,$\mu$m can be even stronger than [\ion{C}{2}] in low-metallicity galaxies (e.g., \citealt{cormier15}), but it is expected to be weaker than the (factor of $\sim$2) lower-redshift [\ion{C}{2}] line at the same observing frequencies. Compared to the 21\,cm \ion{H}{1} signal, which is primarily sensitive to the intergalactic medium (IGM), [\ion{C}{2}] and [\ion{O}{3}] have the advantage that they enable direct mapping of the distribution of the sources of reionization. 

\begin{figure*}[t!]
\begin{center}
\vspace{-0.5in}
\includegraphics[width=7.0in]{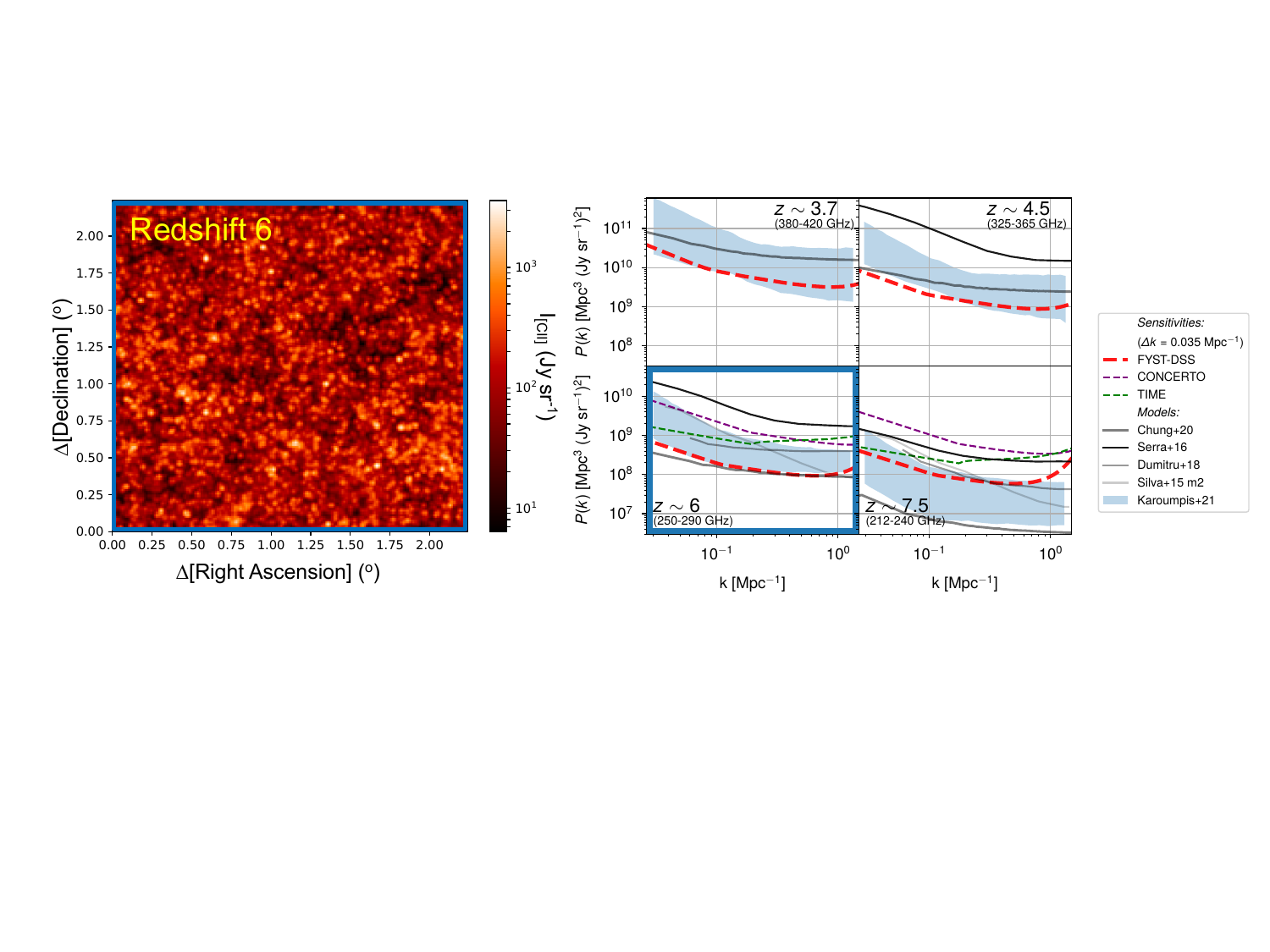}
\end{center}
\vspace{-2.25in}
\caption{\small 
({\it left}) Simulated ``redshift slice'' of the area covered by the [\ion{C}{2}] intensity mapping survey in the epoch of reionization (EoR), representing a small spectral bin (2.7\,GHz wide, or $\Delta\,z$=0.07 at the adopted redshift) within the vast $z=3.5$--8.0 redshift range covered by the EoR-Spec DSS [\ion{C}{2}] measurements. ({\it right})  Power spectrum of the [\ion{C}{2}] emission, revealing the topology of reionization and the [\ion{C}{2}] luminosity functions at redshifts of 3.7 to 7.5. The range of recent model predictions (solid lines; \citealt{silva15,Serra16,2019MNRAS.485.3486D,Chung20}) differ by factors of 10 to 50, and an ensemble of state-of-the-art models from Karoumpis et al.~2021 (in prep.) spans an order of magnitude in power spectrum predictions at each redshift. Overlaid are our predicted EoR-Spec sensitivities. Our proposed survey is as much as 10$\times$ more sensitive than other proposed surveys (TIME; \citealt{crites14,Sun2021}; CONCERTO; \citealt{2020A&A...642A..60C}; raw sensitivities are shown for all experiments) leading to an expectation for detection of the intensity mapping signal at redshifts approaching 8 for most models. ALMA cannot make these measurements, since its field of view is smaller than a single pixel of EoR-Spec on FYST so that it requires enormous numbers of pointings to map the large spatial scales required for these measurements. Furthermore, an interferometer is insensitive to extended low surface brightness emission and individual faint sources below its point source detection limit, which would be recovered by the intensity mapping signal.}
\label{fig:IM}
\end{figure*}

The EoR-Spec DSS survey is designed to obtain a tomographic map of the EoR through fluctuations in the aggregate clustering signal encoded in the [\ion{C}{2}] cooling radiation from faint star-forming galaxies from $z\sim$ 3.5--8.05 (e.g., \citealt{riechers14b,hashimoto18}; see Fig.~\ref{fig:IM}), and through cross-correlation of [\ion{C}{2}] and [\ion{O}{3}] at $z>7$ (and cross-correlation of [\ion{C}{2}] and [\ion{O}{3}] at lower redshift from balloon-based surveys similar to EXCLAIM \citealt{EXCLAIM}; see also \citealt{Pad21}). This can be done most efficiently with intermediate size telescopes like FYST that have a spatial resolution matched to the key structures and very wide fields-of-view and can therefore map much more quickly than larger facilities. We will map a $\gtrsim8\,\rm{deg}^2$ region over at least two deep multi-wavelength extragalactic surveys fields (E-COSMOS and CDF-S/H-UDF;
see Fig.~\ref{fig:surveys}, Table \ref{tab:surveys}, and Sect.~\ref{sec:SS IM}).
The full survey area will be sufficient to obtain a significant detection of the clustering signal in [\ion{C}{2}] out to $z$ $\sim$ 5--8 (e.g., \citealt{silva15,yue15}; 
see Fig.~\ref{fig:IM}), 
and to enable cross-correlation at the high redshift end with 
\ion{H}{1} 21\,cm maps from experiments like HERA covering some of the same fields. While direct detection of [\ion{O}{3}] will likely be more challenging than [\ion{C}{2}], cross-correlation of [\ion{O}{3}] and \ion{H}{1} could be a promising avenue to push to even higher redshifts.

\begin{figure}[t!]
\begin{center}
\includegraphics[width=3.5in]{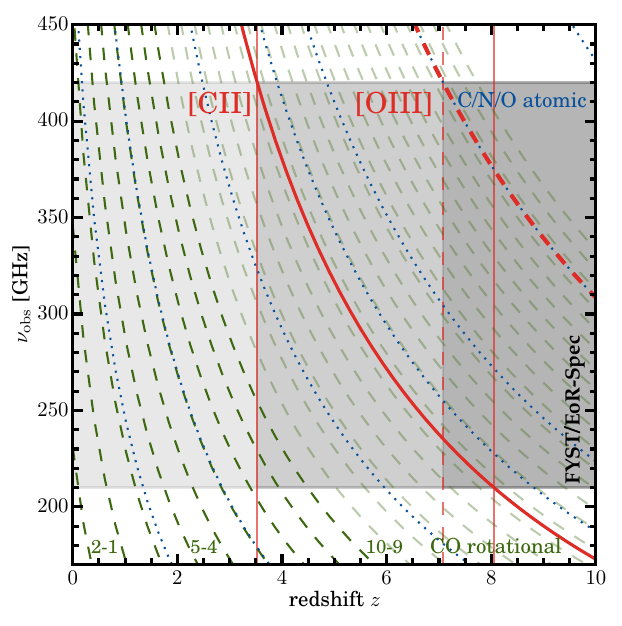}
\end{center}
\vspace{-0.25in}
\caption{\small 
Lines accessible to EoR-Spec at different redshifts (gray shaded regions). Dashed green lines correspond to transitions of the CO rotational ladder, where lighter shading is used for lines that will likely be faint. All these CO lines are the foreground (Section \ref{sec:FR IM}) to the [\ion{C}{2}] and [\ion{O}{3}] lines (solid and dash-dotted red lines; Section \ref{sec:CII IM}) targeted by the main experiment, but they will also provide line intensity maps at intermediate redshifts (Section \ref{sec:CO IM}). Fainter fine-structure lines due to [\ion{C}{1}], [\ion{O}{1}], and [\ion{N}{2}] are indicated as dotted blue lines.
}
\label{fig:IM2}
\end{figure}

\subsection{CO Intensity Mapping from Present Day to the Peak Epoch of Star Formation}
\label{sec:CO IM}

While the bright [\ion{C}{2}] line from the EoR is the primary intensity mapping target for the EoR-Spec DSS survey, many other emission lines from different redshifts fall into the same frequency bands (Fig. \ref{fig:IM2}).  Among the most important of these are lines emitted by carbon monoxide molecules.  CO is the second most abundant molecule in the ISM and has a series of rotational transitions that are easily excited in typical molecular cloud conditions.  The resulting emission lines are evenly spaced in energy, with frequencies $115J_{u}$ GHz, where $J_{u}$ is the upper angular momentum quantum number of the transition.  Because CO lines are both bright and relatively accessible from the ground, they provide the primary means we have of probing molecular gas properties, and since molecular gas is the ISM phase that immediately precedes SF, observations of CO play a crucial role in understanding the ecosystems within and of galaxies across cosmic history \citep[see reviews by, e.g.,][]{Kennicutt2012,Bolatto2013,Carilli2013,Heyer2015}.

EoR-Spec DSS intensity mapping with CO can therefore open a unique window into the molecular universe.  CO lines from the (2$-$1) transition and above fall into the EoR-Spec frequency band, with the brightest lines spanning $z\sim0-2$.  As discussed below, these lines are typically treated as foreground contaminants to the high-redshift [\ion{C}{2}] signal \citep{silva15,Sun2018}, but they can also serve as an interesting science target in their own right.  Dedicated CO intensity-mapping experiments exist, primarily targeting the CO(1-0) transition at $z\sim3$ \citep{Keating2016,Li2016}, but the EoR-Spec DSS offers two distinct advantages over these surveys.  First, it is sensitive to a unique redshift range, enabling continuous tomographic mapping of galaxies from cosmic noon to the present day \citep{Madau2014}.  Second, the wide frequency coverage of EoR-Spec means that there are several different CO transitions accessible at each redshift.  This enables the recovery of more detailed galaxy properties than can be obtained from single-line observations \citep{Kamenetzky2018}.

Just as the CO transitions are a foreground to the [\ion{C}{2}] line, care must be taken to separate out individual transitions from the full ladder of lines.  This is typically done using cross-correlations, using the fact that two populations at the same redshift will trace the same large-scale structure, while interlopers at other redshifts will not be correlated \cite[see, e.g.,][]{Switzer2019}. In addition to removing contaminating emission, these cross-correlations also provide the opportunity for further science output.  Spectroscopic galaxy maps are available in abundance in the EoR-Spec target fields, and correlating these surveys with our intensity maps will allow us to isolate the CO properties of specific galaxy populations \citep{Wolz2017}.  This type of measurement has many uses, such as probing the interaction between AGN and their host galaxies \citep{Breysse2019}.  One can gain additional information through internal cross-correlations within the EoR-Spec data.  By matching up frequency bands corresponding to different CO lines at the same redshift, one can not only isolate specific pairs of lines, but also obtain a measurement of how the intensity of one line varies with respect to another, allowing for comprehensive studies of high-redshift galaxy properties \citep{breysse17,2019ApJ...887..142S,2021JCAP...05..068S}.

\subsection{Survey strategy and ancillary data}
\label{sec:SS IM}
To maximize the potential for cross-correlation with spectroscopy at other wavelengths, it is critical to survey the fields that have both the most sensitive observations across the broadest coverage in wavelength bands from X-ray to meter wavelengths and that have the densest spectroscopic coverage across fields that are many degrees in size on the sky (including, where possible, overlap with the best cosmological \ion{H}{1} 21\,cm surveys). The combination of data sets enables removal of foreground sources as well as the extraction of the ancillary physical properties of some of the galaxies that contribute to the line intensity-mapping signal. These considerations have resulted in the choice of the E-COSMOS and E-CDFS (including the H-UDF) fields as the highest priority survey fields for the line intensity mapping surveys with EoR-Spec. The UDS/SXDS, and \textit{Euclid} SEP fields and a region around 18$^{h}$ in the HERA constant declination strip serve as backup fields. E-COSMOS and E-CDFS will have deep $\it{Y}$+$\it{Z}$+$\it{HK}$ band coverage, including grism spectroscopy across 4.4 and 10\,deg$^2$ from \textit{Euclid}, and deep/ultra-deep optical Subaru/HSC griz\textit{Y} + 4 narrow-band imaging and PFS spectroscopy as part of the HSC strategic program and Hawaii-Two-0 (H20) surveys -- both of which are essential for the selection of the highest-redshift galaxies. The \textit{Euclid} deep fields are expected to lead to the identification of thousands of galaxies at $z \sim 8$. The same E-CDFS area is covered by the 6000\,hour {\em Spitzer}/IRAC SLS survey at 3.6 and 4.5\,$\mu$m in preparation for \textit{JWST}, \textit{Roman} and \textit{Euclid}. The \textit{Euclid} deep field has also been selected for deep imaging in the LMT/TolTEC LSS survey at 1.1, 1.4, and 2.1\,mm and additional deep optical/near-infrared imaging with CFHT and UKIRT. Thus, by the time of first light of FYST, the multi-wavelength coverage in these fields will extend far beyond their current nominal sizes, providing the necessary ancillary data over the full anticipated EoR-Spec field sizes (4\,deg$^2$ each). The E-CDFS field also falls within the much larger HERA 21\,cm constant declination survey strip.

The EoR-Spec DSS survey will scan the full 210--420\,GHz\footnote{Data near 325 and 375\,GHz will likely be lost due to poor atmospheric transmission.} frequency range at a spectral resolution of $R \sim 100$. As shown in Fig.~\ref{fig:IM2}, this will give access to the [\ion{C}{2}] 158\,$\mu$m line at $z \sim 3.5-8.05$, to the [\ion{O}{3}] 88\,$\mu$m line at $z>7$, and to multiple CO lines across virtually the entire redshift range (with bright transitions accessible out to $z \sim 4-5$). Coverage of the 4\,deg$^2$ fields will be achieved by scanning spatially at a specific FPI setting, then every few spatial scans, stepping the FPI, using 13 steps to cover the full redshift range in every spatial pixel every 2 to 3 minutes (see \cite{cothard/2020} for details). The E-COSMOS field will be visible at $>$40\,$^{\circ}$ elevation for 6\,hours each day, while the E-CDFS field will be visible for 7.5\,hours.  The remaining LST range will be filled by either switching to another \pcam survey
or by observing a field at 18$^{h}$ in the HERA 21\,cm strip, contingent on weather and instrument readiness. The highest priority survey fields will be observed for a total of 4000\,hours in 2023-28, with additional observations of the HERA 18$^{h}$ backup "filler" field as time is available. This will result in a nominal sensitivity of 3,500 Jy/sr per spectral resolution element (a 3000 km/sec "bin") in each 58$\arcsec$ beam of the (total) 8\,deg$^2$ field. This corresponds to a representative large-scale signal of 0.02\,MJy\,sr$^{-1}$\,bin$^{-1}$ at 220\,GHz (i.e., [\ion{C}{2}] at $z$=7.6; see Table \ref{tab:sens}).

\subsection{Methods to derive power spectra and cross spectra from raw data}
\label{sec:PS IM}

    While serendipitous detection of individual [\ion{C}{2}] sources may be possible, the central aim of intensity mapping is to statistically capture fluctuations in the integrated [\ion{C}{2}] line intensity across the surveyed comoving volume (cf. \citealt{kovetz17}). Currently, both models and experiments are focused on a possible detection of these fluctuations not in map space but in Fourier space (in much the same way that cosmic background surveys and galaxy surveys measure and analyze angular or 3D power spectra). The main observable is the spherically averaged power spectrum $P(k)$ of the line intensity field across the comoving volume as a function of comoving wavenumber $k$. Single-dish surveys like for \pcam are typically sensitive to $P(k)$ at lower values of $k$, corresponding to large-scale structure and clustering of line emitters.

    The relative lack of information about high-redshift star-formation activity and its relationship to line emission both provides motivation for [\ion{C}{2}] intensity mapping and leads to a wide range of signal forecasts (see Fig. \ref{fig:IM}). All of these forecasts rely on the galaxy-halo connection (as reviewed in~\citealt{WechslerTinker18}) --- relating star-formation activity per galaxy to host dark matter halo properties --- and on a presumed conversion between star-formation rates (SFR) (or IR luminosities) and [\ion{C}{2}] luminosities of galaxies. For instance, \cite{Serra16} use a halo model for CIB anisotropies to connect halo mass to IR luminosity and a local calibration from~\cite{Spinoglio12} to relate IR luminosity to [\ion{C}{2}] luminosity. More recently,~\cite{Chung20} combined the empirical model of~\cite{Behroozi19} with the high-redshift [\ion{C}{2}] simulations of~\cite{Lagache18b} (as opposed to an observational but low-redshift result). Very recently, C.\ Karoumpis et al.~2021 (in prep.) produced a set of alternative $P(k)$ predictions starting from a common dark matter (DM) cone built using the DM halo catalog from the Illustris TNG300-1 simulation \citep{springel18}. They applied different models of occupying the DM halos with mock galaxies as well as multiple SFR-to-$L_{[C\,II]}$ coupling relations, predicting a range of possible values of the $P(k)$ during the EoR.

    The resulting forecasts for the [\ion{C}{2}] power spectrum span at least an order of magnitude, with the \cite{Serra16} forecasts on the more optimistic side, the \cite{Chung20} forecasts on the more conservative side, and the C.\ Karoumpis et al.~2021 (in prep.) forecasts spanning the range between them. However, even with these substantial uncertainties in the expected signal, we expect the fundamental sensitivity of the EoR-Spec survey (as shown in Figure~\ref{fig:IM}; updated from~\citealt{Chung20} based on current estimates for FYST-DSS from Table 1, as well as more recent estimates for CONCERTO based on~\cite{CONCERTO2020}) to enable strong detections of [\ion{C}{2}] emission anisotropies out to $z\sim5$. Non-detections at $z\gtrsim6$ is a possible outcome, but would nonetheless be valuable in constraining a model space that is currently wide open.

\subsection{Foreground rejection and extraction of CO signal}
\label{sec:FR IM}
One of the most critical technical challenges for [\ion{C}{2}] and [\ion{O}{3}] intensity mapping studies is to extract the large-scale information in the presence of foreground line emission. There are two kinds of foregrounds: the continuum; and the line emission interlopers \citep[e.g.][]{yue15}.

The continuum foregrounds, far-infrared (FIR) radiation emitted from interstellar dust grains, are spectrally smooth and limited to the larger-scale Fourier modes. To avoid them, we will initially exclude this confused part of the  Fourier space from our statistical analysis \citep[e.g.][]{Cheng:Chang:Bock:2016}.

The mitigation of the line emission foregrounds is more challenging. In the case of [\ion{C}{2}] intensity-mapping of the EoR, the more important interlopers are the CO transitions. Since the line foregrounds are not spectrally smooth like the continuum foregrounds, we need different approaches to remove them. Several of the proposed techniques focus only on the extraction of the two-point statistics (power spectrum or correlation function), whereas others attempt to reconstruct the individual line maps. Some examples of the proposed methods are as follows:
\begin{enumerate}

    \item One approach is masking out the brightest velocity pixels (voxels) in the survey. The CO line emission comes from galaxies of lower redshift than the [\ion{C}{2}] emission. These galaxies tend to be more massive and thus brighter. As a result, the brightest voxels have a higher-probability of containing CO line emitters and by simply masking them we can decrease the foreground emission \citep[e.g.][]{Breysse2015}.  

    \item Another way to mask the line emission foregrounds is with the help of ancillary data. Using galaxy catalogs with accurate redshift information and proxies for the CO line emission of the interloping galaxies, we can identify the voxels containing the more luminous CO lines emitters and mask them \citep[e.g.][]{yue15,Sun2018}.

    \item \cite{Cheng2020} use the fact that multiple CO lines that are emitted by a source are observable in a LIM survey. They demonstrate that the redshift of a source can be derived by fitting to a set of spectral templates that are unique at each redshift. In this way, individual line maps can be retrieved without using any external tracers or spectroscopic follow-up observations.

    \item An alternative technique is based on projecting the LIM data cube from the intrinsic observing angular and spectral coordinates to the comoving coordinate frame of the targeted [\ion{C}{2}] emission. Because in this projection we assume a redshift for the target, the interloper emission fluctuations are mapped to the wrong spatial scales. This makes the interloper emission highly anisotropic. \cite{Cheng:Chang:Bock:2016} show that this anisotropy can be used to fit out the interloper contamination.

    \item \cite{Moriwaki2020} demonstrate that deep learning networks, after being appropriately trained with a large set of mock observations, can generate accurately the intensity distribution of the targeted line.

    \item Finally, in the frequency range covered by the EoR surveys, there are sets of frequency channels that contain emission from two or more adjacent CO lines, originating from the same redshift. When cross-correlating these channels, since only the two lines emitted from the same redshift will be correlated, we just measure the CO foreground, which we can then remove \citep[e.g.][]{Sun2018}.
\end{enumerate}

As described in Section \ref{sec:SS IM}, our high priority fields will have a broad frequency coverage. This includes both the spectroscopic redshift of the foreground galaxies and indirect tracers of their CO emission, making it possible to mask the voxels that contain them. Using these well-covered fields, we will be able to verify and improve the techniques that do not require external tracers, and adapt them for future use. The EoR-Spec DSS survey strategy thus is optimized for foreground removal to maximize recovery of the main survey signal. \\

\section{Tracing galaxy evolution from the first billion years to Cosmic Noon}
\label{sec:highz}
Much of star formation is hidden from view in the optical bands due to dust in the molecular clouds that envelop the newly formed stars. 
This dust absorbs the starlight, heats up and re-radiates the power at FIR wavelengths. 
This dust-enshrouded star formation is important over cosmic time --  the cosmic optical background (COB) power and cosmic infrared background (CIB) energy density are roughly equal \citep[e.g.,][]{Dole06,2013Burgarella}. 
Consequently, when considering the star formation history of the Universe, both optical and FIR wavelengths must be considered. 
A discrepancy in the expected infrared output, inferred from the dust masses in local galaxies, and the observed measurements of the CIB suggested first that IR sources must become more numerous at higher redshift and be a critical component of early mass build-up \citep[e.g.,][]{Lagache05}, which was further supported by studies of the most luminous ``submillimeter galaxies'' \citep[e.g.,][]{Casey2014}. 
Subsequent \textit{Spitzer} and \textit{Herschel} surveys resolved  $\sim\,$80\% of the CIB at $\lambda$$\,<\,$$160\,\mu$m ($\nu\,>\,1.9$ THz) 
into individual galaxies. 
This work provided robust constraints on the cosmic SF history at $z\lesssim1.5$\ and demonstrated that about $70\,\%$ was dust-enshrouded at these epochs \citep{2013Magnelli}. However, the dust emission from higher redshift galaxies is redshifted to longer wavelength ($\,>250\,\mu$m or $\nu\,<\,1.2$ THz) where the inability of the 3.5$\,$m \textit{Herschel} telescope to separate emission from multiple galaxies within a single spatial beam (source confusion) becomes a serious problem.  Therefore, at $z\gtrsim1.5$\, the 
fraction of CIB resolved into individual galaxies drops to only $\sim$10\%\ \citep{Bethermin12}.
Such limitation is unfortunate and critical, as in this wavelength range the CIB is dominated by $z\gtrsim1.5$ galaxies \citep{Bethermin12} while providing the best dust-obscured star formation rate proxy for this galaxy population (Figure~\ref{fig:SFH} {\it lower right}). 
Thus, to study the source of the CIB at $\lambda$$\,>\,$$160\,\mu$m in detail (understanding the physical properties of high-redshift galaxies contributing to the CIB), high-resolution IR surveys such as those done with \pcam on FYST need to be conducted.

\pcam on FYST provides the instrumentation and platform to enable studies of DSFGs at redshifts from 1 to 5 and beyond.
With the 6\,m aperture of FYST, \pcam reaches about $3$ times deeper into the source confusion than was possible with the {\em Herschel} 3.5\,m telescope, resulting in a difference in luminosity limit of a factor of $\approx$2--3, depending on which {\em Herschel}-SPIRE bands are considered (see Fig.~\ref{fig:SFH}). 
Using the model of \citet{2017Bethermin}, we predict that confusion-limited \pcam surveys (DSS and CIB-Mid) will securely detect hundreds of thousands of galaxies, with a small fraction of these ($<$0.5\%) at redshifts potentially as high as 6 (Fig.~\ref{fig:SFH}, {\it upper right}) and resolve into individual galaxies up to $\sim\,$40\% of the CIB at 850\,GHz. Specifically we estimate
$440{,}000$ individual galaxies above a signal-to-noise ratio of 5 within each of our confusion-limited ${\sim}\,100\,\deg^{2}$ survey (e.g., CIB-Mid), $1{,}300$ of which we expect to be at $z>5$ and detected primarily by the lower frequency Prime-Cam modules (220-410\,GHz).
Compared with {\em Herschel} at 350 $\mu$m ($\sim 850$ GHz), this represents $\approx10\times$ the number of sources per square degree detected above the confusion limit.

\subsection{Dusty Star-Forming Galaxy Surveys}
\label{sec:DSFG}
In its early pilot phases, our DSFG survey will cover the rich, multi-wavelength GAMA fields (${\sim}\,100\,\deg^{2}$ total) to the confusion limit at $350\,\mu$m
(850 GHz (Figure~\ref{fig:SFH}, {\it upper left inset}), making use of the best quartile in PWV. Ultra-deep integrations covering $4{-}16\,\deg^{2}$ over key galaxy evolution survey fields (COSMOS and E-CDFS) will also be obtained, in parallel alongside the EoR intensity mapping fields, and will reach far below the confusion limit, providing a precise measurement of the confusion noise on the scales most relevant to \pcam. The full survey will also cover the ${\sim}\,100\,\deg^{2}$ extended SDSS Stripe-82 area (previously covered only shallowly by \emph{Herschel}-SPIRE) to comparable depths. 

Its deep, yet wide-field ($100\,\deg^2$; e.g., CIB-mid) survey will sample environments on large scales that cannot be mapped with ALMA, and reach down to luminosities that remained inaccessible to {\em Herschel} (Fig.~\ref{fig:SFH}, {\it upper right}). 
The 850\,GHz band is critical to extract the infrared luminosities and subsequently the dust-obscured star formation rates to far greater precision than possible with long-wavelength surveys (e.g., LMT; Fig.~\ref{fig:SFH}, {\it lower right}) alone.
For example, the first followup of 350\,GHz-selected galaxies with {\em Herschel}-SPIRE at these high frequencies constrained the peak of their dust spectral energy distributions (SEDs) and allowed their dust temperatures and far-infrared luminosities to be measured for the first time, enabling accurate star formation rate constraints \citep{2010Chapman,2012Magnelli}. 
Today, measuring the peak of the dust SEDs is a prerequisite to precision constraints on dusty and highly star-forming galaxies at $1\gtrsim z\gtrsim6$, and putting them in context with the cosmic history of star formation \citep{2018Elbaz}.

In summary, the large and comprehensive sample of DSFGs that will result from \pcam galaxy evolution survey will: (1) lead to robust constraints on the the luminosities of bright submillimeter sources from $z=1$ out to $z>5$;
(2) statistically constrain the cosmic star formation history out to $z\sim5$; (3) study the impact of environment on the population of high luminosity DSFGs (SFR$\,\gtrsim 200\,M_{\odot}\,$yr$^{-1}$) galaxies and (4) reveal their evolutionary link to today's ellipticals; and (5) allow for the study of ``exotic'' sources, such as protoclusters and strongly lensed sources, extending luminosity limits downwards by 2--3 times that achieved by {\it Herschel} at high ($z>3$) redshifts. 
We describe these predicted science results in more detail below.

\subsection{Cosmic History of Star Formation}

\begin{figure*}[t]
\begin{center}
\includegraphics[width=0.85\textwidth]{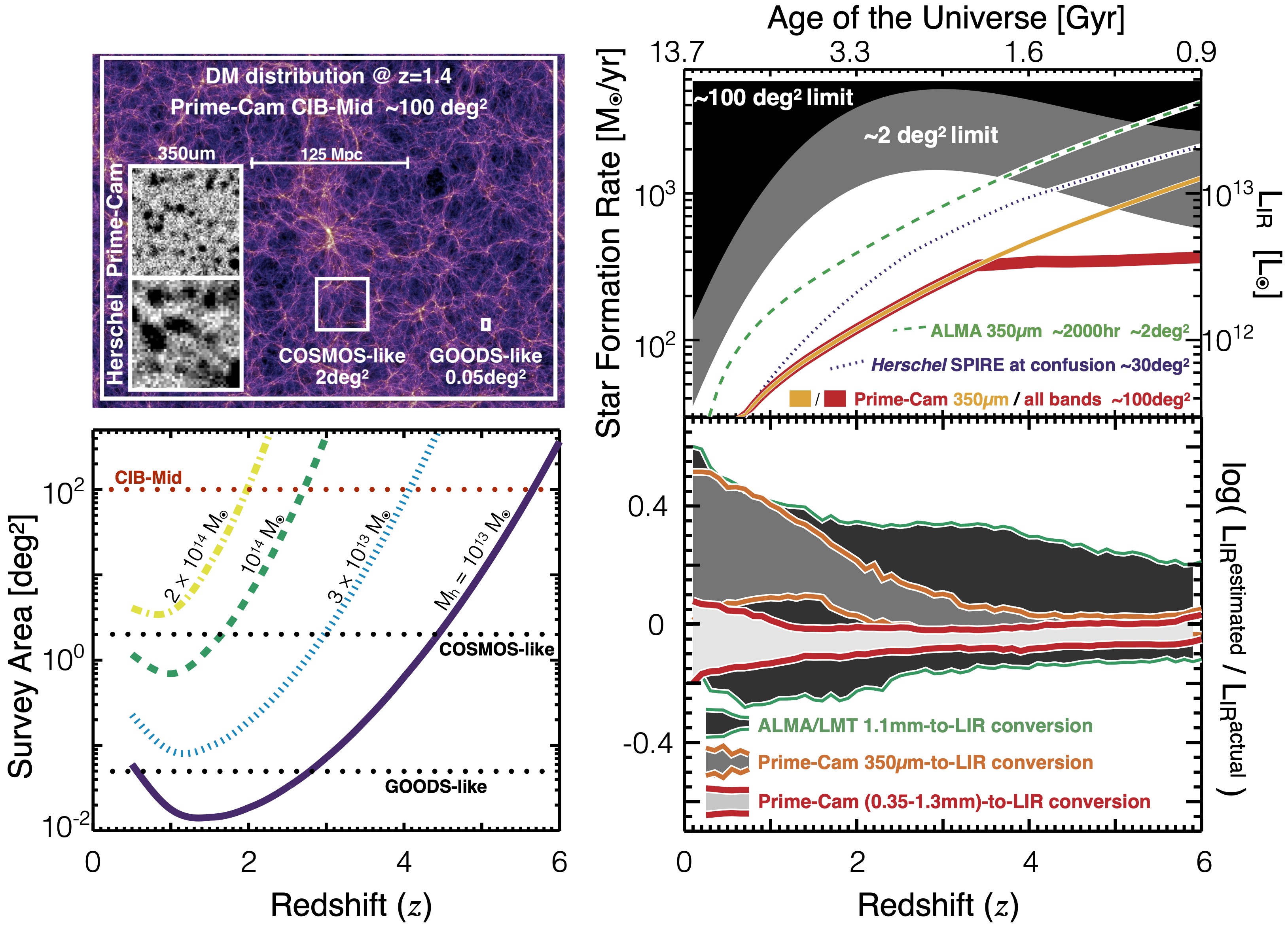}
\caption{Contiguous wide-area multi-wavelength \pcam surveys will measure the dust-obscured star formation properties of hundreds of thousands of galaxies out to $z\sim6$ with high precision. This will greatly enhance our picture of cosmic galaxy evolution, probing $2{-}3$ times lower luminosities than \emph{Herschel}-SPIRE at $z>1$ (due to source confusion; see insets in {\it upper left} panel) while sampling large-scale environments (voids, average density regions, groups, and clusters) over 100 times larger areas than feasible to map with ALMA (the field of view of ALMA is a fraction of a single \pcam pixel) and providing more accurate dust-obscured star formation measurements than current or future millimeter-only surveys. {\it Upper left:} Sky coverage of the CIB-Mid survey (i.e., $\sim\,$100\,deg$^{2}$) compared to the dark matter distribution at $z\sim1.4$ from the Millennium simulation \citep{springel18}.  Insets illustrate how the improved angular resolution of \pcam helps going deeper into the source confusion than was possible with the {\em Herschel} 3.5\,m telescope. {\it Lower left:} Survey area needed to detect at least one halo of mass, $M_{\rm h}$, at a redshift, $z$, within a redshift bin, $\Delta z = 0.1\times z$, as inferred using the halo mass function of \citet{2013Murray}. Different lines correspond to different halo masses. This plot only includes structures that will likely be within our field of view, not which ones that would be detectable, as that depends on the occupation of these structures by DSFGs. These structure will by identified by other means (e.g., X-ray, optical) and studied by \pcam through either direct detection or stacking analysis. {\it Upper right:} Simulated \pcam survey luminosity limits, compared to surveys feasible with existing instrumentation covering the $250-500\,\mu$m wavelength range, which is critical to measure accurately the infrared luminosity of $1\lesssim z \lesssim 6$ galaxies.	\pcam will probe up to $2{-}3$ times deeper than \emph{Herschel}-SPIRE, while sampling luminosities inaccessible to small (e.g., ALMA) surveys because of the rarity of luminous objects. These limits are illustrated by the grey and black areas, i.e., there will be less than 10 galaxies more luminous than these limits within a redshift bin, $\Delta z = {\rm max}(0.15\times z, 0.5)$ in a survey of 
2 or 100 deg$^2$, as inferred using the model of \citet{2017Bethermin}. {\it Lower right:} Predicted reliability ($16^{th}$ to $84^{th}$ percentiles) of far-infrared luminosity measurements, highlighting the vast improvements with the short submm wavelength \pcam $350\,\mu$m (850 GHz) band and wide spectral coverage. These reliabilities were inferred by fitting (or renormalizing) single modified blackbody function with $\beta=1.5$ and $T_{\rm dust}\in[20{-}50\,{\rm K}]$ \citep[e.g.][]{2012Magnelli} to the variety of SEDs from the \citet{2007Draine} template library. See text for detailed description of this procedure. At all redshifts, we use the same set of SEDs - we assume no redshift evolution in L$_{IR}$. Errors improve at higher redshift because we probe closer to the SED peak, and thus any dust temperature dependency becomes less problematic (see also Fig.~9 of Schreiber et al.\ (2015)).}
			\label{fig:SFH}
\end{center}
\end{figure*}

The science legacy of these Prime-Cam surveys will be rich with a better understanding of the cosmic history of star formation.
The cosmic history of star formation, galaxy assembly, and matter content are encoded in the variation of the physical properties of galaxies with their spatial and redshift distributions \citep{2014Madau}. 
In order to properly trace and understand the formation and evolution of galaxies large statistical studies are required over representative cosmic volumes that adequately sample the entire range of cosmic environments (${>}$\,100\,$\deg^2$; Fig.~\ref{fig:SFH} {\it upper and lower left}), and map the (unknown) large scale structure at redshifts 2 to 4.
Over the next decade, short-wavelength surveys will be carried out to commensurate depths and areas at X-ray to near-infrared wavelengths (e.g., \emph{eROSITA}, \emph{Rubin}, DES, \emph{Euclid}, and \emph{Roman}) tracing unobscured star formation. 
Combining our surveys with synergistic work in the optical/near-infrared (DES, {\em{Rubin}}, {\em{Euclid}} and {\em{Roman}}) Prime-Cam will help identify key parameters that regulate star formation (such as environment, AGN, and matter content) 
in high luminosity DSFGs (SFR$\,\gtrsim 200\,M_{\odot}\,$yr$^{-1}$; Figure~\ref{fig:SFH} {\it upper right}) over cosmic time.
Such constraints are paramount to shed light on the formation of local massive ellipticals, likely produced from gas-rich mergers at high redshift that trigger violent and short-live star formation activities \citep{2006Hopkins,2007Faber}. 

However, the dust-obscured half of star formation is only robustly accessible at rest-frame FIR wavelengths, which are redshifted into the $250{-}500\,\mu$m wavelength range ($1200{-}600$ GHz) where no commensurate survey capabilities exist yet. 
Since the properties of dust in galaxies appear to evolve significantly with cosmic time \citep{Casey14}, short-wavelength surveys (rest-frame UV and optical) have to rely on order-of-magnitude extinction corrections that are too large for measuring reliable SF properties. 
Without accurate measurements of FIR luminosities, the total (i.e., direct + obscured stellar light) star-formation rates and SF properties in galaxy evolution studies therefore remain out of reach. {\it Herschel} has certainly pushed our understanding considerably here.  Thanks to its increased depth and wide coverage, \pcam surveys will further our understanding of rest-UV selected populations, statistically constrained through stacking the mean dust-obscured star formation rates of galaxy population not directly detected but selected from large optical/near-infrared surveys. 
Using this approach, it will be possible to estimate the \textit{total} star formation activities of ``normal'' (i.e., $L\ast$) galaxies out to $z\sim5$ \citep[galaxies which have $L_{\rm IR}^{L_\ast} \sim 0.1 \times L_{\rm IR}^{\rm \pcam\ limit}$ at $z\sim1-5$;][]{2013Magnelli,2013Gruppioni,2020Gruppioni}, and provide improvements in a robust determination of the cosmic star formation history. 

We note that the $z>5$ far-infrared luminosity function is still highly uncertain \citep{casey18} 
and a factor of $\approx$2 uncertainty exists on the number of $z>5$ sources that will be found in these surveys.
Figure~5 illustrates these model surveys and luminosity limits. The model includes those galaxies detected at a given band, so indeed the highest redshift galaxies constrained at 850\,GHz alone need to be detected at this frequency and  therefore have larger SFRs ($>1000$M$_\odot$/yr) as shown in the upper left panel. Including all the Prime-Cam bands, there are a much larger number of galaxies extending to lower luminosities, and the majority of these anticipated $z>5$ galaxies will be undetected dropouts at 850 GHz. 
At $1<z<6$, the 850\,GHz band of \pcam will probe the FIR emission of galaxies near the rest-frame of the dust emission SED peak, thereby providing an excellent proxy for estimating FIR luminosities and dust obscured star-formation rates in the high-redshift Universe. 
This can be accomplished with typically $2{-}3$ times more accuracy than is possible with, for example, LMT/ALMA mm-wave surveys alone, and $3{-}5$ times more accurately for galaxies also detected by \pcam lower frequency (down to 220 GHz) bands ($5{-}8\%$ according to the model of B\'ethermin et al. \citeyear{2017Bethermin}; Fig.~\ref{fig:SFH}, {\it lower right}). Our procedure for estimating these curves in Fig.~\ref{fig:SFH} is as follows. We take a subset of {\it realistic} \cite{2007Draine} SED templates, limiting to MW-type PAHs, fixing radiation fields U to  U$_{max}$=$1\times10^6$ and $0.7<U_{min}<25$ (see Magnelli et al.\ 2012), corresponding to 2700 templates. We renormalized all these templates to L$_{\rm IR}$ = 1 L$_\odot$, and then for each redshift, we calculate their observed flux densities. For the 270 GHz (1.1\,mm)-to-L$_{\rm IR}$ conversion, we renormalized to the observed 270 GHz flux density of each template and look at the entire distribution in L$_{\rm IR}$/(1 L$_\odot$). We do the same for the 850 GHz-to-L$_{\rm IR}$ conversion. Finally, for the 860 GHz to 230 GHz (0.35-1.3\,mm)-to-L$_{\rm IR}$ conversion, we fitted the flux densities (850, 410, 350, 280, 220 GHz)
of each template with single grey-bodies, and again look at the distribution in L$_{\rm IR}$/(1 L$_\odot$).
While Prime-Cam does not provide a unique ability to make such measurements (for instance, LMT 270 GHz (1.1\,mm) detections could be followed up one-by-one with ALMA at 850 GHz, 
it does facilitate such measurements efficiently over an unprecedented large numbers of galaxies.
\pcam therefore fills a critical gap with its unique combination of deep 850\,GHz coverage, significantly improved spatial resolution (${\sim}\,15$\,arcsec at 850\,GHz) compared to previous facilities operating at these frequencies \citep[in particular \emph{Herschel},][]{Lutz2014}, and a very large field-of-view ($0.78\,\deg^2$; ${>}\,200{,}000$ times larger than ALMA at 850\,GHz).
Through deep, large-area contiguous submillimeter wave surveys \citep[critically complementing ultra-deep ``pencil
beam'' surveys with ALMA;][]{Walter2016,Aravena2016,Dunlop2017,Franco2018,2019GonzalezLopez}, 
\pcam surveys will in its lifetime securely detect millions of star-forming galaxies via many 100 deg$^2$, confusion-limited fields across a broad range of cosmic environments out to $z\sim5$, probing typically about 3 times deeper than the most sensitive {\em Herschel} surveys \citep{Lutz2014}. 

From direct observations of the FIR spectral energy distribution peak, accurate dust-obscured SFRs, can be inferred for the vast majority of high luminosity DSFGs (SFR$\sim$ hundreds of M$_{\odot}$\,yr$^{-1}$) galaxies out to $z\sim5$ within the wide-field optical/near-infrared survey fields that will become available at first light for FYST. 
For about $5{-}10\%$ of these galaxies, detection in several \pcam lower frequency (down to 220 GHz) bands will be also available, that will yield accurate dust masses and temperatures.
Additional physical parameters of these massive and intensely star-forming systems may be determined for a subset of the sample by combining \pcam surveys with these synergistic optical/near-infrared (O/IR) surveys to obtain their photometric redshifts, stellar masses, and unobscured fractions of the SFR, which is ${\le}\,50$\,\% globally \citep{2013Burgarella,2015Pannella,2020Gruppioni}. The ability to obtain photometric redshifts from O/IR surveys will however depend on redshift and obscuration \citep{2003Chapman,2005Chapman,2012Casey,2017Danielson}. Far-infrared based photometric redshifts will need to be used for a substantial fraction of the sources, especially at higher redshifts, with the caveat made in these references that these methods have errors of d$z$$\sim$0.5. In these cases the L$_{\rm IR}$ and M$_{\rm dust}$ estimates can be made, but M$^{*}$ and unobscured SFR fraction won't be
accessible.

We note that cross identification with O/IR-selected galaxies is not trivial or in some cases even unique. However, so called {\it super-deblending} studies have made significant progress in this regard for {\it Herschel} observations (Jin et al.\ 2019), and Prime-Cam will  help further these studies. We estimate the fraction of Prime-Cam detections which will have a O/IR counterpart using the model of Schreiber et al.\ (2017) 
which includes both O/IR and FIR. 
First, we used this model to predict the number of Prime-Cam detections (above confusion) in our 100 deg$^2$ field and found very consistent numbers with Bethermin et al.\ (2017). 
Then, we assumed that our 100 deg$^2$ field will be covered by the Euclid 15,000 deg$^2$ wide survey and thus will have a ($g, r, i, z$)-coverage from the ground down to AB$\sim$24 and ($Y, J, H$)-coverage from Euclid down to AB$\sim$24.
The final model predicts that about 68\% of our \pcam survey will have a counterpart in one of the ($g, r, i, z$) images and 36\% in all ($g, r, i, z$) images; 96\% will have a counterpart in one of the ($Y, J, H$) images and 85\% in all ($Y, J, H$) images; 96\% will have detection in one of the ($g, r, i, z, Y, J, H$) images and 36\% in all ($g, r, i, z, Y, J, H$) images. 
Therefore, the restframe-UV images will be biased against highly obscured SFGs, but observed frame optical/NIR images will provide counterparts for most of the Prime-Cam sources. 

From this, it is then possible to statistically identify the key parameters regulating matter assembly (cosmic epoch verses environment) in these extreme systems which are likely the progenitors of local massive ellipticals.
\pcam wide-area surveys will particularly being advantageous in this context as it will allow for accurate measurement of the clustering properties of these extreme systems and thereby test their evolutionary link to today’s massive ellipticals which are known to reside in the central region of massive groups and clusters. 
Through stacking of hundreds of galaxies, \pcam will also statistically constrain the mean dust-obscured star formation rates of galaxy population not directly detected but selected from large optical/near-infrared surveys ($S/N^{\rm stack} \propto S/N^{ind.} \times \sqrt{N_{\rm stack}}$). 
From this, it is then possible to obtain a significantly improved determination of the cosmic star formation history. 

Because of the large survey areas, it will also be possible to discover large samples of strongly-lensed dusty galaxies for detailed follow-up studies with ALMA and {\em{JWST}} \citep[e.g.,][]{canameras2015,Hodge2020} and IR-bright galaxy protoclusters at high redshift as signposts of early structure formation \citep{oteo18,Miller2018,2019GomezGuijarro,Hill2020}.
\pcam surveys will also be suitable to uncover very dusty sources (including the most intense starbursts in the Universe) that are entirely missed by even the deepest optical/near-infrared surveys, over a broad luminosity range; out to $z>7$ \citep[e.g.,][]{riechers13b,riechers17,Strandet2016,Brisbin2017,Marrone2018,Wang2019,Reuter2020}. 
These sources are very rare, and only very small numbers are currently known at $z>5$, due to the limited sensitivity, wavelength coverage, and/or areas of past and current (sub)millimeter surveys \citep[see ][for a recent overview]{riechers20a}.
While such sources are bright in mm-wave surveys, they can be discriminated from the dominant lower-redshift foreground with the inclusion of the shorter submillimeter wave band of \pcam \citep{riechers17}.
As illustrated in Figure~\ref{fig:SFH} ({\it upper right}), applying such a 850 GHz ``dropout'' technique will indeed be a powerful tool to select $z>4$ candidates. While {\it Herschel} has made significant progress in studying submillimeter-wave dropouts and $500\,\mu$m (600 GHz) risers (e.g., Lewis et al.\ 2018,
 Donevski et al.\ 2018), 
the additional depth afforded by the smaller beamsize of \pcam\ will lead to both more stringent selection criteria and fainter sources being selected at these high redshifts.

By obtaining good statistics on this population of dusty, highly star-forming galaxies at higher redshifts than probed by {\it Herschel}, one can address important issues of galaxy evolution. For example: when and how did local ``red-and-dead'' ellipticals form? 
Two leading theories that link the evolutionary tracks are through major mergers, which trigger intense, yet short-lived starburst leaving behind a passively evolving elliptical, and massive star-forming galaxies that are starved of gas by a ``too-hot'' halo. 
Overall, in the \pcam galaxy evolution survey, through the detection of large samples of dusty, highly star-forming galaxies over a broad range of redshifts, the mechanism of evolution into local ellipticals can be studied in more statistical detail.

\section{Measuring CMB foregrounds to aid the search for primordial gravitational waves}

\label{sec:dust}
In recent years, enormous progress
has been made using CMB temperature and polarization 
measurements to constrain cosmological parameters and characterize the large-scale structure of the Universe \citep{komatsu/etal:2014,bicep/keck/2018,PlanckCosmo:2020,adachi/etal:2020,aiola/etal:2020,choi2020JCAP,dutcher/etal:prep}. 
In addition to upcoming projects such as Simons Observatory \citep[SO,][]{Ade19} and BICEP Array \citep{bicep/keck/2018}, the CMB research community is developing plans for a next generation ground-based ``Stage IV" CMB survey (CMB-S4, \citealt{abazajian/2019})
and for the \emph{LiteBIRD} space mission \citep{hazumi/etal:2019,Hazumi_2020}. These facilities promise to achieve dramatic improvements in constraints on the amplitude of the primordial gravitational waves \citep{grishchuk:1975,starobinsky:1979} which are imprinted on the CMB `\emph{B}-mode' polarization \citep{kamionkowski/kosowsky/stebbins:1997,seljak/zaldarriaga:1997}, and on the effective number of light relic species
\citep{bashinsky/seljak:2004,hou/etal:2013}. 

FYST is a potential telescope platform for CMB-S4, and it also offers unique capabilities for important advances in high-frequency polarization science before CMB-S4. 
\pcam on FYST will help address fundamental questions about the origins of the Universe by testing theories of early conditions through improved constraints on primordial gravitational waves (or tensor perturbations). Scalar perturbations cause energy density fluctuations at the surface of last scattering that only generate the even-parity \emph{E}-mode polarization \citep{kamionkowski/etal:1997,zaldarriaga/seljak:1997}, which has been well characterized \citep{komatsu/etal:2014,bicep/keck/2018,PlanckCosmo:2020,adachi/etal:2020,aiola/etal:2020,dutcher/etal:prep}. 
Inflation models predict a period of rapid expansion in the early Universe \citep{guth:1981,sato:1981,linde:1982,albrecht/steinhardt:1982} and generation of both scalar \citep{mukhanov/chibisov:1981,starobinsky:1982,hawking:1982,guth/pi:1982,bardeen/steinhardt/turner:1983} and tensor perturbations \citep{grishchuk:1975,starobinsky:1979}, the latter of which would leave odd-parity \emph{B}-mode imprints in the CMB polarization \citep{kamionkowski/kosowsky/stebbins:1997,seljak/zaldarriaga:1997}.
Measuring the primordial \emph{B}-mode polarization to constrain the amplitude of tensor perturbations and hence the tensor-to-scalar ratio $r$ will test many inflationary and other early Universe theories \citep[see][for a review]{kamionkowski/kovetz:2016}. In Section \ref{sec:CMBexp} we describe \pcam measurements in the context of current and upcoming CMB experiments, and in section \ref{sec:CMBmethods} we present forecasts for improving constraints on $r$ with the addition of \pcam data.

\subsection{\pcam in the context of current and upcoming CMB experiments} \label{sec:CMBexp} The current best constraint on the primordial \emph{B}-mode signal is $r<0.07$ (95\%~CL, from \citealt{bicep/keck/2018}), which improves to $r<0.044$ when using all the CMB temperature and polarization data (from \emph{Planck}, \citealt{tristram/etal:2021}). This is limited by the uncertainty on the polarized Galactic dust emission. Several experiments that are planned or under construction aim to better constrain $r$ from more precise measurements of both the CMB and the Galactic dust through observations in multiple frequency channels \citep{Ade19,abazajian/2019, schillaci/2020,hazumi/etal:2019}. In particular, SO aims to achieve $\sigma(r)=0.003$ with multiple small aperture telescopes (SATs) and a large aperture telescope (LAT) like FYST observing at six frequencies between 27 and 280\,GHz. 

\pcam will make the most sensitive measurements at frequencies $>300$\,GHz of polarized dust foregrounds yet, with the wide field survey improving on the current state-of-the-art \emph{Planck} 353\,GHz map by a factor of $>2$ over more than half the sky (also see Section~\ref{sec:bfield}). We illustrate here the value of \pcam high-frequency observations for constraining the polarized dust emission and mitigating potential bias on $r$ from SO. At the current sensitivities, the polarized dust emission is well-described by a single-temperature modified blackbody model, $D_\nu(\beta_{\rm{d}},T_{\rm{d}})\propto\nu^{\beta_{\rm{d}}-2}B_\nu(T_{\rm{d}})$ (in Rayleigh-Jeans temperature units),
where $\beta_{\rm{d}}$ is the dust spectral index, $T_{\rm{d}}$ is the dust temperature, and $B_\nu$ is the Planck function \citep{planck/2020}. \pcam will play a major  role in constraining of $\beta_{\rm{d}}$ and $T_{\rm{d}}$, thereby improving the constraint on $r$ achievable by SO (and other experiments with overlapping fields). 

\subsection{Methods: Improving constraints on $r$ and Galactic dust} \label{sec:CMBmethods} 

The possible improvement from \pcam on the constraint on $r$ beyond the SO-SATs is analyzed as follows.\footnote{The Julia codes and documentation for this analysis are available in \url{https://github.com/komatsu5147/CleanCMB.jl}.} We first simulate realistic foreground sky maps at \pcam and SO frequencies using the \texttt{PySM} simulation package \citep{thorne/2017}. We then add Gaussian simulations of CMB (without the primordial \emph{B}-mode but with the scalar \emph{E}-mode and the lensing-induced \emph{B}-mode) and noise \citep[including white and $1/f$ components extrapolated from on-sky measurements; see][] {choi2020JCAP,Ade19} modulated by the realistic survey hit-count maps of SO-SAT and FYST, and compute the auto- and cross-frequency power spectra of all the simulated maps using the {\tt NaMaster} package \citep{alonso/etal:2019}\footnote{\url{https://github.com/LSSTDESC/NaMaster}} with the ``purify B'' option and the ``C2'' apodization scheme (10 deg). We use the maximum likelihood method \citep{stompor/etal:2009} for finding the best-fitting values of $\beta_{\rm{d}}$, $T_{\rm{d}}$, and $\beta_{\rm{s}}$ (the spectral index for the synchrotron emission). In the fitting, we assume that these parameters are homogeneous over the sky patch observed by SO-SAT, though the simulation does contain spatially-varying foreground parameters. We impose Gaussian priors of $\beta_{\rm{d}} = 1.6\pm 0.5$, $\beta_{\rm{s}} = -3.0\pm 0.5$ and $T_{\rm{d}} = 19.6 \pm 5~{\rm K}$. 
Finally, we deproject the best-fitting synchrotron and dust emission using the constrained internal linear combination (cILC) method \citep{remazeilles/delabrouille/cardoso:2011a,remazeilles/delabrouille/cardoso:2011b,hurier/etal:2013}, estimate the foreground-cleaned \emph{B}-mode polarization power spectrum, and calculate the tensor-to-scalar ratio $r$ from the multipole range of $30<\ell<260$. Since the cILC is a blind method, we can further reduce the residual foreground contamination that is not modeled by the parametric method; thus, our approach is a hybrid of the parametric maximum likelihood method and the cILC.
We repeat this proceedure for 1500 realizations of CMB and noise and calculate the mean and standard deviation of $r$. 

\begin{figure}[t]
\begin{center}
\includegraphics[width=0.5\textwidth]{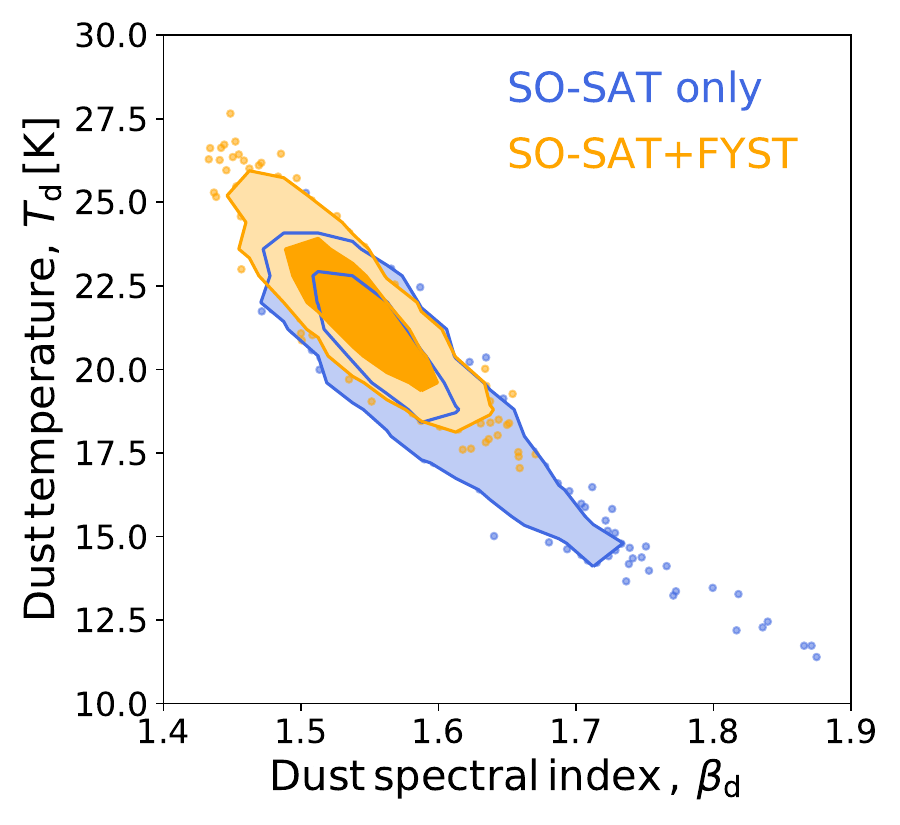}

\end{center}
\caption{Distributions of the best-fitting values of the dust parameters, $\beta_{\rm{d}}$ and $T_{\rm{d}}$, from 1500 realizations of the simulation with SO-SAT only (blue) and SO-SAT + FYST with \pcam (orange). Shaded contours show the approximate 1$\sigma$ and 2$\sigma$ intervals, while the dots represent the simulation results outside of the 95\% confidence range. This shows a significant reduction of the degeneracy when combining the two data sets.}
\label{fig:beta_improve}
\end{figure}

In Fig.~\ref{fig:beta_improve}, we show the distributions of the dust parameters, $\beta_{\rm{d}}$ and $T_{\rm{d}}$. The SO-SAT-only results show a strong degeneracy between the two parameters; thus, the constraints are dominated by the prior. This degeneracy is reduced substantially by adding the data of FYST with \pcam. We note that \pcam's highest frequency channel, 850\,GHz, is crucial for reducing the degeneracy of the two parameters. 

\begin{figure}[t]
\begin{center}
\includegraphics[width=0.45\textwidth]{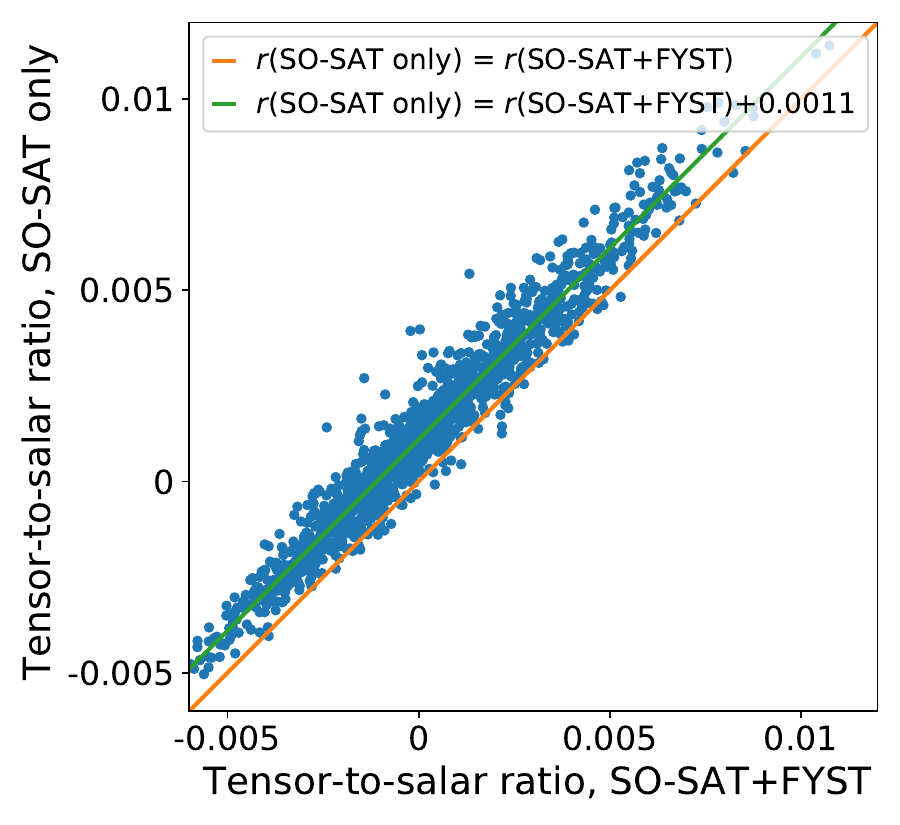}
\includegraphics[width=0.45\textwidth]{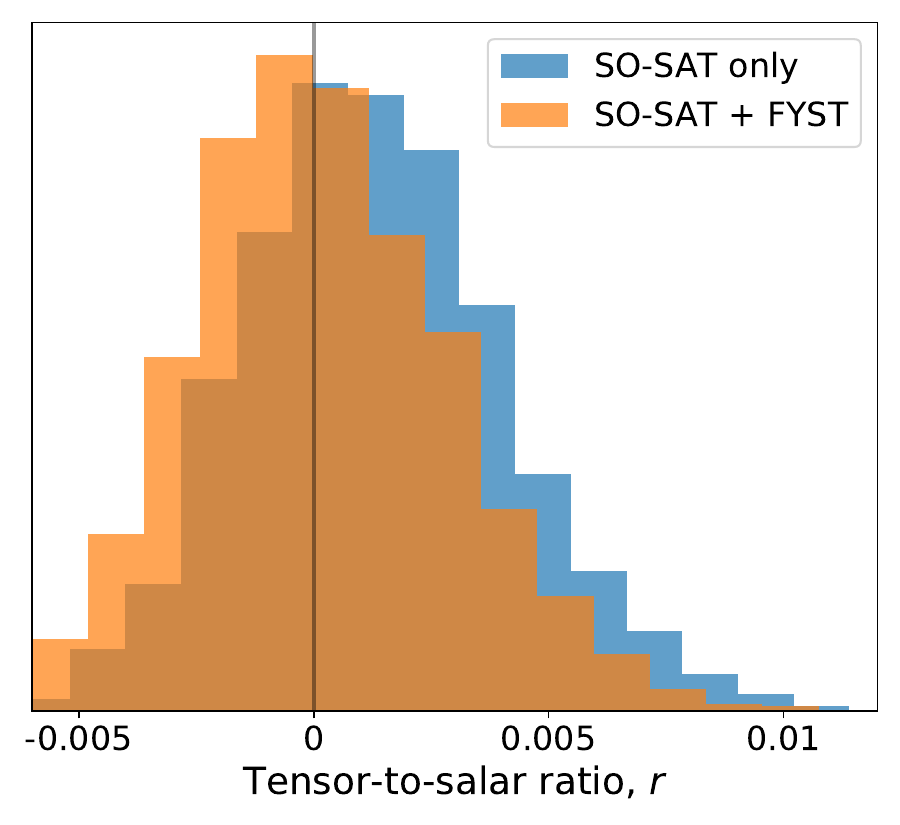}
\end{center}
\caption{(Left) Scatter plot of the distributions of the best-fitting values of $r$, showing that the SO-SAT only results have a systematic bias at the level of $\delta r=1.1\times 10^{-3}$ (green line) compared to the SO-SAT + FYST results. The orange line shows the case with no bias.
(Right) Histograms of $r$ constraints from 1500 realizations of the simulation with SO-SAT only (blue) and SO-SAT + FYST with \pcam (orange), showing a significant reduction of the bias on $r$ when combining the two data sets. Note that the simulation does not have the primordial \emph{B}-mode signal, i.e., $r_{\rm input}=0$ (vertical line).}
\label{fig:r_improve}
\end{figure}

For $r$, SO finds a  bias due to the residual dust foreground emission that is on the order of the statistical uncertainty \citep{Ade19}. This bias can be reduced by marginalizing over the foreground model parameters; however, as mentioned in \citet{Ade19}, this reduction (1) comes at the expense of increased $\sigma(r)$ by about $70\%$; and more importantly, (2) is possible only with simulations where the actual foreground residual is known. By combining the \pcam high-frequency data with SO, we find that $r=(0.2\pm 2.7)\times 10^{-3}$ is achievable. 
Comparing this to $r=(1.3\pm2.8)\times 10^{-3}$, which we obtained for the SO-SAT-alone case with their `pess-$1/f$' noise in the baseline design \citep{Ade19}
and without marginalization over the residual foreground, we find that the combination with \pcam can reduce the 
foreground residual bias by a factor of more that 6 while maintaining a similar level of $\sigma(r)$ (Fig.~\ref{fig:r_improve}). The 95\%~CL upper limits thus improve from $r<6.8\times 10^{-3}$ to $r<5.5\times 10^{-3}$. Even if we remove the \pcam data at low multipoles ($\ell<60$), we still find a significant reduction in the bias, $r=(0.5\pm 2.8)\times 10^{-3}$. In other words, \pcam measurements have the potential to aid the search for primordial gravitational waves in a significant manner. 

Our results do not rely on one specific frequency channel. We find that each of the 350, 410, and 850\,GHz data contribute approximately equally to the reduction of bias in $r$. While the degeneracy between the dust parameters is reduced largely by the 850\,GHz data, the other frequency channels help reduce the bias via the cILC step, which highlights the value of having measurements at multiple submillimeter frequencies. In summary, \pcam measurements of Galactic dust combined with SO-SAT measurements are predicted to reduce bias and improve constraints on the primordial gravitational waves.

\section{Galactic Polarization Science: Magnetic Fields and Dust Properties}
\label{sec:bfield}
\newcommand{\msun}{\ensuremath{M_{\odot}}}
\providecommand{\sorthelp}[1]{}

Interstellar dust represents most of the solid material in our Universe. Despite this importance, many questions remain about its composition and structure \citep{Hensley2021}.
Thermal dust emission is a multi-purpose tool in the study of clouds and star formation. It is commonly used as a density tracer to obtain the total mass and gravity of gas-dominated systems \citep[e.g.,][]{stutz15,Stutz2016, sadavoy2016}. Polarized dust emission is used to map magnetic fields in all phases of the interstellar medium \citep{PlanckXIX2015,PattleFissel2019,Tahanietal2018,Tahanietal2019,sadavoy19,fissel16}. Non-spherical dust grains in the ISM align with the magnetic field \citep{andersson_2015, reissl18,reissl21}. Hence, thermal dust emission is imprinted with a linear polarization signal.  This polarization signal traces the Plane-Of-the-Sky (POS) magnetic field geometry, while the degree of polarization depends on the properties of the dust grains \citep{Hildebrand1988,Draine+Fraisse_2009,Draine+Hensley_2020}, as well as the 3-D geometry of the magnetic field \citep{chen2019}. \pcam observations will measure polarized dust emission in resolved systems in both the Galaxy and the Large Magellanic Cloud (LMC) over wide fields, accessing a broad range of scales.  These data will enable tests of dust models, probe the properties of interstellar turbulence, and investigate the effects of the magnetic field on cloud, star, and cluster formation in both high- and low-mass 
star-forming regions.

\begin{figure*}
    \centering
    \includegraphics[width=\textwidth]{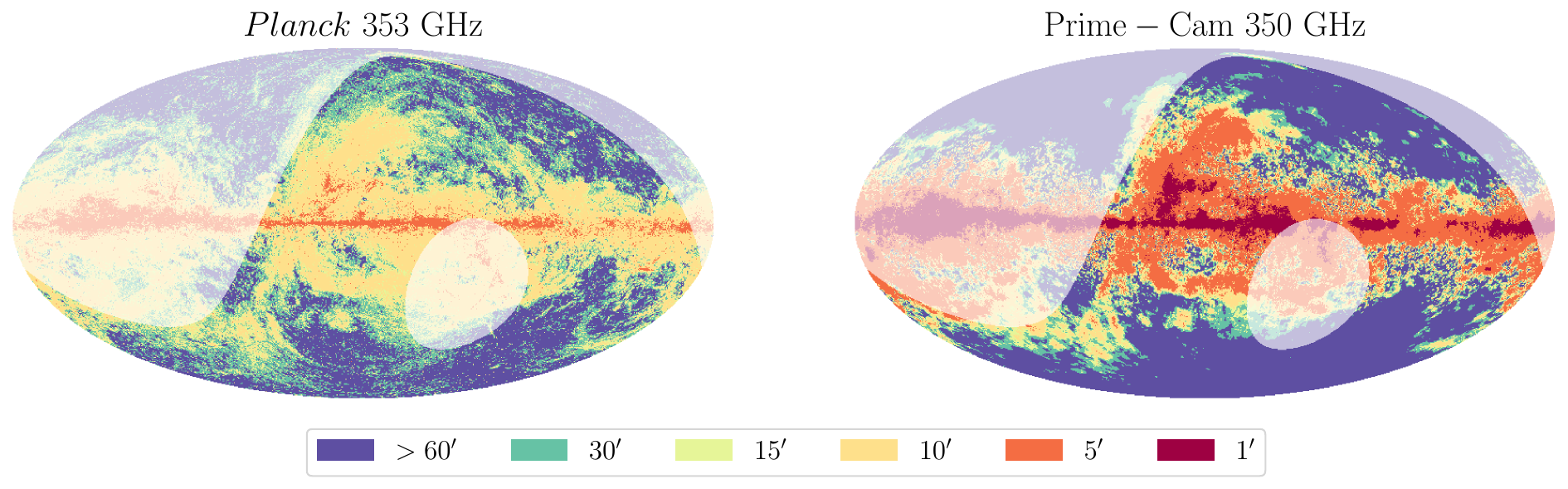}
    \caption{Effective resolution for SNR $>$ 3 measurements of dust polarized intensity with \textit{Planck} at 353 GHz (left) and FYST with Prime-Cam at 350 GHz (right), for the 4000 hour Wide Field Survey (WFS) at different levels of spatial resolution. Translucent contours indicate the boundaries of the FYST observable region. \textit{Planck} measured 3$\sigma$ detections of the polarized intensity at $5'$ resolution only in the inner Galactic plane. Prime-Cam will make comparable measurements over much more of the sky, and along bright sightlines will make high-fidelity measurements at higher angular resolution, as demonstrated here by the additional contour at $1'$. }
    \label{fig:dustSNR}
\end{figure*}

In this section we describe our survey design. The survey is driven by the requirement for \pcam dust polarization observations of both diffuse regions and denser molecular clouds. Of the latter it covers a relevant range of cloud masses. 
Our planned surveys have two main components: (1) the 4000-hour Wide Field Survey (WFS), which is comparatively shallow but will cover more than half of the Galactic plane and will provide $<$\,1$^{\prime}$ resolution maps of higher mass molecular clouds (see Figure \ref{fig:dustSNR}); and (2) targeted deeper wide-field mapping observations of individual systems.  Specifically, \pcam will make deep dust polarization maps of seven nearby molecular clouds, the nearby translucent ($N_\mathrm{H} \sim 10^{21} \mathrm{cm}^{-2}$) cloud Pyxis, and the Large Magellanic Cloud (LMC); Table \ref{tab:poln_surveys} lists the polarization survey targets.  With both high sensitivity and resolution that is $\gtrsim20\times$ that of the {\em Planck} satellite, \pcam will be the first polarimeter able to continuously trace polarized dust emission with high fidelity from entire cloud scales down to the scales where individual stars and and their planetary systems form.   

Many of our target regions will also be observed by CHAI in the submillimeter [\ion{C}{1}] fine-structure and the CO(4-3)/CO(7-6) CO rotational lines or will be covered by other Galactic spectral surveys, providing complementary surveys of turbulence and gas kinematics. With our Prime-Cam surveys and this ancillary data we will study the MHD turbulent power spectrum, dust properties, and magnetic field properties across a range of ISM phases, cloud environments, clouds masses and inferred evolutionary stages.  Hence our survey design enables evaluation of the role of various fundamental physical processes as a function of both mass and time \citep{Stutz2016}.  
\pcam will map dust emission and polarization in hundreds of molecular clouds in both the Milky Way and the LMC, forming a large basis set for launching new investigations into interstellar dust and the role of turbulence and magnetic fields in the formation of stars and clusters from the gas phase.

In the following, Section \ref{sec:dust_models} describes how \pcam polarization data constrains dust grain models, Section \ref{sec:turbulence} describes how Prime-cam polarization addresses the physics of the diffuse ISM, Section \ref{sec:BinLMC} describes a multi-scale survey of magnetic fields in the LMC, and Section \ref{sec:bfield_sf} describes the roles of magnetic fields in SF outcomes.

\begin{table*}[h!t]
\begin{center}
\setlength{\tabcolsep}{8pt}
\caption{\pcam targeted polarization surveys. \label{tab:poln_surveys}}
\begin{tabularx}{0.9\textwidth}{l r r c c c c c c}
Target & R.A. & Decl. & Dist. & Band$^a$ & $\theta^b$ &  $p_{\mathrm{353GHz}}^c$ & Area & Depth\\
 & [deg] & [deg] & [pc] & [GHz] & [pc] &  [\%] & [deg$^2$] & [hrs deg$^{-1}$]\\
\hline
\hline
\multicolumn{9}{l}{\em Low mass molecular clouds ($M\,<\,10^4\,$M$_{\odot}$)}\\
Lupus I      & 235.56 & -33.92 & $\phn\phn\phn$160 & 350$\phn$ & 0.029 &  $\phn$3 & $\phn$4 & 10.0\\
Pipe Nebula  & 258.43 & -27.33 & $\phn\phn\phn$180 & 350$\phn$ & 0.032 & $\phn$3 & $\phn$4 & 10.0\\
Musca        & 187.73 & -71.58 & $\phn\phn\phn$200 & 850$^d$ & 0.015 &  12 & $\phn$5 & 12.5\\
\hline

\multicolumn{9}{l}{\em Intermediate mass molecular clouds ($10^4\, $M$_{\odot}
\,<\,M\,<\,10^5\,$M$_{\odot}$)}\\
Ophiuchus    & 247.93 & -24.47 & $\phn\phn\phn$140 & 850$\phn$ & 0.010 &  $\phn$4 & $\phn$8 & $\phn$8.0\\
Aquila       & 277.78 & -2.16 & $\phn\phn\phn$400 & 850$\phn$ & 0.029 &  $\phn$2 & $\phn$4 & $\phn$8.0\\
\hline
\multicolumn{9}{l}{\em High mass molecular clouds ($M\,>\,10^5\,$M$_{\odot}$)}\\
Orion A      & 84.62  & -6.99 & $\phn\phn\phn$400 & 850$^d$ & 0.029 &  $\phn$4 & 11 & $\phn$8.0\\
Vela C       & 135.10 & -44.01 & $\phn\phn\phn$900 & 850$\phn$ & 0.065 &  $\phn$3 & $\phn$5 & $\phn$8.0\\
\hline

\multicolumn{9}{l}{\em Translucent cloud}\\
Pyxis        & 134.52 & -23.14 & $\phn\phn\phn$175 & 350$^d$ & 0.013  & $\phn$5 & 20 & $\phn$3.0\\
\hline
\multicolumn{9}{l}{\em Nearby Galaxy}\\
LMC          & 80.00  & -68.50 &        50,000& 350$^d$ & 9.0$\phn\phn$  & $\phn$ 3 & 30 & $\phn$ 3.3\\

\hline
\end{tabularx}
\end{center}
\noindent \footnotesize{$^a$ Primary band for observations, possibly distinct from e.g., early science observation band(s).} \\
\noindent \footnotesize{$^b$ Physical beam size at target distance.} \\
\noindent \footnotesize{$^c$ \emph{Planck} 353\,GHz polarization fraction.}\\
\noindent \footnotesize{$^d$ High priority targets with coverage at both  850 GHz and 350 GHz.  The coverage depth will only be uniform for the primary band.}
\end{table*}

\subsection{Testing Dust Grain Models with \pcam Polarization Data}
\label{sec:dust_models}

Dust emission at far-infrared and microwave frequencies arises from grains heated by UV/optical starlight that radiate thermal vibrational emission. Interactions with photons and gas atoms exert torques on the grains, inducing rapid rotation about their axis of greatest moment of inertia, i.e., their short axis. A rotating grain can develop a magnetic moment antiparallel to its rotation axis through the Barnett effect if the grain material is paramagnetic \citep{Dolginov76,Purcell79}. Interaction between the grain's magnetic moment, radiative torques, and the local magnetic field dissipates the component of the rotational kinetic energy perpendicular to the magnetic field, aligning the rotation axis with the local magnetic field orientation. Since the rotation axis of each grain is also its short axis, the total emission from the ensemble of dust grains is linearly polarized perpendicular to the field direction \citep[see][for a recent review of grain alignment]{andersson_2015}.

The processes of extinction, emission, and alignment are all sensitive to the composition of grains. For instance, grains made predominantly of hydrocarbon materials attain higher temperatures than those made from silicate materials, even in identical radiation fields \citep{Mathis+etal_1983,Li+Draine_2001}. Furthermore, polarization has been robustly detected in interstellar extinction features at 9.7 and 18\,$\mu$m that arise from silicates \citep{Dyck+etal_1973,Aitken+etal_1989,Smith+etal_2000}, whereas no polarization has been observed in the 3.4\,$\mu$m hydrocarbon feature, even on the same sightlines where 9.7\,$\mu$m polarization is observed \citep{Chiar+etal_2006,Mason+etal_2007}. Thus, dust polarization is not only a probe of Galactic magnetic fields but also the composition of interstellar grains. Different assumptions on dust composition can result in stark discrepancies between dust masses inferred from infrared emission, whether it be from high-redshift galaxies, molecular clouds, or protoplanetary disks.

Within the last decade, multi-frequency FIR polarimetry has accessed for the first time diffuse regions of molecular clouds and the large-scale diffuse ISM. These observations have challenged long-held notions about the nature of interstellar grains. For instance, the {\it Planck} satellite uncovered regions of sky having 353\,GHz polarization fractions $\gtrsim 20\%$ \citep{Planck_2020_XII}, well in excess of the maxima predicted by pre-{\it Planck} dust models \citep{Draine+Fraisse_2009}. Also unexpected was the finding from both {\it Planck} at wavelengths $\nu \leq 353$\,GHz and the BLASTPol balloon telescope between 353 and 1200\,GHz that the polarization fraction is essentially frequency-independent \citep{Planck_Int_XXII,Ashton+etal_2018,planck/2020}. Models employing separate populations of silicate and carbonaceous grains generically predict that the ratio of polarized to total intensity changes as the relative contribution of the more polarized silicate-bearing grains to the less polarized hydrocarbon-bearing grains increases toward long wavelengths \citep[e.g.,][]{Draine+Fraisse_2009,Guillet+etal_2018,Draine+Hensley_2020}. Although this is consistent with the dichotomy seen in the polarization properties of the respective extinction features, a frequency-dependent FIR polarization fraction is not observed on diffuse lines of sight.

New types of dust models have been created to explain these findings. \citet{Guillet+etal_2018} proposed a suite of models with highly elongated grains coming in distinct silicate and carbonaceous varieties. However, these different dust species have sufficiently similar FIR emission properties to push the wavelength-dependence of the polarization fraction below current observational limits. In contrast, \citet{Draine+Hensley_2020} proposed that the long-held silicate/carbonaceous dichotomy is incorrect and that interstellar dust exists as a comparatively homogeneous composite. While both models are consistent with current FIR polarimetry, they have starkly different implications for the lifecycles of dust grains and the extent to which they are homogenized in the diffuse ISM. They also make different predictions for measurements of dust polarization at higher frequencies than those accessible by {\it Planck}.

\pcam has the frequency coverage and sensitivity to discriminate among these models and thus elucidate the nature of interstellar dust. Tests of dust composition are best conducted on diffuse sightlines, where complicating factors like line-of-sight temperature gradients can be mitigated. We focus here on a nearby translucent cloud in Pyxis, inaccessible to previous experiments lacking the requisite sensitivity, but well-matched to the capabilities of \pcam. As demonstrated in Figure~\ref{fig:PyxisPS}, \pcam observations of Pyxis can discriminate among dust models at high frequencies where such comparisons have not yet been possible.

Frequency coverage is especially critical for tests of dust models. At long wavelengths, dust emission is in the Rayleigh-Jeans regime, and so different dust populations with different temperatures have similar spectra. The 850\,GHz \pcam band is sufficiently close to the peak of the dust spectrum that temperature effects are non-linear, allowing the spectrum to be decomposed into distinct components. The power of \pcam to discriminate among models in the Pyxis region, especially at 850\,GHz, is illustrated in Fig.~\ref{fig:PyxisPS}. Uncertainties are quoted for a single 15$^\prime$ region at the mean brightness of Pyxis. We compare these forecasted data to dust models from \citet{Guillet+etal_2018} (Models A and B), \citet{Draine+Hensley_2020} (``astrodust''), and a parametric model from \citet{Meisner2015} assuming relative polarization fractions of 1.6:1 for the two components (``MF15-like''). We include a comparison at 220\,GHz for the SO-LAT survey \citep{Ade19}, highlighting synergies with SO for whom constraining dust composition is also a principal Galactic science goal \citep{SO_GS_2021}. The sensitivity and frequency lever arm of the \pcam observations will allow us to assess which of these models, if any, correctly predict the polarization spectrum up to 850\,GHz.

The question of dust homogeneity is one of potential importance for CMB experiments. The existence of multiple dust species each with unique emission and polarization properties are a known challenge for parametric component-separation methods \citep[e.g.,][]{Kogut2016,Remazeilles2016,Hensley2018}. Also, multiple dust components, each having a unique spatial distribution and SED, leads to frequency decorrelation, a key challenge for widely used foreground mitigation algorithms \citep[see discussion in, e.g.,][]{bicep/keck/2018,CMBS4_2020,Pelgrims:2021}. Using sensitive, high-frequency polarimetry of a diffuse region like Pyxis to constrain the presence of multiple dust components thus directly informs dust mitigation strategies, such as appropriate parameterizations of the dust SED.

A diffuse region like the nearby translucent cloud Pyxis is the ideal laboratory for establishing the composition of dust in atomic gas. Nevertheless, \pcam will also target a number of star-forming molecular clouds (see Section~\ref{sec:bfield_sf}), which are denser environments than Pyxis. In the densest regions of these molecular clouds, specifically toward filaments and dense cores, dust grains are expected to grow in size via coagulation and to form ice mantles, thereby changing their grain chemistry and emission properties. While molecular clouds are a more complex environment than diffuse clouds, they provide an important test of dust grain models under denser conditions that lead to star and planet formation. The scales over which these dust properties begin to substantially change are not well constrained. Thus, by comparing \pcam observations of dust emission and polarization as a function of column density from the diffuse edges of molecular clouds to their denser interiors, we can chart the evolution of dust properties as grains are processed.

Another test of dust grain models in molecular clouds is the efficiency with which grains align with the local magnetic field. Dust grains in the cold dense cores are shielded from UV radiation from the interstellar radiation field and may therefore lack the suprathermal rotation necessary to keep the dust well aligned with the magnetic field. As a result, dust grains at the centers of dense cores may have lower alignment efficiencies and thus collectively produce less polarized emission.  Evidence for this phenomenon can be found in the tendency of the total emission to increase toward high column densities within clouds, without a corresponding increase in the polarized intensity \citep[e.g.,][]{andersson_2015,PattleFissel2019}. This effect complicates inference of the magnetic field geometry from the polarized emission, since the densest regions may contribute minimally to the polarization signal. \pcam will resolve polarization across the molecular cloud from large-scale diffuse emission to emission from many high density starless cores embedded deep within the molecular clouds (see Section~\ref{sec:bfield_sf}). Thus, we can quantify the grain-alignment efficiency and test theories of dust grain alignment mechanisms as the dust grains themselves evolve over different column densities and temperatures within star-forming regions.

\subsection{Measuring Properties of Magnetic Turbulence in the Diffuse ISM with Prime-Cam}
\label{sec:turbulence}

The structure of the ISM is influenced on many scales by magnetohydrodynamic (MHD) turbulence. Energy injected into the ISM by stellar winds, supernovae, and large-scale Galactic processes cascades down a range of scales until it is dissipated. The properties of MHD turbulence affect a wide range of physics --- including mediating phase transitions in interstellar gas --- that shape the formation of interstellar structures and the properties of star-forming clouds \citep{Elmegreen:2004}. Despite its ubiquitous importance in ISM processes, the properties of interstellar turbulence are poorly understood. To understand turbulence, which correlates magnetic field and density structures across a range of scales, it is necessary to make high spatial dynamic range observations of the diffuse ISM. With \pcam we will be able to make high-resolution observations of the polarized dust emission in intermediate- to high-column density regions of sky ($N_\mathrm{H} \gtrsim 10^{21}$ cm$^{-2}$). 

\begin{figure*}
    \centering
    \includegraphics[width=\textwidth]{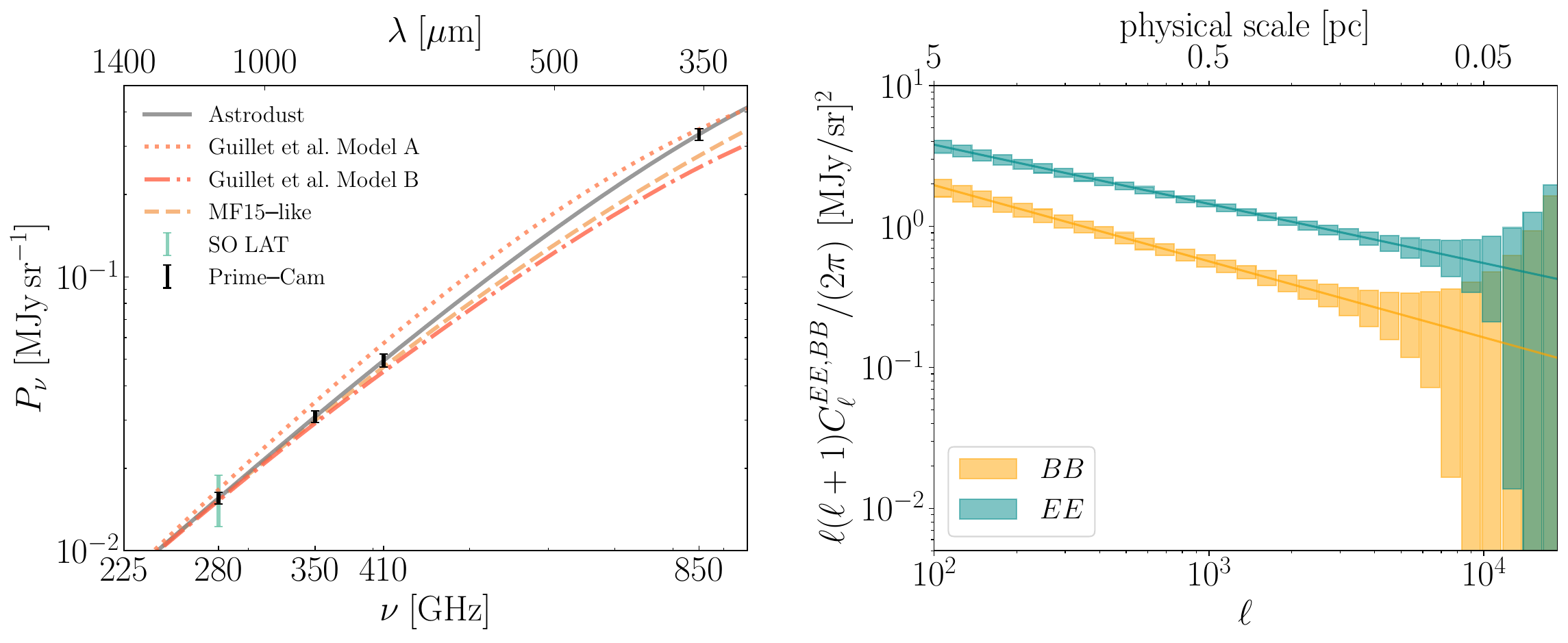}
    \caption{Forecasted dust polarization spectrum (left) and power spectrum (right) of dedicated observations of the Pyxis Cloud. In the left panel, the forecasted observations of a single 15$^{\prime}$~(0.8\,pc) region are compared with several dust models, all consistent with {\it Planck} data, but able to be differentiated among with \pcam observations. A forecast for the SO LAT survey at 220\,GHz is presented for comparison. Even more stringent constraints can be obtained by averaging over independent regions within the cloud. The right panel shows the forecasted $EE$ and $BB$ power spectra of a 350 GHz map of Pyxis. The spectral indices match the \textit{Planck} fits to 353 GHz power spectra computed over $70\%$ of the high-Galactic-latitude sky. Error bars are computed from the Prime-Cam noise model from \citet{choi20} for logarithmically-spaced multipole bins. }
    \label{fig:PyxisPS}
\end{figure*}

Rich information on the structure of polarized dust emission at low- and mid-Galactic latitudes will be obtained as a component of the \pcam Wide Field  Survey (WFS). Fig. \ref{fig:dustSNR} shows the effective resolution of these dust polarization measurements at 350 GHz, compared to \textit{Planck} 353 GHz measurements. We show contours of the resolution to which the WFS data must be degraded in order to achieve 3$\sigma$ measurements of the polarized dust intensity. The \textit{Planck} effective resolution is computed directly from the publicly available R3.01 Stokes parameters and noise covariance information \citep{planck/2020}. We forecast the sensitivity of the \pcam system to Galactic dust by comparing a PySM simulation of polarized dust emission at 350 GHz \citep{Thorne:2017} to the rms noise at each resolution using the \citep{choi20} noise model for a nominal 4000 hour survey over $60\%$ of the sky. At its highest resolution ($5'$), \textit{Planck} made $3\sigma$ measurements of the 353\,GHz polarized intensity over only a small strip of the inner Galactic plane. At equivalent sensitivity, \pcam will map nearly $1/3$ of the sky, with even higher effective resolution measurements in the inner Galactic plane and in targeted observations of selected regions, as described in Table \ref{tab:poln_surveys}. These data can be compared to maps of neutral hydrogen and other high-resolution gas tracers \citep[e.g.][]{Peek:2018}, to determine the small-scale structure of the dust polarization \citep[][]{Clark:2015, ClarkHensley:2019}.

\begin{figure}
    \centering
    \includegraphics[width=0.5\textwidth]{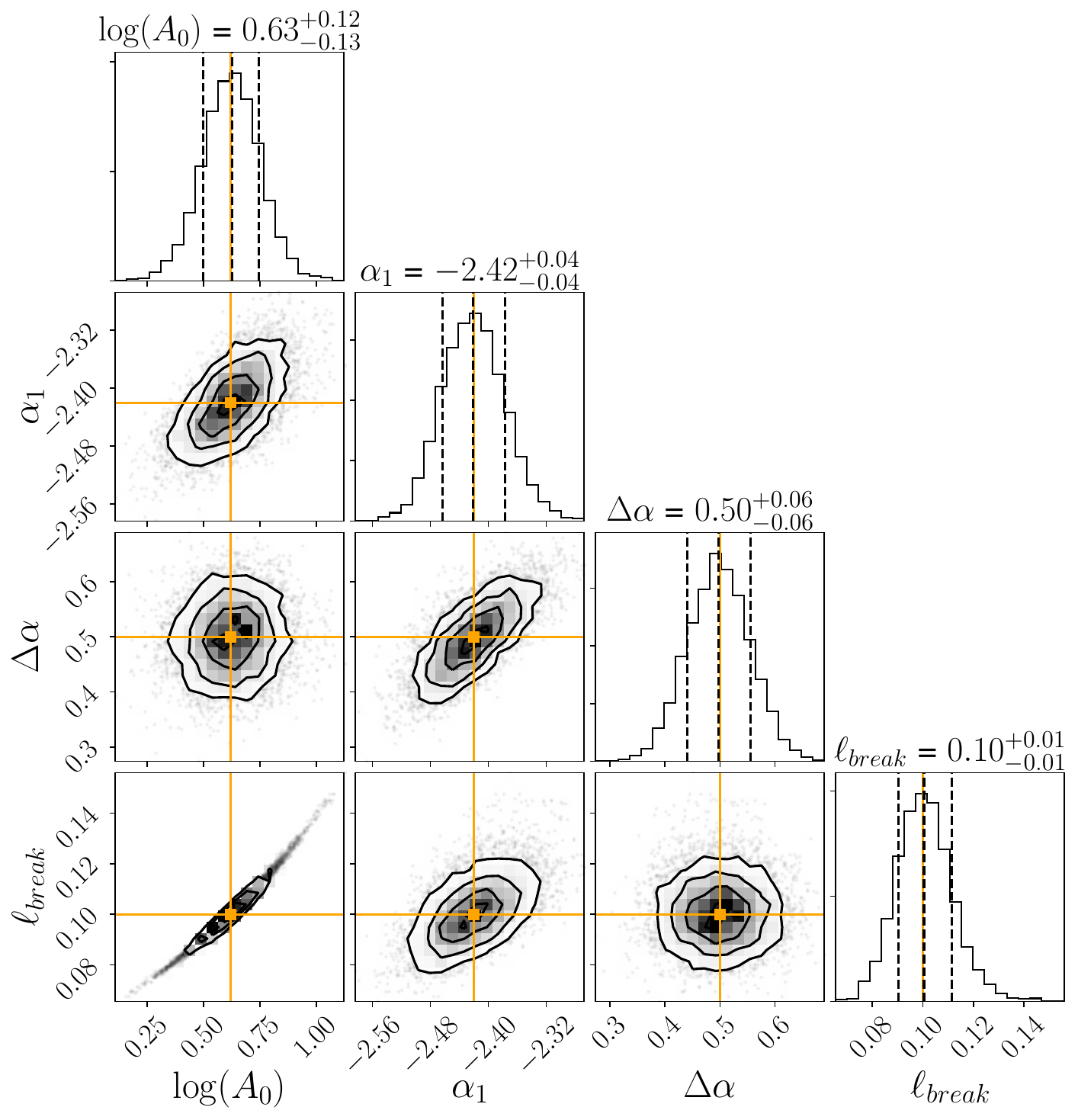}
    \caption{Distributions of broken power law parameters for forecasted measurements of a dust polarization $EE$ power law spectrum toward the Pyxis Cloud (see Figure \ref{fig:PyxisPS}). We fit four parameters: the power spectrum amplitude $A_0$, the slope of the power spectrum before the break $\alpha_1$, the change in the power spectrum slope $\Delta\alpha$, and the power spectrum break scale $\ell_{break}$. The parameters shown correspond to a break in the $EE$ power spectrum at 0.1 pc, which we will constrain to within 0.01 pc.}    
    \label{fig:cornerplot}
\end{figure}

The Prime-Cam map of Pyxis will provide excellent data for studying power spectra of polarized dust emission. The distance to Pyxis is approximately 175\,pc, based on 3D dust reddening measurements \citep{Capitanio:2017}. The sensitivity and angular resolution of \pcam polarization measurements at 350\,GHz will allow us to make high-fidelity measurements of the polarized dust power spectra toward Pyxis. An open question in interstellar turbulence is how (and at what scale) energy from the turbulent cascade is dissipated into the ISM \citep[e.g.][]{Elmegreen:2004}. The dissipation scale should be associated with a break in the power spectrum on small scales. This break is expected to occur at $0.01-0.5$\,pc, depending on the properties of the medium, if energy dissipation is driven by turbulent ambipolar diffusion \citep[e.g.][]{Momferratos:2014, Miville-Deschenes:2016}. The known distance to Pyxis will allow the power spectra to be associated with a physical scale (see the right panel of Figure \ref{fig:PyxisPS}), and critically, the proximity of Pyxis will enable a search for this break in the power spectrum. Figure \ref{fig:cornerplot} is a corner plot \citep{corner} showing the posterior distributions of broken power law parameters for forecasted measurements of the $EE$ power spectrum.

\subsection{The Large Magellanic Cloud: A Multi-scale View of Magnetic Fields}
\label{sec:BinLMC}
The role of magnetic fields in SF and the lifecycle of galaxies is a complex outstanding problem, made difficult by the paucity of high-resolution polarimetric observations. With \pcam we will make a dedicated, high-resolution map of the polarized dust emission in the Large Magellanic Cloud (LMC). Observing the LMC with \pcam affords a unique opportunity to study magnetic fields across different SF environments. On large scales, our dedicated high-resolution polarization survey of the LMC will allow us to study magnetic fields on the scales associated with turbulent energy injection. 
Maps of the magnetic field structure in the dust can be compared with magnetic field tracers of other ISM phases, such as Faraday rotation measurements with the POSSUM survey \citep[][]{Gaensler2010}, and used in conjunction with atomic and molecular line observations of the LMC at comparable angular resolution, e.g. with the Galactic Australian SKA Pathfinder \citep{Dickey2013}, the SKA \citep{McClureGriffiths2015}, and CHAI surveys of the LMC.

\begin{figure}
    \centering
    \includegraphics[width=0.45\textwidth]{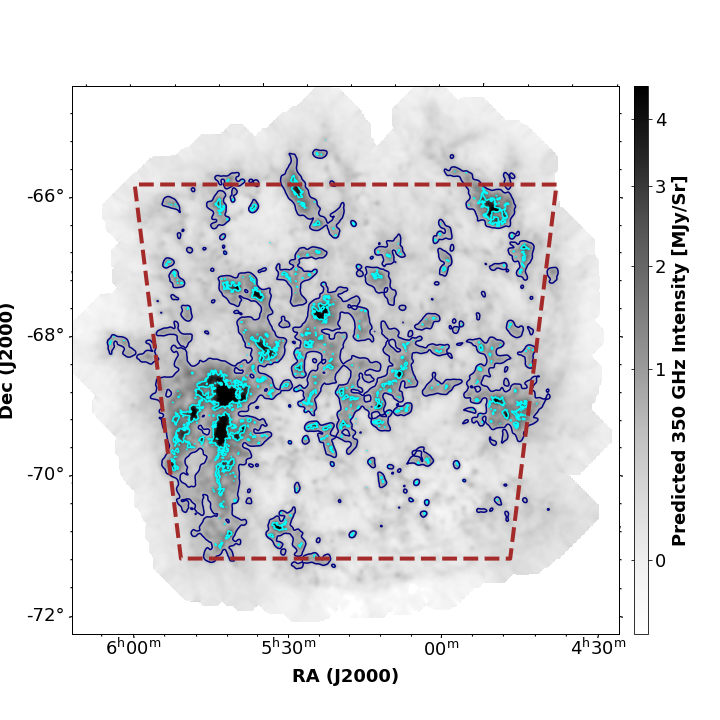}
    \includegraphics[width=0.45\textwidth]{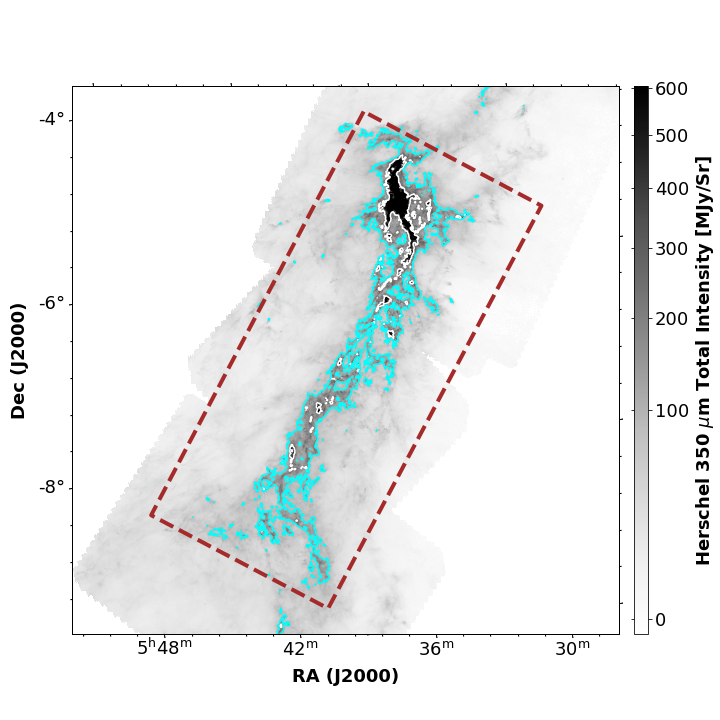}
    \caption{\textit{Left Panel:} Predicted regions of the Large Magellanic Cloud where Prime-Cam's 100-hour survey of a 5.5$\times$5.5$^{\circ}$ region would be expected to obtain $\geq$3-sigma 350\,GHz polarization detections  at full 9\,pc resolution (cyan contours) and smoothed to 50\,pc resolution (blue contours), assuming the dust in the LMC is 3$\%$ polarized.  The background image shows a predicted 350 GHz total intensity map based on the \textit{Herschel} HERITAGE Survey of the LMC. \textit{Right Panel}: \textit{Herschel} 350 $\mu$m (850 GHz) intensity map of the Orion A molecular cloud, with contours showing the regions where our planned 88-hour, 850 GHz Prime-Cam survey will make $\geq$3-sigma polarization detections assuming 3$\%$ (cyan) or 1$\%$ (white) polarization levels. In both panels dashed brown polygons show the area that we plan to survey with Prime-Cam.}
    \label{fig:lmc_oriA}
\end{figure}

At a distance of ($50\,\pm\,1$)\,kpc, full-resolution \pcam observations of the LMC at 350\,GHz will correspond to a physical resolution of 9\,pc.  The left panel of Figure \ref{fig:lmc_oriA} shows a map of the predicted 350\,GHz total intensity of the LMC based on the \textit{Herschel} SPIRE 500 $\mu$m intensity scaled assuming a dust spectral index $\beta\,=\,1.8$, and using the publicly available dust temperature maps from the \textit{Herschel} HERITAGE project archive \citep{meixner2013}.  Contours indicate where we estimate that \pcam will detect polarized intensity with $\geq$3-sigma significance (corresponding to polarization angle uncertainty of $<\pm$10$^{\circ}$) at the full 9\,pc resolution (cyan) and also when smoothed to 50\,pc resolution (blue). For this estimate we have assumed a constant polarization level of 3$\%$, which is the median \textit{Planck} 353 GHz  polarization level towards the LMC (see Table \ref{tab:poln_surveys}).  Based on these predictions we expect to detect polarization in at least 96 out of the 272 molecular clouds identified in \citet{fukui2008}.

With this unprecedented resolution we will study the origins of magnetic fields, testing whether their magnetic field direction is consistent with the ordered component of the galactic magnetic field, or alternatively whether there are indications that feedback and galactic turbulence have decoupled the cloud fields from the large-scale fields.  Previous studies have had to rely on observations of the molecular clouds in the Galactic disk, which are complicated by line-of-sight confusion and viewing geometry \citep{Stephens2011}, or with CO polarization observations of a small number of clouds in a nearby galaxy \citep{Li2011}.  Our detailed survey will also complement ongoing far-IR polarization surveys of nearby bright galaxies with $>$\,100\,pc resolution,  such as the SOFIA Legacy program {\em Magnetic Fields in Galaxies}  \citep{Borlaff2021}, which is targeting 17 galaxies but will not resolve individual molecular clouds. With deep \pcam maps covering almost the entire LMC, we will map magnetic fields in hundreds of individual molecular clouds and connect the cloud fields to detailed large-scale magnetic field maps.

\subsection{Measuring the connection between cloud magnetic field properties and SF outcome}
\label{sec:bfield_sf}

Within clumps and filaments of dense gas, strong magnetic fields are proposed to inhibit movement of material across field lines (while the motion parallel to the field lines does not face any resistance), possibly delaying the onset of SF \citep{inutsuka_2015,walch_2015}. \citet{Stutz2016} proposed that magnetic fields, if they become concentrated enough, could lead to pinch instabilities that may aid in cluster formation and may be an important component in driving the dynamical state of high line-mass filaments.  Regardless of the proposed model, the degree to which magnetic fields affect the SF efficiency within molecular clouds is poorly constrained \citep{Krumholz2019}, primarily due to the difficulty in observationally tracing the magnetic fields that thread star-forming clouds   \citep{PattleFissel2019}. 
To address these central issues in our understanding of SF, we require: (1) observations tracing cloud-to-core magnetic field properties: and (2) maps of magnetic fields in a large sample of clouds over the relevant range in cloud masses and evolutionary stages. 

Figure \ref{fig:dustSNR} shows that the Wide Field Survey (WFS) \pcam 350\,GHz survey is expected to achieve high signal-to-noise detections of most of the Galactic plane at better than 1$^{\prime}$ resolution. Most Milky Way molecular clouds are observed to be located near the Galactic plane.  Dust emission is typically optically thin, therefore we would like to initially study clouds that have minimal line-of-sight confusion, where the magnetic field inferred from Prime-Cam polarization maps can be unambiguously attributed to one cloud. To investigate, we examined the molecular cloud catalog presented in \cite{miville-deschenes2017}, which uses all-sky 8$^{\prime}.5$ arcmin resolution observations of $b \pm 5^{\circ}$ from the \cite{dame1987} all-galaxy $^{12}$CO J\,=\, 1$-$0 survey.  Using the cloud data made available by the authors on Dataverse,\footnote{\url{https://dataverse.harvard.edu/dataset.xhtml?persistentId=doi:10.7910/DVN/QR9CFW}} we find that 354 clouds in the Galactic longitude range covered by the WFS have less than half their sightlines with more than one cloud at different distances, and also better than 1\,pc resolution (assuming a\,1 arcmin FWHM beam).  Future investigations with higher-resolution cloud catalogs such as the \textit{Herschel} Hi-GAL survey \citep{molinari2016}, may reveal that a larger fraction of nearby clouds have little line of sight confusion and can be used in our analysis of cloud magnetic fields.

With this sample we will investigate the ratio of turbulent and gravitational potential energy to magnetic energy for each cloud, using standard polarization analysis techniques, such as comparing the relative orientation of magnetic fields and cloud column density structures \citep{soler_2013, soler_2017, planck2016-XXXV} and by analyzing the joint probability distribution functions (PDFs) of polarization level and disorder in the magnetic field \citep{king2018}. We will also apply the Davis Chandrasekhar Fermi (DCF) Method, which uses the disorder in the inferred magnetic  field orientation angle to estimate the balance between turbulent and magnetic energy \citep{davis1951,chandra_1953,hildebrand_2009,houde_2009}.  We will apply these statistical measurements to clouds observed as part of the WFS survey, our targeted high-resolution \pcam maps of nearby clouds (see Table~\ref{tab:poln_surveys}), and also to synthetic polarization observations of numerical simulations \citep[e.g.,][]{king2018,seifried2019,reissl21} in order to determine which simulations best reproduce the properties of our observations. 

Dust polarization maps are only sensitive to the component of the magnetic field parallel to the plane of the sky. Our analysis of the magnetic field properties would be greatly enhanced by constraining the 3D structure of the magnetic field in each cloud. For clouds with lower levels of confusion (i.e., at least 5$^{\circ}$ degrees away from the Galactic plane), we plan to use a new technique developed by \citet{Tahanietal2018} based on Faraday rotation measurements (RMs) to probe the line-of-sight magnetic field (\blos).

Combining the \pcam dust polarization maps with  \blos\ 
maps will enable us to model the 3D magnetic field morphology of dozens of 
molecular clouds. For example,  \citet{Tahanietal2019} combined the obtained \blos\ observations from  \cite{Tahanietal2018}, with the available {\em Planck} plane-of-sky magnetic fields (\bperp ) to map the 3D morphology of magnetic fields in the Orion A cloud. They concluded that a bow-shaped magnetic field morphology around Orion A is the most probable candidate among the other possible 3D magnetic field morphologies in this region.  
This bow-shaped magnetic morphology is consistent with
the elongated molecular cloud formation scenario of \citet{inutsuka_2015} and MHD simulations of \citet{Inoueetal2018}. We will compare our 3D observations
, which will include a range of physical properties and observations in different environments, with the predictions of cloud-formation models. We can further test the accuracy of the determined 3D morphologies by analyzing the level of polarization in dust polarization observations \citep{chen2019}. 

High-resolution, large-area dust polarization maps from \pcam, combined with existing and future \blos\ observations will constrain the 3D magnetic field morphology of star-forming clouds, across a range of cloud masses and evolutionary states. The higher resolution of \pcam will allow for a more accurate comparison between \blos\ and \bperp\ than is possible using Planck data, particularly for more nearby clouds (distances $<1$\,kpc), where the \blos\ resolution at each point is typically $1'$ or less. We note that future RM catalogs obtained by the new and next generation radio surveys, such as the new VLA All Sky Survey (VLASS), the POSSUM\footnote{\url{http://www.dunlap.utoronto.ca/~askap.org/possum/}} survey~\citep{Gaensler2010}, and future SKA\footnote{\url{https://www.skatelescope.org/}} observations~\citep[][]{Healdetal2020}, will allow for more detailed \blos\ maps of molecular clouds, with at least 10 times more \blos\ detections per cloud. This increased source density in \blos\ maps and the improved resolution in \bperp\ observations (by \pcam) will enable more accurate and detailed 3D magnetic field observations of molecular clouds. We also plan to use expected SKA Zeeman measurements \citep{Robishawetal2015} to better constrain our 3D magnetic field models of molecular clouds.


 On smaller scales, molecular clouds host a rich range of substructure, including filaments, the densest of which are the preferred locations for most SF, and  dense cores, which are the precursors to individual stellar systems \citep{andre2014,Dunham14,Stutz2016}. With 15\,$^{\prime \prime}$~resolution at 850\,GHz (350\,$\mu m$), ~\pcam will be the first polarimeter able to directly observe the connection between magnetic fields within cores (at $\lesssim\,0.05$\,pc scales), through filament scales, and up to cloud scales (at $\gtrsim\,10$\,pc), directly tracing the polarization signal over a factor of more than 200 in size.  That is, the \pcam observations  bridge the present-day and critical observational gap between the low resolution (but all-sky) {\em Planck} maps and the maps of magnetic fields in protostars and disks on the much smaller scales observable with ALMA.  
\pcam will make an unbiased survey of seven nearby molecular clouds, spanning a wide range of masses and evolutionary states (see Table~\ref{tab:poln_surveys} for the planned targets), spending 30-90 hours per cloud for a total of 352 hours.  As an example, the contours in the right panel of Figure \ref{fig:lmc_oriA} shows the areas where we expect to make $\geq$3-sigma detections of polarized intensity, assuming 3$\%$ or 1$\%$ polarization levels (cyan and white contours respectively) in our planned 850 GHz survey of the Orion A molecular cloud.  These estimates are based on the survey depth from Table~\ref{tab:poln_surveys}, sensitivity estimates from Table \ref{tab:sens} and a Level-3 \textit{Herschel} SPIRE 350\,$\mu$m map downloaded from the \textit{Herschel} Science Archive that was first published in \cite{polychroni2013}.  If the dust is at least 3$\%$ polarized then we predict $>$100,000 independent polarization detections for our Orion\,A survey.
These surveys will allow us to probe the role of the magnetic field in regulating SF relative to that of turbulence, gravity, and feedback from previous generations of SF. Furthermore, our surveys target a range of cloud masses, densities, and levels of SF activity, enabling \pcam to explore how magnetic fields impact SF as a function of cloud properties.

\pcam will also investigate whether or not magnetic fields regulate SF on the scales of filaments and dense star-forming cores.  In strong field models, the magnetic field plays a fundamental role in the formation of dense cores and their subsequent collapse to form stars, disks, and planets \citep{mouschovias_1991,price2007, basu2009}.  In weak magnetic field models,  turbulence determines how cores and stars are able to form (e.g., \citealt{maclow_klessen_2004,offner2010,Li14}). Connecting clouds to core scales, recent observational studies proposed that the magnetic field in massive clouds has an effect on the cloud and filament gas dynamics \citep[e.g.,][]{Stutz2016,soler_2017,fissel_2019,alvarez21,gonzalez2019}.  \pcam provides a unique tool set with which one can test theoretical models through its large, unbiased, and sensitive dust polarization surveys of filaments and dense cores within molecular clouds. 

The combination of observing at 850\,GHz, where the cold dust emission peaks, and the high instrument sensitivity, will enable \pcam to conduct large surveys of entire core and filament populations in nearby clouds that cannot be achieved with existing or upcoming polarimeters (e.g., see \citealt{PattleFissel2019} for a review). In our seven targeted clouds we expect to resolve over a thousand cores and hundreds of filaments. As an example the \cite{mairs2016} JCMT/SCUBA-2 study of the Orion A cloud found 431 cores, which are all within our 11\,$deg^2$ survey region shown in the right panel of Figure \ref{fig:lmc_oriA}.  Even if the core polarization level is only 1$\%$, 201 of these cores should be detected in polarization at greater than a 3-sigma level, which corresponds to a measurement uncertainty in the inferred magnetic field orientation of $\leq$10$^{\circ}$. With this large sample we
 will investigate the connection between magnetic fields and molecular gas kinematics.  For example, the orientation of protostellar outflows are commonly used as an indicator of the angular momentum axis of accretion within dense cores. In magnetically regulated collapse, magnetic braking is expected to align the core angular momentum axis with the field orientation \citep[e.g.,][]{Allen2003,Hull2019}. To date studies of dozens of protostellar cores have found no statistical preference for an alignment of outflow directions and magnetic field orientations \citep[e.g.,][]{Yen2021BISTRO,Hull2019}.  However, searching for a 3-D dimensional alignment trend is complicated by the fact that both the magnetic field and outflow direction are observed in projection on the plane-of-the-sky. \pcam's polarization surveys will cover hundreds of protostellar cores, including both isolated cores as well as cores clustered within dense environments.  We will also use our observations to probe how magnetic fields affect accretion onto dense cores and dense filaments by searching for correlations between line-of-sight velocity gradients and magnetic field orientation \citep{Hu2020,Gomez18}.

Theoretical models also suggest that gravitational contraction could drag magnetic fields inward thus altering their orientation with respect to the larger-scale magnetic field.  For example, some strong-field models predict a pinched ``hourglass'' shape in dense cores \citep{GalliShu93, Myers18}. However, such profiles have only been seen for a handful of cores \citep{Girart06, Monsch18}. \pcam will enhance these statistics by both resolving the magnetic field structure of over a thousand dense cores, and also by quantifying any change in orientation between core, filament and cloud scale magnetic fields in these cores.  
Such observations will probe the interplay between magnetic fields and gravity, and how they might affect the initial mass of stars and their ability to form 
disks that eventually produce planets.  
Finally, \pcam uniquely offers the opportunity to resolve the magnetic field structure toward both starless cores and prestellar cores, i.e., cores that have not yet formed stars.

In summary, FYST/\pcam has both the sensitivity and resolution to map dust linear polarization across cloud scales, while sampling the conditions within clouds, filaments, and cores.  These observations will enable direct evaluation of the plane-of-the-sky magnetic field directions in the cradles of star cluster formation and its connection to the diffuse ISM.  Combined with modeling and independent information on the gas radial velocity field and density structure, the FYST/\pcam data will be a keystone in the testing of theoretical frameworks addressing the role of the magnetic field in star forming systems, 
as a function of fundamental cloud parameters, such as mass, mass per unit length, gravitational potential, and evolutionary stage.

\section{Probing Galaxy and Galaxy Cluster Evolution with Sunyaev-Zeldovich effects}

\label{sec:sz}

\begin{figure*}[t!]
\begin{center}
\includegraphics[width=0.485\textwidth]{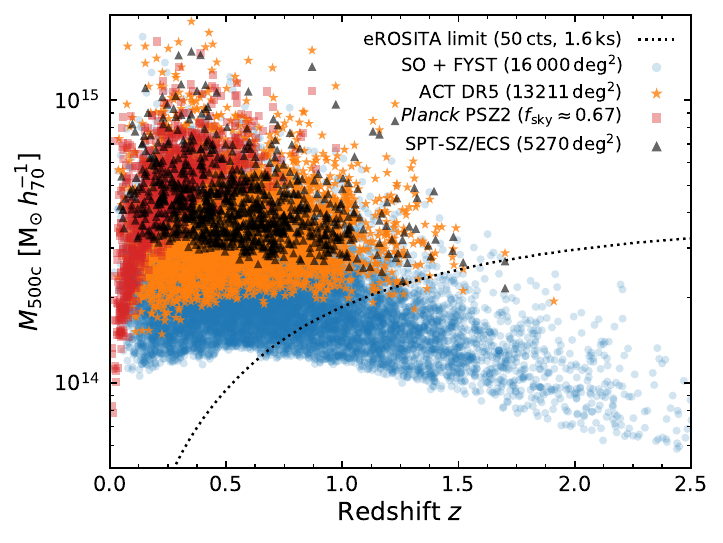}
\quad
\includegraphics[width=0.485\textwidth]{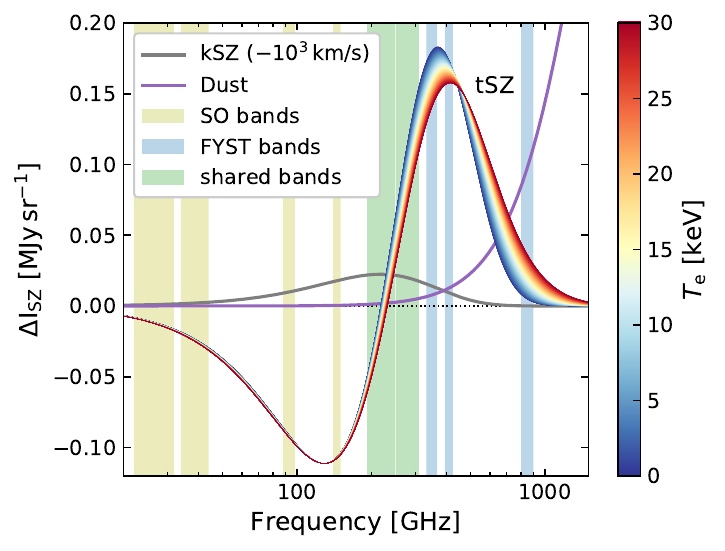}
\end{center}
\caption{\small
FYST is planned to perform a sub-millimeter follow-up survey of the sky area surveyed by AdvACT and SO, providing sub-millimetre observations of around 16{,}000 galaxy clusters (\textit{left}). Compared to current SZ-selected cluster samples like those offered by \textit{Planck} \citep{Planck_PSZE, Planck_PSZE2}, SPT \citep{Bleem15, Bleem20, Huang20} and ACT \citep{Hasselfield13, Hilton20}, the predicted SO+FYST cluster sample includes galaxy clusters down to lower masses of approximately $10^{14} \, \mathrm{M}_\odot$ and out to higher redshifts of up to $\sim 3$, which is attributed to the lower noise and spectral coverage of SO+FYST. We note that the addition of FYST measurements to SO measurements will not yield significantly more clusters than SO measurements alone. The combination of AdvACT/SO data with data from FYST allows for complete coverage of the spectra of the thermal and kinematic SZ effects (\textit{right}) at vastly improved sensitivity and spatial resolution compared to \textit{Planck}. Observations with FYST above the tSZ null at around 220\,GHz are especially valuable for temperature measurements of the ICM via the rSZ (colour-coded spectra as function of electron temperature $T_\mathrm{e}$) and for characterising the FIR emission from galaxy clusters that is attributed to warm dust grains in cluster galaxies or the ICM.}

\label{fig:SZ1}
\end{figure*}

CMB photons encounter two effects on their way from the last scattering surface to our telescopes: they are deflected by the gravitational potentials of the large-scale structures, and scattered by free electrons and neutral/partially ionized atoms. The phenomenon of scattering of CMB photons by unbound and highly energetic electrons is called the Sunyaev-Zeldovich (SZ) effect \citep{SZ1970, Sunyaev72}, which is the main scientific focus of this section. The SZ effect comes in multiple flavors, carrying the imprint of various contributing factors of the electron velocity distribution, and disentangling these effects promises new breakthroughs in astrophysics and cosmology. With its high-frequency leverage, FYST will pioneer some of these studies, and in so doing, will also probe the distribution and time-evolution of dust grains within galaxy cluster environments.

In Section \ref{sec:subsec-szmain} the main themes of the SZ science are discussed, along with the impact that FYST will have on improving the galaxy cluster number counts, determining cluster temperatures via relativistic-SZ spectral distortions, and measuring the cosmic growth of structures via the kinematic effect. All these science targets will be accomplished in synergy with lower-frequency data from the current and upcoming CMB experiments. Section \ref{sec:subsec-szpower} discusses the specific contribution of FYST in constraining the thermal-SZ angular power spectrum and the leverage from cross-correlation studies; in Section \ref{sec:subsec-dustmass} it is outlined how our understanding of the distribution and composition of dust grains in galaxy-cluster environments will be improved; and  lastly, in Section \ref{sec:subsec-ntsz}, an example of a new method for detecting the nonthermal SZ effect for galaxy clusters is described. 

\subsection{SZ Science with Prime-Cam}
\label{sec:subsec-szmain}

Galaxy clusters stand at the nexus between astrophysics and cosmology. Large, statistical samples of clusters have been produced by several observational techniques across multiple wavelength regimes, from X-ray to optical to millimeter wave. Among these various methods to find and characterize galaxy clusters, probably the most rapidly developing is the Sunyaev-Zeldovich (SZ) effect \citep{SZ1970, Sunyaev72, Birkinshaw99, Carlstrom02, kitayama:2014, Mroczkowski19}. The SZ effect has two main variants, the thermal SZ (tSZ) effect and the kinematic SZ (kSZ) effect. The tSZ effect arises from inverse Compton scattering of the CMB photons by hot free electrons, has a unique spectral signature, and its amplitude connects to the total cluster thermal energy. The kSZ effect comes from the Doppler shift of CMB photons that have scattered off free electrons that have a non-zero line-of-sight peculiar velocity, to first order has a blackbody spectrum like the CMB, and its amplitude is proportional to the line-of-sight momenta of these electrons. There is also the full, relativistic spectrum of the tSZ effect (referred also as the relativistic SZ, hereafter the rSZ effect) that potentially carries information about the mean temperature of the scattering electrons \citep[e.g.,][]{Wright79, Nozawa97,Itoh98, Chluba12}, as well as higher-order polarized effects that are probably too weak to detect.

Several thousand galaxy clusters have been identified from observations of the tSZ \citep[e.g.,][]{Hasselfield13, Bleem15, Planck_PSZE, Planck_PSZE2, Hilton20, Bleem20}, enabling cosmological studies of structure growth based on their number counts and angular correlations; however, current cosmological constraints are limited by systematic uncertainties in cluster properties. Further characterization and understanding the cluster gas properties such as the pressure, temperature, density profiles from tSZ, kSZ, and rSZ measurements aid in reducing the systematic uncertainties in cluster properties and provide valuable information on the detailed astrophysical properties within clusters, like the role that active galactic nuclei (AGN) play in heating of the intracluster gas. A current-generation CMB experiment, Advanced ACTPol \citep[AdvACT] {Henderson16}, and next-generation Simons Observatory \citep{Galitzki18, Ade19} provide complete sky overlap with \pcam's large area survey and are expected to find 16,000 clusters through their tSZ signal (see left panel of Fig.~\ref{fig:SZ1}). \pcam's 220--850\,GHz measurements, combined with 30--270\,GHz data from AdvACT and SO is 
forecasted to enable the detection of all three SZ components (cluster optical depth, bulk velocity, and temperature) \citep{Mittal18} and thermal dust emission from individual clusters \citep{Erler18}, for a large, statistically significant cluster sample. The significance of such a detection will depend on the achieved sensitivities of Prime-Cam. Among planned experiments, only FYST will provide observations with sufficiently broad coverage of the SZ increment (see right panel of Fig. \ref{fig:SZ1}), and will do so with 5 to 20 times better resolution (depending on frequency) than \textit{Planck}, the current standard for panchromatic SZ science \citep{Planck15_SZ}. In addition, spectral imaging of the brightest clusters with EoR-Spec will be possible and could enable SZ component separation. This would represent an evolution of this field towards using spectroscopic SZ science to understand clusters and cosmology. One example of such higher-order spectral distortions that could be unlocked with SZ spectral imaging is the thermal-kinetic SZ effect (which is the relativistic correction of the kSZ effect), whose detection possibility is discussed in a recent paper by \citet{Coulton2020}.


Advances have been made recently in our theoretical and numerical modeling of how galaxies form and evolve over cosmic time. These theoretical constructs are challenged and verified through observations of the physical and thermodynamical properties of the baryons in galaxies and clusters. High signal-to-noise cross-correlation measurements of the tSZ and kSZ effects provide independent windows into the thermodynamic profiles of ensemble-averaged groups and galaxies~\citep{Battaglia17,Schaan2020,Amodeo2021}. These measurements will quantify and constrain the  processes that make star formation globally inefficient, such as energetic feedback and non-thermal pressure support. Thus, cross-correlations between FYST observations and galaxy, group, cluster, or quasar samples will probe the baryonic processes important for galaxy formation, such as energetic feedback and non-thermal pressure support.

\subsection{\pcam leveraging ongoing experiments}
\label{sec:subsec-szpower}

\begin{figure}[t]
    \centering
    \includegraphics[width=0.5\columnwidth]{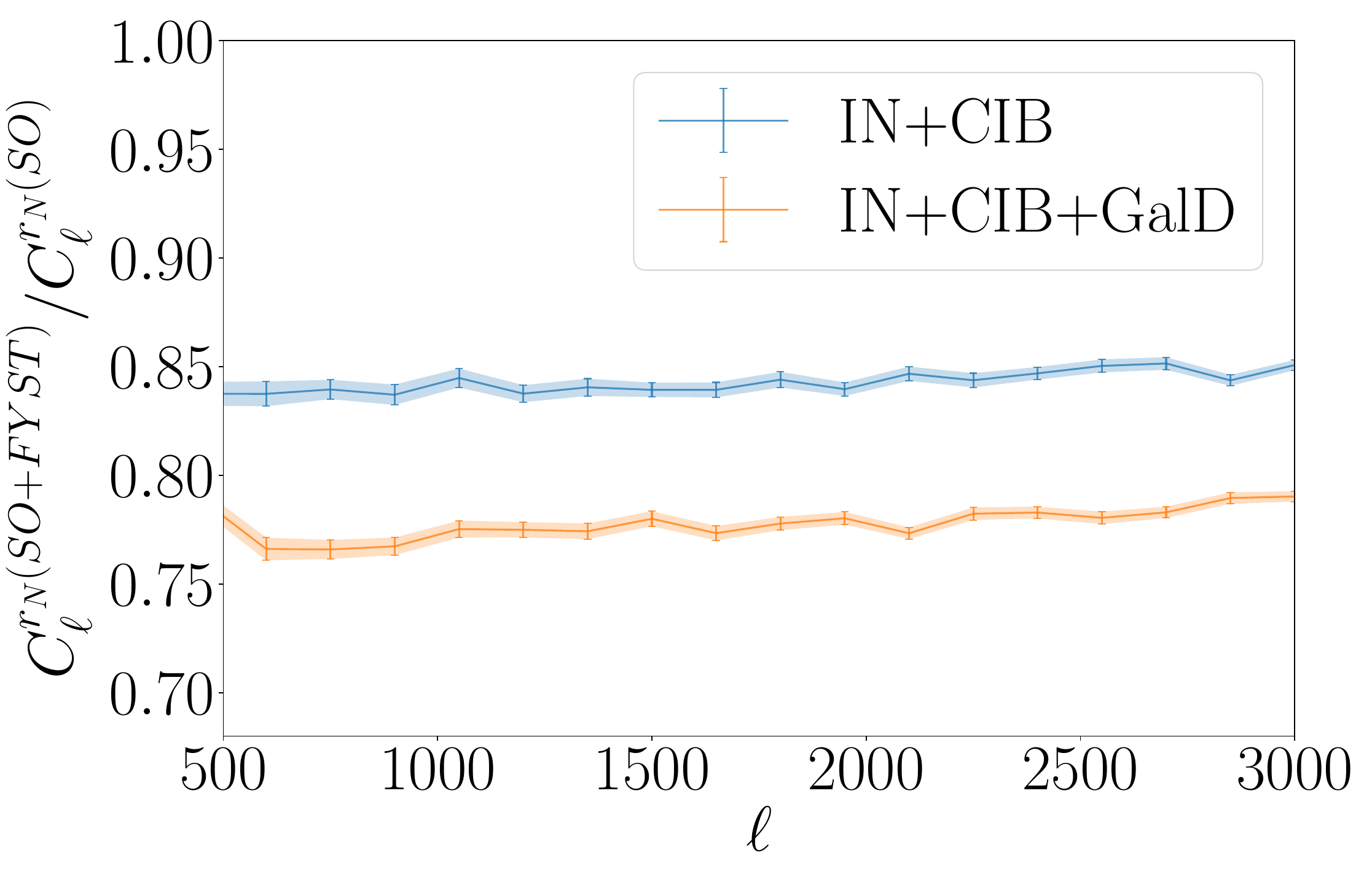}
    \caption{Ratio between the tSZ power spectrum of the cumulative ILC residual(noise $C_{\ell}^{r_N}$) for SO+FYST combined and SO alone. The ILC residual noise, is the noise leftover by contaminants in the reconstructed map, the difference between the recovered signal and the expected signal. The blue trace is the case where the cumulative residual noise is composed of instrumental noise (IN) and CIB. The orange trace is the case where the cumulative residual noise is composed of IN, CIB, and Galactic Dust (GalD). We see that the ratio of SO+FYST versus SO is around $0.84$ for IN+CIB and $0.78$ for IN+CIB+GalD. The error bars represent the standard error of the binned power spectra over a window $\Delta \ell = 150$.}
    \label{fig:11}
\end{figure}

A major hurdle for these SZ measurements will be their clean separation from the intrinsic thermal dust emission from galaxies and clusters, which we will refer to as "CIB leakage". This separation requires the multi-wavelength coverage from millimeter to submillimeter wavelengths that FYST provides to avoid biased SZ measurements \citep{Basu2019}. For example, in cluster cosmology, an uncontaminated tSZ map will allow for a more accurate cluster selection function \citep[e.g.,][]{Melin18}. Quantifying the improvement FYST high frequencies coverage brings to better separate tSZ and CIB is therefore crucial. Simulations of a a mock sky, using templates maps from the WebSky simulations \citep{Stein2019, Stein2020}, containing tSZ, CIB, and instrumental noise at the frequencies, sensitivities, and resolution of SO alone or SO and FYST combined (SO+FYST) show that, when using a simple map-based ILC to retrieve the tSZ, the power spectrum of the cumulative residual noise ($C_{\ell}^{r_N}$), i.e. the total noise left by the contaminants in the reconstructed ILC map, is $16\%$ less important for SO+FYST than it is for SO alone. Moreover, when adding Galactic dust to the mock sky, the gain in combining SO with FYST is even more pronounced:  the power spectrum of the cumulative residual noise in tSZ is $22\%$ less than when SO alone is used (see Fig. \ref{fig:11}). This gain is due to the high-frequency coverage of FYST that better probe and constrain the Galactic dust that is the dominant contaminant (for more information on the method, modeling and some complementary results see Charmetant et al. in prep.). A harmonic-space, constrained ILC procedure, which nulls the CIB contribution, leads to a further 10\% improvement (that is $\sim 25$\% overall) for SO and FYST combined compared to SO alone when including 1/$f$ noise and all extragalactic foregrounds. The tSZ power spectrum is a highly sensitive non-linear probe of cosmology and astrophysics. For example, its amplitude scales like $\sigma_8$ to the eighth power so that a $20\%$ improvement in signal-to-noise ratio results in about a factor of 5 improvement in measuring $\sigma_8$. Improvements on tSZ power spectrum signal-to-noise can be roughly converted to real space measurements, for example a cross-correlation analysis, by taking their square-root of the power spectrum, assuming a fixed sample size. FYST therefore opens a new unbiased observational window into the thermodynamic properties of galaxies and clusters.



A limitation of the tSZ and kSZ measurements is the lack of redshift information. This can be overcome easily by cross-correlating the FYST-derived SZ maps with the distribution of galaxies with known spectroscopic (as well as good photometric) redshifts. This SZ {\it tomography} is a powerful technique, enabling the study of the thermodynamical properties of groups and clusters as a function of cosmic time \citep{vikram/lids/jain:2017,makiya/ando/komatsu:2018,pandey/etal:2019,koukoufilippas/etal:2020,chiang/etal:2020,yan/etal:prep} as well as its connection to the theory of the structure formation \citep{chiang/etal:2021}. The current cross-correlation measurements show a clear sign of the CIB contamination at $z>1$ \citep{chiang/etal:2020}, which FYST can help mitigate. 

\subsection{Measuring the dust mass and spatial distribution of dust grains in galaxy clusters}
\label{sec:subsec-dustmass}

Clusters are formed hierarchically through mergers and the accretion of unbound intercluster gas. Infalling galaxies experience tidal distortions and ram pressure, leading to increased rates of star formation and the stripping of enriched gas and dust \citep{Sarazin88}. This process can potentially enrich the ICM with warm dust grains, which are stochastically heated by collisions with hot electrons and re-emit the absorbed energy in the far-infrared \citep{Ostriker73, Dwek1990}. However, the lifetime of dust grains in the ICM is highly uncertain. In the cores of galaxy clusters, dust grains can be destroyed efficiently by thermal sputtering \citep{Draine79} while estimated grain lifetimes in the outskirts of clusters reach several billion years \citep{Dwek92,Vogelsberger19}. Dust in the ICM has been observed statistically by stacking FIR images of large cluster samples \citep[e.g.,][]{Montier05,Giard08,planck_dust,Planck_dust_SED}. These observations point to a low dust-to-gas mass ratio of around $10^{-4}$ and an SED similar to those of local star-forming galaxies \citep{Planck_dust_SED, Erler18}. However, there is no conclusive evidence for the presence of dust in the ICM of local galaxy clusters. Recent theoretical work suggests that the majority of dust grains in clusters at low and intermediate redshifts is instead bound in starforming cluster galaxies \citep{Gjergo20}. 

Understanding the nature and properties of FIR emission from galaxy clusters is a major challenge for precise measurements of the SZ effects at sub-millimeter wavelengths. Recent spectral studies of the SZ effect highlight the importance of including a dedicated spectral component for the cluster FIR emission and constraining it with high-frequency observations above $220 \, \mathrm{GHz}$ \citep{Erler18, Schaan2020}. Other works suggest that up to $9\%$ of clusters between redshift $0.5$ and $0.8$ have been missed in the creation of the \textit{Planck} cluster catalog due to FIR contamination \citep{Melin18}. Insights into the properties of dust in clusters will be especially important for the analysis of proto-clusters at redshift $z>2$ \citep{Casey16, Cheng19, Smith19}. Clusters at these redshifts are not fully virialized yet, so their ICM has not reached the temperature necessary for efficient sputtering of dust grains \citep{Gjergo20}. 
With its access to submillimeter frequencies combined with having more than five times better angular resolution than \textit{Planck}, FYST will help to measure the spatial distribution of diffuse intergalactic dust in the dense cluster environments, and will inform the simulations regarding the origin and lifetime of such dust grains.

\subsection{Observations of the non-thermal SZ effect}
\label{sec:subsec-ntsz}

\begin{figure}[t]
\begin{center}
\includegraphics[width=0.485\textwidth]{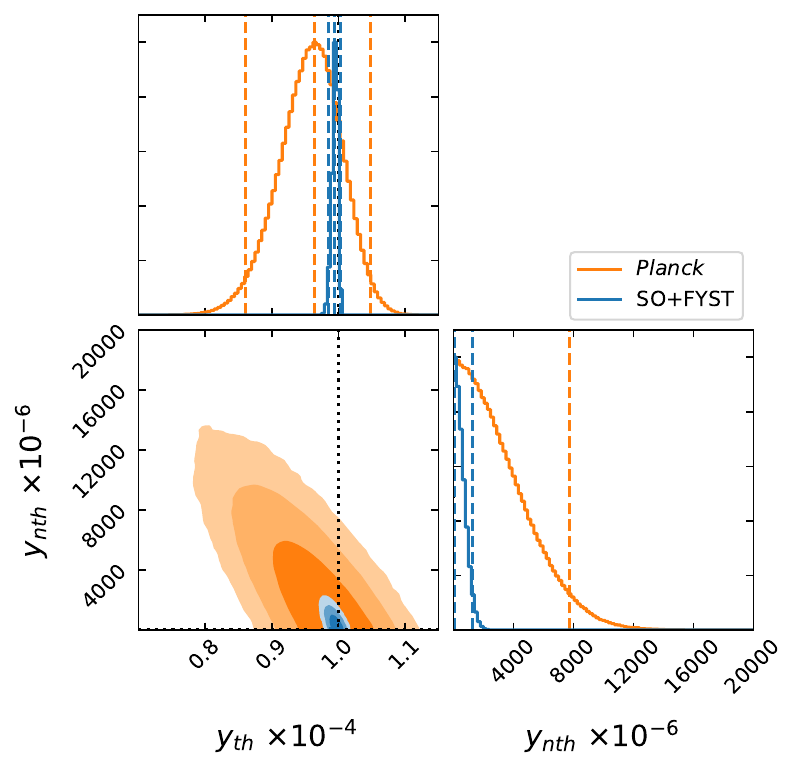}
\quad
\includegraphics[width=0.485\textwidth]{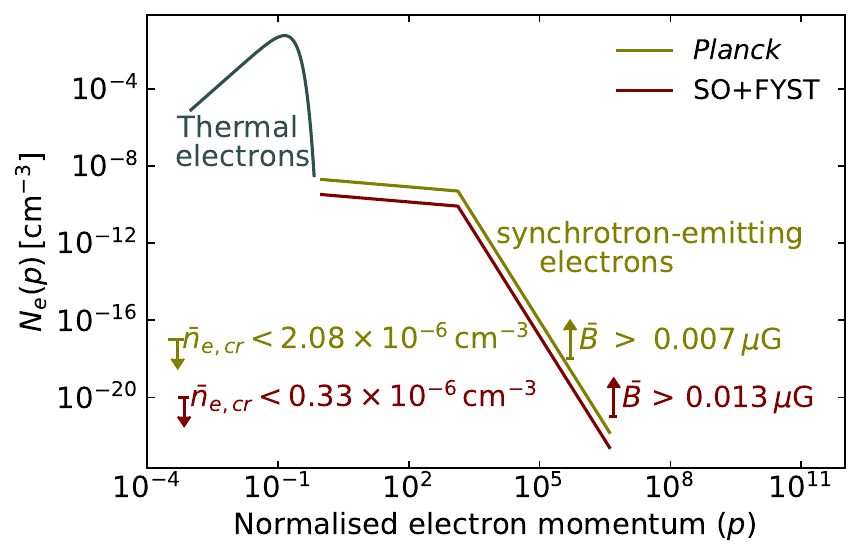}
\end{center}
    \caption{Left: Two-dimensional posterior and the one-dimensional marginalised posteriors of the thermal and non-thermal Compton-\textit{y} parameters ($y_{th}$ and $y_{nth}$, respectively) are computed using simulated spectra of the SZ effect considering the combined sensitivities of SO+FYST, in comparison with the posteriors computed using \textit{Planck} sensitivities from data. The contours in the 2D distribution correspond to the 68.3\%, 95.4\% and 99.7\% credible intervals. Right: Constraints on the volume-averaged number densities of cosmic-ray electrons, and the corresponding magnetic field strengths, for a broken power-law electron distribution with $p_\mathrm{break}=1000$. The {\it Planck} limits are obtained from stacking the spectra of 62 known radio-halo clusters, and the SO$+$FYST limits are a forecast for 200 clusters with similar radio halo properties. For comparison, the average magnetic field, using the assumed density profile and spectral index of electrons and a central value of 1$\,\mu$G, will be 0.09$\,\mu$G. 
    }
    \label{fig:ntsz}
\end{figure}

The non-thermal SZ (ntSZ) effect provides information about the highly relativistic cosmic-ray electron populations in the ICM \citep{Ensslin2000,Colafrancesco2003,Colafrancesco2013}. The spectral signature of the ntSZ effect is similar to that of the rSZ effect but is shallower and smaller in amplitude than the rSZ spectrum. Since the ntSZ null is different from the rSZ null, one can exploit this difference in the spectral distortions to distinguish between the thermal and cosmic-ray electron populations. Typically, the overall contribution to the energy budget in clusters from these non-thermal electrons is only about 1\% \citep[e.g.,][]{Zandanel2014}. This makes a direct detection of the ntSZ effect incredibly difficult, even with the sensitivities of FYST. However, one can place meaningful upper limits on the mean number densities of relativistic electrons by stacking the spectra of a large number of clusters. When this exercise is carried out for clusters hosting radio halos (RHs), the known average synchrotron flux then provides a lower limit on the volume-averaged magnetic field. The limits are obtained under the assumption that the non-thermal electrons causing the ntSZ effect are the same population of electrons that emit synchrotron radiation in radio halos. In Muralidhara et al. (in prep.), this analysis was performed for a sample of known RH clusters using the multifrequency data from the {\it Planck} satellite, and then a forecast was made for 200 RH clusters using simulated sky maps for SO and FYST. The posterior probability distributions of the amplitudes of the tSZ and ntSZ effects, and the resulting limits on the cluster-averaged non-thermal electron densities (as well as \textit{B}-field values) are shown in Fig. \ref{fig:ntsz}. The SO$+$FYST data can provide highly competitive constraints on the average magnetic field within clusters, and when combined with results from Faraday rotation measurements \citep[e.g.,][]{Bonafede2011}, can rule out certain simplistic power-law models for cosmic-ray electrons. The value of adding FYST data in this regard is to gain better control over foregrounds, especially dust emission, which will in turn enable a better success with the matched filtering and spectral fitting techniques to get constraints on the ntSZ signal.

\section{Rayleigh Scattering a New, Blue Surface of Last Scattering}

\label{sec:rs}
\begin{figure*}[t!]
\begin{center}
\includegraphics[width=\textwidth]{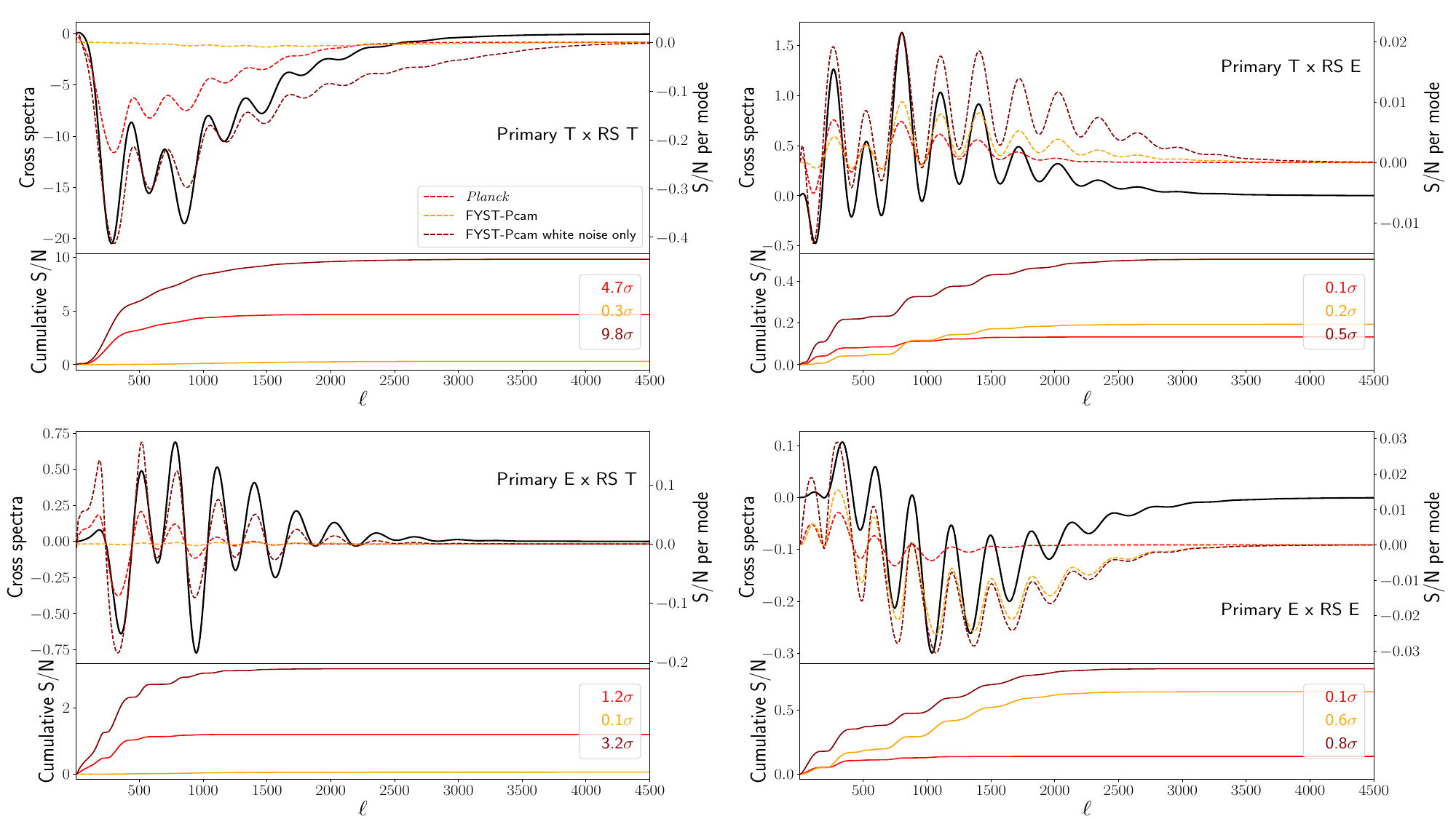}
\end{center}
\caption{Detectability of the four primary $\times$ Rayleigh cross-power spectra for three experiments:  {\it Planck}  (red);  \pcam (orange); and \pcam white noise only (dark red). Black lines show the cross-spectra wile coloured dashed lines show the signal-to-noise per $\ell$-mode. The bottom panels display the cumulative signal to noise. Note that these forecasts do not include any foreground contamination.}
\label{fig:SN_CCAT_PLANCK}
\end{figure*}

 Just after recombination, Rayleigh scattering of neutral species can generate a secondary CMB signal. Unlike the primary signal, generated by Thomson scattering, this signal is frequency dependent. The unique frequency coverage of \pcam could potentially help us detect this signal, which eventually will benefit cosmological parameter inference. This section explores the feasibility of a first detection of the Rayleigh signal using \pcam on FYST.  

In Section \ref{sec:RS_intro} we briefly review the physics of CMB Rayleigh scattering and cosmological constraining power enabled by its measurement. In Section~\ref{sec:RS_detectability} we discuss the detectability of the Rayleigh signal. We will comment on the role of the atmosphere and show how this affects detectabilty prospects with \pcam. Finally, we will show the impact of foregrounds in Section~\ref{sec:RS_foregrounds}.

\subsection{Rayleigh scattering of the CMB}\label{sec:RS_intro}

Prior to recombination around redshift 1100, the Universe was optically thick due to frequent Thomson scatterings of CMB photons by free electrons. As electrons became bound into neutral hydrogen, the rate of Thomson scattering dropped, and the Universe became nearly transparent. However, the neutral species formed during recombination (primarily hydrogen and helium) also scattered CMB photons through Rayleigh scattering, the classical scattering process of long wavelength photons by the induced dipole of polarizable particles~\citep{rayleigh_lord_1881_1431155,takahara/sasaki:1991,Yu_01,Lewis_13,Alipour_14}. Rayleigh scattering exhibits a strongly frequency-dependent cross-section ($\propto \nu^4$)~\citep{rayleigh_lord_1881_1431155}. This additional scattering produces a frequency-dependent shift of the visibility function towards later time (smaller redshifts). At the power spectrum level, this results in the following effects:
\begin{itemize}
    \item A suppression of small-scale anisotropies both in temperature and $E$-mode polarization, caused by an increase of diffusion damping.
    \item On large angular scales, Rayleigh scattering primarily affects the  $E$-mode polarization signal. By shifting the last-scattering surface towards lower redshifts, where the local temperature quadrupole is larger, Rayleigh scattering boosts the large-scale $E$-mode signal. 
    \item Rayleigh scattering introduces frequency dependence in the size of the sound horizon, leading to a shift in the location of the acoustic peaks, both in temperature and $E$-mode polarization spectra.
\end{itemize} 

Rayleigh scattering of the CMB is a definite prediction of standard cosmology. However, a first detection has yet to be achieved and is further made challenging by the small amplitude of the signal. With sufficient sensitivity and frequency coverage, the additional cosmological information carried by the Rayleigh scattering signal can improve constraints on cosmological parameters. For example, forecasts show that an experiment like PICO~\citep{PICO_19} would be able to use Rayleigh scattering to improve constraints on the sum of the neutrino masses $\sum m_\nu$ by up to $50\%$. Similarly, the constraint on the number of relativistic species $N_{\rm eff}$ would be improved by $\sim 10\%$~\citep{Beringue_20}. The fact that the Rayleigh signal is effectively a second last-scattering surface could potentially also benefit the search for primordial non-Gaussianities \citep{Coulton_2021}, which could lead to an improvement greater than a factor of two compared to an analysis using only primary CMB modes. 

The broad frequency coverage ($270$--$850$\,GHz)  and wide survey area (Fig.~\ref{fig:surveys}) of \pcam could potentially contribute to a first detection of CMB Rayleigh scattering, especially in combination with surveys covering lower frequencies. Furthermore, with the location of FYST at an elevation at more than~$5600$ m above sea level, the assumptions made in the forecasts presented here are likely to be somewhat conservative with respect to the impact of the atmosphere on the observation of large-scales anisotropies. 

\subsection{Detectability of the Rayleigh scattering signal with CCAT-prime} \label{sec:RS_detectability}


The first detection of the Rayleigh signal will be in cross-correlation with the primary CMB. 
Despite several experiments measuring CMB anisotropies on large scales from space (e.g.~{\it Planck};~\cite{planck/2020}), as well as on smaller scales from the ground (e.g.~ACT~\cite{Thornton:2016wjq} and SPT~\citep{Austermann:2012ga}), there has not been any reported detection of Rayleigh scattering of CMB photons. As highlighted in Fig.~\ref{fig:SN_CCAT_PLANCK} {\it Planck} observed the CMB with sufficient depth that in principle a statistical detection of the CMB-Rayleigh cross-correlation could have been possible. The presence of astrophysical foregrounds, as well as limited sensitivity of {\it Planck} in polarization, are likely to have prevented a first detection. 


Similarly, Fig.~\ref{fig:SN_CCAT_PLANCK} shows that the raw sensitivity of \pcam would be sufficient for a detection of the Rayleigh scattering signal. Because \pcam will produce polarization maps with unprecedented resolution and sensitivity and it will measure foreground contamination on small scales at very high frequencies (see Section \ref{sec:RS_foregrounds}), theoretically this would benefit detectability prospects. However, FYST is a ground-based observatory, and the atmosphere will hinder the extraction of the cosmological information from large-scale anisotropies. This is especially true in temperature, as highlighted in Fig.~\ref{fig:SN_CCAT_PLANCK}. Atmospheric effects should have less of an impact on the observation of polarization fluctuations and \pcam will provide complimentary information in currently unobserved regimes (high frequencies and small angular scale polarization anisotropies).

Finally, we stress that the CCAT-prime noise model is currently calibrated on available ACT measurements. Its knee frequency $\ell_\mathrm{knee}$, must be less than 200 in order for \pcam to obtain a detection.  Based on the expected scaling of the ACT-measured atmospheric noise rms with PWV (and thus altitude) and its power law spectrum $PSD \sim$ $k^{-11/3}$, we expect $\ell_\mathrm{knee} \sim 800$ compared to the $\ell_\mathrm{knee} \sim 1000$ at the ACT site and used in these forecasts. It seems therefore optimistic to expect such a low knee frequency.  

\subsection{Mitigation of the impact of foregrounds} \label{sec:RS_foregrounds}

Besides the impact of the atmosphere, and similar to observations of primary CMB anisotropies, the Rayleigh scattering signal will be affected by astrophysical foregrounds. Fortunately, several unique properties of the Rayleigh scattering signal can be leveraged to mitigate the impact of foreground contamination. In this analysis we include CIB, SZ, Radio point sources, dust and synchrotron foregrounds. We find that the CIB and dust and radio point sources dominate in foreground mitigation. 

\begin{figure}[ht!]  
\begin{center}
\includegraphics[width=0.55\columnwidth]{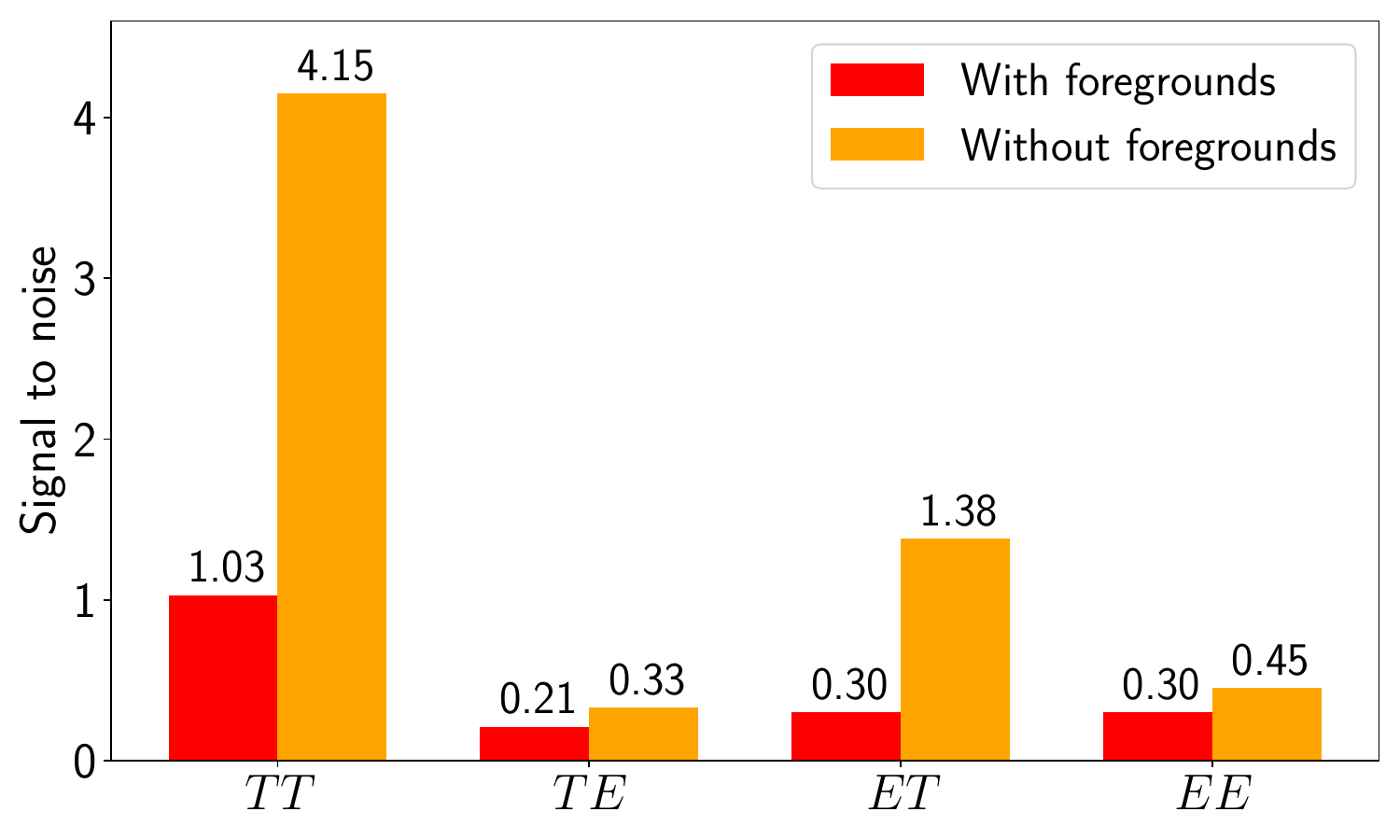}
\end{center}
\caption{Effect of foregrounds on the forecasted signal to noise of the detection of the Rayleigh scattering cross spectrum using \pcam \ combined with {\it Planck}. As expected, foregrounds have a larger effect on temperature, making a detection of Rayleigh scattering challenging. These forecasts have been carried out using a constrained-ILC~\citep{Remazeilles11_CILC} algorithm in order to avoid residual bias.}
\label{fig:RS_foregrounds}
\end{figure}

First, the frequency scaling of Rayleigh scattering ($\propto \nu^4$ at lowest order) is unique, since no foregrounds exhibit such a scaling. This makes Rayleigh scattering suitable to use with \emph{blind} component separation methods such as internal linear combination (ILC) approaches. Second, Rayleigh scattering is a robust prediction of standard physics so that once the cosmological parameters are fixed, the Rayleigh scattering signal can easily be modelled, and included in \emph{parametric} foreground cleaning techniques. 

Measuring the cross-correlation between the primary CMB and the Rayleigh signal will require the removal of foreground contamination from two maps: a map of the primary CMB; and a map of the Rayleigh-scattering signal (at a reference frequency). In order to avoid biasing the detection, one has to be careful to avoid residual primary CMB in the Rayleigh scattering map (and vice-versa). This demands the use of a constrained-ILC method \citep{remazeilles/delabrouille/cardoso:2011a}.

Figure~\ref{fig:RS_foregrounds} shows the impact of the foregrounds on the detection of the Rayleigh scattering cross-correlation using a constrained-ILC. As expected, a larger penalty has to be paid in temperature where foregrounds are brighter than in polarization. However, we should note that resorting to a constrained-ILC increases the noise in the reconstructed maps. Using more advanced component-separation techniques such as SMICA~\citep{Planck_2018_IV} could further improve the signal-to-noise ratio of the cross-correlations, while also making use of the known $\ell$-shape of the Rayleigh scattering signal.

In conclusion, atmospheric noise and foregrounds will make it unlikely that \pcam alone will be able to make a detection of the Rayleigh signal. Future work will be undertaken to determine if combining \pcam with other experiments, such as Simons Observatory and {\it Planck}, as well as resorting to more advanced foreground mitigation techniques, will make a first detection with currently planned experiments realistic. Within the collaboration, current efforts are focused on improving the forecasts by injecting more realism while also exploring new methods to clean foregrounds. 

\section{A new submillimeter window into time-domain astrophysics}

\label{sec:YSO}

Compared to other parts of the electromagnetic spectrum, relatively few submillimeter studies have focused on the time domain. Nevertheless, systematic submillimeter observations in the time domain open up a wide open discovery space for variable and transient phenomena, and the opportunity to quantify the physical properties of variable sources captured in this regime. These phenomena stretch across timescales from seconds to years, arising from young stellar objects (YSOs) through super-massive black holes and with both thermal and non-thermal origins.

We begin in Section \ref{sec:TD-intro} with examples of the types of transient events expected in the submillimeter and the typical timescales associated with their variation. Next, in Section \ref{sec:TD-commensal} we describe the opportunities for commensal time-domain investigations during the WFS and DSS large surveys. Finally, in \ref{sec:TD-focussed}  we present focused time-domain observations for protostars, explosive sources, tidal-disruption events, and fast transients.

\subsection{Submillimeter Time-Domain Overview}
\label{sec:TD-intro}

We already know that FYST will be able to monitor the variability of thousands of known AGN, as well as tracking the motion of hundreds of asteroids.  However, there will also be new classes of submillimeter sources probed with a cadence of minutes to months.
As a specific example, deeply embedded YSOs exhibit submillimeter time-variable behavior, typically due to changes in the mass accretion rates adjusting the equilibrium temperature of the enshrouding envelope \citep[e.g., ][]{2020ContrerasPena,Lee+20}, with observed timescales 
from weeks to many years.
Magnetic reconnection events in stellar flares have been detected at millimeter and submillimeter wavelengths from 
low-mass stars \citep[][]{2020MacGregor,2021Guns,2021Naess}
and T Tauri stars \citep[][]{2003Bower,2019Mairs} with timescales as short as 15 minutes through a few hours and with multiple detections over 6 days \citep[e.g., a chromospherically active binary whose dwarf star may
be about to arrive, or recently has arrived, on the main sequence;][]{2021Naess}.
Cycles of dust formation and destruction can lead to periodic variability in asymptotic giant branch stars, while free-free emission with origins in the radio photosphere may also contribute \citep[][periods and optical-to-submillimeter lags on a few year timescales]{2019bDharmawardena}. 

As a second example, although relativistic jets are often probed at lower frequencies, submillimeter observations have revealed the synchrotron emission from relativistic jets in black hole X-ray binaries on timescales significantly less than an hour \citep[][]{2017Tetarenko}, quiescent supermassive black holes like Sgr A* \citep[][]{2017Subroweit}, and gamma-ray-bright blazars \citep{2014Fuhrmann}, probing the  jet physics at scales close to where electrons are first accelerated. 
Energetic (and often explosive) transients like supernovae, gamma-ray bursts, merging neutron stars, and tidal disruption events produce shocks in the surrounding circumstellar medium and interstellar medium on hour-to-day timescales \citep[][]{2019Ho}.
At the extreme mass end, recently, Event Horizon Telescope measurements of the wobbling shadow of the black hole in M87 have been analysed on timescales of days to years \citep{2020Wielgus}.

While limited time-domain studies utilizing the submillimeter have been ongoing for years (e.g., the transient sources illustrated in Fig.~\ref{fig:extragaltransients}), the lack of sensitive facilities has kept these studies in a state of infancy compared to the radio, UVOIR, and X-ray regimes.
Moreover, most submillimeter studies to date have arisen from small, directed observational campaigns, as opposed to larger, more general surveys {\citep[see][for recent examples of the latter]{2016Whitehorn,Herczeg+17,2021Naess,2021Guns}.}
As time-domain surveys in the optical take center stage in the upcoming decade, time-domain measurements at other wavelengths will be increasingly important for determining the physical processes behind the wide variety of transient and variable events uncovered in our Universe.  In this regard the submillimeter is particularly important because of its ability to unveil the dust-enshrouded and non-thermal regimes.

FYST has the capacity to dramatically improve our knowledge of the submillimeter time domain Universe through a combination of broad, general-survey approaches and specially crafted, targeted campaigns.
With these approaches FYST will lead the systematic study of submillimeter variability, on timescales from seconds to many years, particularly at the highest submillimeter frequencies. 

\subsection{Capturing the Time Domain through Commensal Observations}
\label{sec:TD-commensal}

As a predominantly survey-oriented camera simultaneously observing continuum emission at five frequencies between 220 and 850\,GHz (1.3\,mm and 350\,$\mu$m), 
\pcam offers an exceptional platform for commensal submillimeter transient studies.
The two large survey programs, WFS and DSS, described earlier in this paper, will monitor an extremely broad range of areal coverage and observing depths over hundreds of nightly epochs.

The primary science drivers for the WFS and DSS are described in Section \ref{sec:survey_details} above. We will utilize data obtained during these 4000-hour surveys for relatively shallow but extremely wide-field monitoring of transients in the WFS, and very deep but narrow-field monitoring during the DSS. Table \ref{tab:TD-1}
lists the cadence and depths that we will obtain at each of the five continuum frequencies. 

\begin{table}[th]
{\centering
 \caption{Primary Survey areal coverage and sensitivity per epoch}\label{tab:TD-1}}
 \begin{tabular}{lcccccc}
Survey &   Coverage& Frequency& Integration& Detection Q1& Detection Q2& Detection Q3\\
       &(per epoch)&                 &  (per beam) & ($5\,\sigma$)   & ($5\,\sigma$)  &  ($5\,\sigma$)  \\
       & [deg$^2$]& [GHz]    & [s] & [mJy-beam$^{-1}$] & [mJy-beam$^{-1}$] & [mJy-beam$^{-1}$] \\
\hline\hline
WFS\tablenotemark{a} %
        & 10000&  850 & 0.6& 1500 & 3500 & 10000 \\
        &      &  410 & 2.7& 300  &  450 & 850   \\
        &      &  350 & 3.7& 150  &  200 & 350   \\
        &      &  280 & 2.8& 70   &  85  & 100   \\
        &      &  220 & 3.4& 55   &  65  &  85 \\
\tableline
DSS\tablenotemark{b} %
        & 12&  850 & 360 & 65 & 150 & 500 \\
        &   &  410 & 1600& 12 & 20  & 35 \\
        &   &  350 & 2100& 5.5& 8.0 & 15\\
        &   &  280 & 1600& 3.0& 3.5 & 5.0\\
        &   &  220 & 2000& 2.0& 2.5 & 3.5\\
\hline
\end{tabular}
\par
\tablenotetext{a }{The WFS will observe for roughly 8 hours each epoch, covering half its full extent. Each location within the final map will be observed over 250 epochs.}
\tablenotetext{b }{The DSS will observe for roughly 6 hours each epoch per field, during which time the continuum cameras will scan over a fixed arc covering roughly 12 deg$^{2}$. Each region will be observed over 360 epochs.}
\end{table}

The WFS aims to uniformly cover $20{,}000\,{\rm deg}^2$ over 4000\,hr (see {Table~\ref{tab:surveys}}) by scanning at fixed azimuth and allowing sky rotation to fill in the observable field during approximately 8 hrs of observations per night. The survey provides an unprecedented opportunity to uncover extremely rare but strongly brightening events in the submillimeter regime (see Table \ref{tab:TD-1}) between individual epochs.  Co-adding observations over multiple epochs will allow for deeper searches of longer duration events (e.g., on daily, weekly, and monthly timescales). Given the proposed scanning strategy, \pcam will provide close to simultaneous (separated by $\sim 10$ minutes) brightness measurements at all five frequencies covered by the continuum cameras.

The DSS dedicates 2000 hours each to two $4\,{\rm deg}^2$ fields observed with the EoR-Spec modules.  As the DSS survey unfolds, the other \pcam instrument modules will scan neighboring swaths of sky (see Section \ref{sec:scanpatterns}).  Each camera band will  observe approximately
the same patch of sky from night to night, but  the overlap of coverage between frequency bands will be 
minimal, resulting in deep coverage maps at up to five frequencies that have little overlap on the sky. 
For each of the continuum cameras these smeared out patches of the sky (compared to the DSS), each covering $\sim 12\,{\rm deg}^2$, will provide an unprecedented opportunity to uncover brightening events at submillimeter frequencies over a wide range of flux densities 
(see Table \ref{tab:TD-1}) between individual nightly epochs.

Together the WFS and DSS will enable a wide range of commensal transient science, from Galactic stellar flares through distant blazars. For example, stellar flares are an important tracer of stellar activity, particularly in young stars and chromospherically active binaries. These flares may play a role in limiting the habitability of planets around young stars, making them an important target.
The stellar flares are thought to arise from magnetic reconnection, leading to either synchrotron or gyrosychrotron emission \citep[][]{2003Bower,2006Massi}.
Approximately 23 stellar flares have been detected in the submillimeter regime
\citep[see figure~4 in][]{2021Guns}, with 17 of these coming from dedicated general submillimeter transient surveys \citep{2019Mairs,2021Naess,2021Guns} where the stars were not specifically targeted.  
In some of the brighter stellar flares, where spectral information was available, the flares roughly followed $S_\nu\propto\nu^{0}$ to $S_\nu\propto\nu^{1}$ behavior; however, the faint stellar flares from M dwarfs showed a variable spectral index 
$S_\nu\propto\nu^{-2}$ to $S_\nu\propto\nu^{2}$ on timescales under a minute.
This raises the question as to whether most stellar flares will have a rising or falling flux density with increasing submillimeter frequencies --- an issue that FYST's large surveys at a variety of flux density levels can address.
We predict of order 1000 stellar flares will be detected commensally during FYST's two primary surveys, increasing the known numbers by two orders of magnitude. If the spectral shape of $S_\nu \propto \nu^{-1}$ is more representative, we will likely detect at least a factor of 3 fewer sources at 220\,GHz, and even fewer at 280\,GHz. The combination of the WFS and DSS make FYST an excellent tool for disentangling the submillimeter properties of stellar flares.

Compared to stars, the number densities of blazars are significantly lower. However, in blazars, we aim to characterize day to week timescale variability around a relatively bright flux density as opposed to just searching for transient phenomena. There are approximately 600 blazars with measured 143-GHz flux densities over the whole sky \citep{Massaro2015}. Since these sources have a (non-simultaneous) 1.4--143\,GHz spectral index of  $0.06^{+0.18}_{-0.43}$ (90\,\% confidence interval), blazars are best observed at 220\,GHz with the WFS. Assuming a median flat spectral index, and extrapolating the 143-GHz flux densities of blazars from  
\citet{Massaro2015} and the 1.4-GHz flux densities from the blazar samples of  \citet{DAbrusco2019}, the WFS should detect 300--600 blazars above 400\,mJy in each nightly epoch. This flux density is sufficient to measure (3$\,\sigma$) excess variance at the roughly 10\,\% level. This would generate a rich data set to explore AGN science. For instance, neutrino experiments like the IceCube Neutrino Observatory are now routinely detecting new astrophysical neutrinos above 0.1\,TeV and localizing them to a few square degrees \citep[e.g., the neutrino identified with the blazar TXS 0506+056;][]{IcecubeMMA2018}. The FYST variable blazar sample will provide a powerful real-time resource for identifying flaring blazars that might be generating these neutrinos.
\\

\subsection{Focused Time-Domain Observations}
\label{sec:TD-focussed}

While the commensal time-domain observations discussed previously will present an opportunity to analyse the variability rate and amplitude of rare brightening events across roughly half the sky, there are several situations where dedicated monitoring for submillimeter variability is preferred. This is especially true for cases where the sources to be monitored are known to reside in discrete, clustered, locations, such as Galactic YSOs located within star-forming molecular clouds. Furthermore, there will be time-limited opportunities where either FYST or other telescopes identify a varying source, and dedicated submillimeter follow-up monitoring will provide important and unique information. Below we focus on three specific examples.

\subsubsection{Scheduled Monitoring of  Variable Protostars}

For 5 years, the JCMT Transient Survey has monitored eight nearby low-mass star-forming regions at submillimeter frequencies, with a monthly cadence \citep{Herczeg+17}. The JCMT survey remains the only long-term dedicated time-domain study of deeply embedded protostars, significantly increasing our understanding of accretion variability by carefully constructing the data reduction and analysis procedures to achieve $\sim$2\,\% relative flux calibration across epochs \citep{Mairs+17}. The Transient Survey has revealed that on timescales of years the dominant submillimeter variability is secular --- approximated by either linear or sinusoidal brightness curves --- with scant evidence of stochastic, epoch-to-epoch variability above the measurement noise \citep{Johnstone+18}. After 4 years, the number of confirmed protostellar variables is 17 out of 51 sources brighter than 350 mJy beam$^{-1}$ at 350 GHz (33\,\%; Lee et al. submitted). From this sample, EC\,53 in Serpens Main (also known as V371 Serpentis) shows a quasi-periodic light curve, allowing for a detailed quantitative investigation into the location and source of the underlying accretion instability \citep{Lee+20}. 

Many JCMT Transient Survey sources show significant time-localized structure within the brightness curves. Small number statistics and the short time duration of the JCMT survey, however, limit a statistical analysis coupling these observations directly to theoretical models. Comparison with multi-epoch observations of the same sources taken at mid-IR frequencies \citep{2020ContrerasPena}, and utilizing radiative transfer models \citep{2019bMacF, 2019aMacF, 2020Baek}, confirms that the observed submillimeter brightness scales with the time-varying temperature of the enshrouding envelope, itself responding to the changing accretion luminosity. Thus, the relative  submillimeter response is {\it weaker\/} than the mass accretion response (the temperature variation in the envelope scales approximately as $T \propto L_{\rm acc}^5$) \citep{2020ContrerasPena}. Thus, the typical 10\,\% submillimeter brightness variation detected over few year timescales relates to an order unity variation in the mass accretion --- these observed events are not insignificant perturbations.

Such variability of deeply embedded protostars is expected. Within dense molecular clouds, initial over-densities of gas become gravitationally unstable to collapse and seed the formation of stars \citep{Shu77, Shu+87, Masunaga+00}. As this material accretes onto each protostar, a fraction of the released gravitational potential energy is converted into radiation, which in turn heats the optically thick surrounding envelope. Thus, throughout the earliest stages of mass assembly, the young stellar object (YSO) is observable only indirectly through the warmth of its enshrouding material, referred to as the ``protostellar core'' \citep[e.g.][]{Johnstone+13}. The accretion history is expected to be time variable, both on million-year stellar-mass assembly timescales and over short times associated with the many potential dynamical instabilities within the system \citep{Armitage15}. 

In detail, the core's initial angular momentum leads to infalling material forming a disk rather than plummeting directly onto the YSO. Without appropriate viscous dissipation within the disk, the infalling material will accumulate until the disk triggers gravitational instabilities, either within the disk itself (disk fragmentation) or between the disk and the YSO (spiral torques). 
Timescales for triggering these outer-disk gravitational instabilities are typically thousands of years \citep[][]{Vorobyov+18}. Once the instability takes over, disk material is flushed onto the YSO on Keplerian timescales, tens to hundreds of years, and the cycle may repeat multiple times. 
Alternatively, viscous dissipation within the disk should lead to on-going disk accretion flow, although there is no a priori expectation that this dissipation rate should match the accretion rate onto the disk from the core or that the disk dissipation should vary smoothly throughout the disk
\citep[][]{Zhu+10, Bae+14, Nayakshin+12}. Mismatched processes will manifest as accretion variability with timescales related to either the viscous dissipation timescale at the discontinuity or the local Keplerian timescale, the latter of which becomes days near the inner edge of the disk \citep{Lee+20}.

There are few empirical studies detailing the time dependence of mass accretion, despite the theoretical expectation. Most observational evidence for protostellar variability comes from rare but extreme brightening events --- FU Ori and EX Ori sources --- and a handful of multi-epoch experiments \citep[][]{Hartmann+96, Audard+14, Pena+17}, many of which target the most evolved YSOs. Monitoring the youngest, deeply embedded protostellar YSOs (protostars) is complicated since the accretion itself is not directly observed but must be inferred through its heating effect on the enshrouding protostellar core \citep{Johnstone+13}. Determining the underlying timescales and amplitudes related to the accretion process is fundamental to understanding protostellar assembly, as they constrain the physical conditions within the inner envelope and disk.

Aside from the JCMT survey, limited YSO variability studies have also utilized {\it Spitzer\/} and {\it Herschel\/} \citep[e.g.][]{Rebull+15, Billot+12}, confirming that young stars and YSOs vary across a wide range of timescales. At these higher frequencies, however, one must untangle source variability from changing optical depth conditions \citep{2019bMacF, 2019aMacF, 2020Baek}. More recently, the mid-IR {\it WISE\/} satellite has monitored YSOs (as part of its all-sky survey) with a 6-month cadence, yielding exceptional leverage for submillimeter monitoring surveys by linking directly the submillimeter and mid-IR variations and by providing access to less embedded, older YSOs \citep[][Park et al.\ in prep]{2020ContrerasPena}. Combined, the submillimeter and mid-IR time domain surveys will recover the history of variability across the main accretion stage of stellar assembly.

FYST offers the best opportunity for achieving sufficient statistical samples of YSOs across epochs. With FYST, each star-forming field will be $4\,{\rm deg}^2$ (Table \ref{tab:TD-2}), an order of magnitude larger area than with the JCMT, while the map depth will be 5\,mJy at 350\,GHz (850\,$\mu$m), 2.5 times deeper than the JCMT Transient Survey. Combined, FYST will monitor an order of magnitude more sources (Table \ref{tab:PS-var}). The FYST survey benefits from the uniqueness of its highest frequency, 850\,GHz (350\,$\mu$m), observations which probe closer to the peak of the protostellar envelope spectral energy distribution (SED) where the response to variability is stronger \citep[][]{Johnstone+13, 2020ContrerasPena}.  The 850 GHz sensitivity per epoch, 60\,mJy\, beam$^{-1}$, provides a fixed signal-to-noise ratio between 350 and 850 GHz for source spectral indices as shallow as $S_\nu \propto \nu^{2.5}$. Finally, FYST is designed for surveys and can effectively schedule epochs at a two week cadence, twice the cadence of the JCMT, to better match the heating timescale in the envelope \citep{Johnstone+13}. 
\begin{deluxetable*}{lcccccc}[ht]
 \tablecaption{Per epoch observing time required to reach targeted depth.\label{tab:TD-2}}
\tablehead{ 
Survey &   Coverage& Frequency& Detection & Time Q1& Time Q2& Time Q3\\
       &(per epoch)&          &  ($5 \sigma$)&        &        &   \\
       & [deg$^2$] & [GHz]    &  [mJy/bm] & [s] & [s] & [s]
       }
\startdata
PS Monitor\tablenotemark{a} & 4&  850 & 300& 320 & 1500 & \nodata  \\
    &     &  350 &  25& 340 & 710 & 2000 \\
\tableline
Localized ToO\tablenotemark{b}  %
        &\nodata  &  850 &  55 &  610 & 2900 & \nodata\\
                & &  410 &  15 &  980 & 2400 &  8600\\
                & &  350 &   7 & 1400 & 2900 &  8000\\
                & &  280 &   3 & 1500 & 2300 &  4600\\
                & &  220 &   3 & 1100 & 1600 &  2800
\enddata
\tablenotetext{a}{The Protostar Monitoring Survey will observe each region every two weeks.}
\tablenotetext{b}{The Localized Target-of-Opportunity (ToO) calculations are appropriate for sources known to within a quarter of the field of view of the camera module. These numbers assume that a portion of the camera remains on the source for the entire observation, increasing the observing efficiency significantly compared with mapping and allowing for rapid sub-scan analysis.}
\end{deluxetable*}

The FYST monitoring plan requires 10 hrs per year per field. Thus, with 500 hrs we will monitor 10 fields for 5 years each. Table \ref{tab:TD-2} presents the observing times required to reach the depth requested while Table \ref{tab:PS-var} lists the 10 target fields. These fields include 7 low-mass star-forming regions presently observed by the JCMT, extending the 850\,$\mu$m (350 GHz) timeline beyond a decade of monthly, or better, cadence. Broadening the range of star-formation conditions being explored, we add two relatively nearby (within 2\,kpc) high-mass star-forming regions, where the individual sources will be more tightly packed, but the opportunity to observe rare events, such as FU Ori bursts, will be significantly enhanced. Lastly, the Galactic center, at a much greater distance (8\,kpc), will provide an extreme Galactic environment that has not yet been monitored in the submillimeter.
\begin{table*}[ht]
\begin{center}
 \caption{Protostellar variability monitoring fields and source counts}
 \label{tab:PS-var}
 \begin{tabular}{lcccccc}
  \hline
Target Field &  RA (J2000) &  Dec (J2000) & Class 0/I$^a$ & Class II$^a$&Submm Peaks$^b$& Class 0/I Targets$^c$\\
             &   [hh:mm]   &   [dd:mm]    & (Total)   & (Total)  &  ($> 0.1\,$Jy-beam$^{-1}$) & (Submm bright)   \\
\hline

Perseus           & 03:29& $+$31:06& 69&  92& 62& 35\\
Orion\,A (OMC)    & 05:36& $-$05:42&190&1600&287& 47\\
Orion\,B (South)  & 05:42& $-$01:54& 46& 287& 28& 10\\
Orion\,B (North)  & 05:46& $-$00:06& 60& 238& 71& 22\\
Ophiuchus         & 16:27& $-$24:24& 52& 152& 52& 11\\
Serpens (North)   & 18:29& $+$00:30& 52& 180& 33& 15\\
Serpens (South)   & 18:29& $-$02:02&115& 589&122& 21\\
Carina            & 10:44& $-$60:04& ...& ...& ... & ...\\
Galactic Center   & 17:46& $-$28:56& ...& ...&$>$1000& ...\\
M17               & 18:19& $-$16:30& ...& ...&181& ...\\
\hline

\end{tabular}
\end{center}
$^a$ Young stellar object source counts for Gould Belt Regions based on catalogues by \citet{Dunham15} and \citet{Megeath16}.\\
$^b$ Submillimeter (Submm) peak source counts based on partial observations of region by JCMT SCUBA-2 at 850\,$\mu$m.\\
$^c$ Class 0/I Target counts determined by coincidence of known Class O/I sources and bright ($> 0.1\,$Jy-beam$^{-1}$ at 850\,$\mu$m) submillimeter peak emission.\\
\end{table*}

\subsubsection{Target-of-Opportunity Potential for Explosive Transients and Tidal Disruption Events}

\begin{figure}[t]
\begin{center}
\includegraphics[width=0.9\columnwidth]{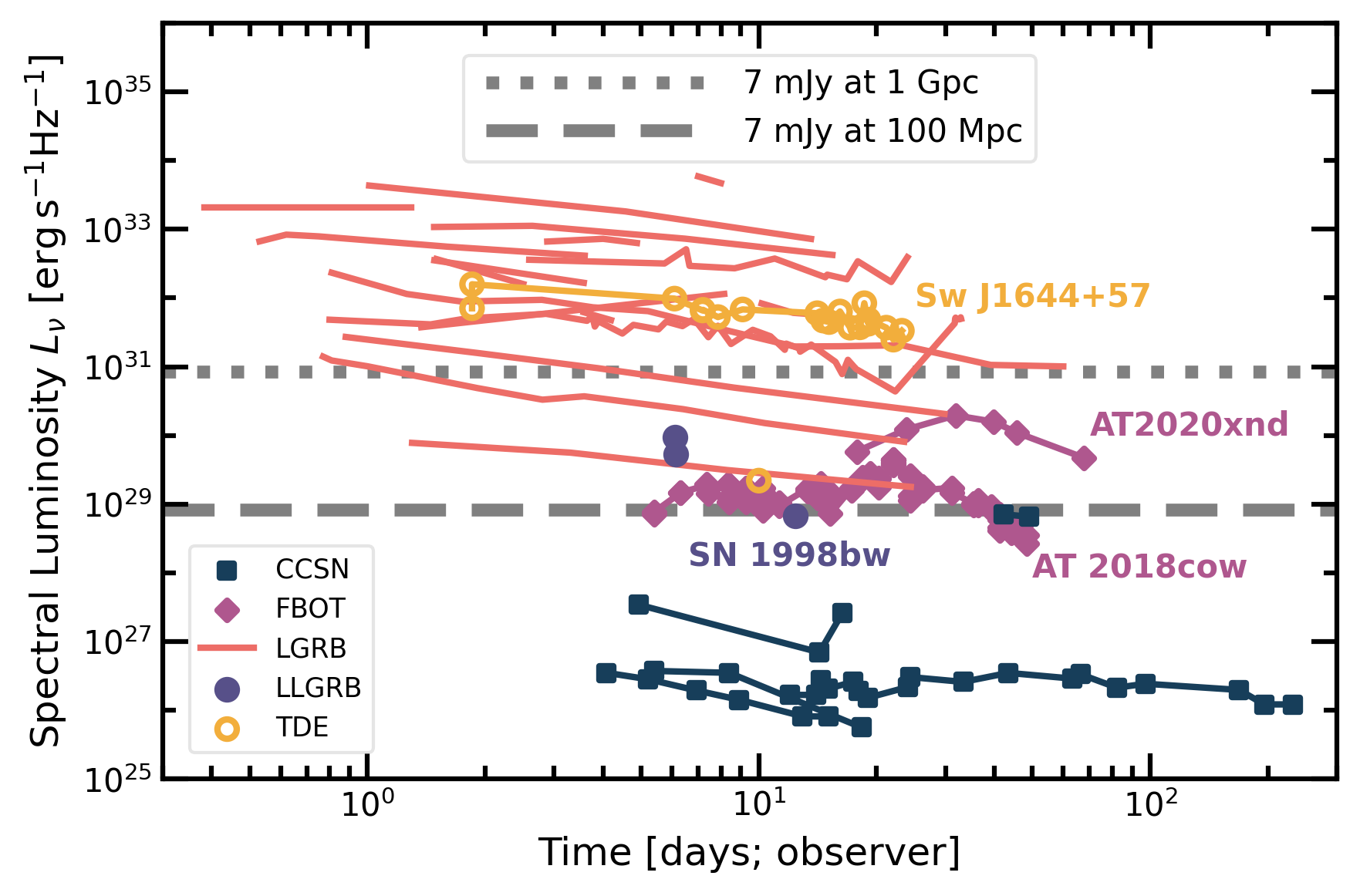}
\caption{
Sample lightcurves of extragalactic transients detected at (sub)millimeter wavelength. Five class types are displayed: core-collapse supernovae (CCSNe); faint blue optical transients (FBOTs); long-duration gamma-ray bursts (LGRBs); low-luminosity gamma-ray bursts (LLGRBs); and tidal disruption events (TDEs). In 1 hour in the top 50\,\% driest conditions, FYST observations at 350 GHz will be able to detect: the brightest (long) gamma-ray bursts (GRBs) with $z\lesssim0.2$ (1 Gpc), with sensitivity to LLGRBs within $100\text{--}300\, {\rm Mpc}$; tidal disruption events (TDEs) like SwiftJ1644$+$57 with $z\lesssim0.2$; and FBOTs like AT2018cow and AT2020xnd within $100\text{--}300$ Mpc. These detection limits, indicated by horizontal lines, assume a $5\,\sigma$ detection, a 1.4\,mJy\,beam$^{-1}$ sensitivity, and a flat spectral index for the (typically longer) wavelengths where these sources were detected.  At 220 GHz and 280 GHz, FYST will be sensitive to objects $\sim 50\%$ more distant. This image was adapted from lightcurves collated in \citet{2021Ho}, with additional information from \citet{2012deUgartePostigo} and \citet{2015Urata}.}

\label{fig:extragaltransients}
\end{center}
\end{figure}

Explosive transients (i.e., transient events that can only occur one time per source, such as supernovae and gamma ray bursts --- SNe and GRBs) and tidal disruption events (TDEs) have both been detected at (sub)mm frequencies (see Fig.~\ref{fig:extragaltransients}).  The majority of the explosive transients arise due to shock waves generated as a relativistic shell expands into a medium. This shell is thought to be driven by a collimated jet, as opposed to a truly spherical shell. In the sense of having a relativistic jet drive shock waves into an external medium, TDEs can act similarly to GRBs.

Most (sub)mm detections in this field have arisen from variations in the synchrotron emission as the forward shockwave intersects the circumstellar medium or interstellar medium. However, a reverse shockwave may propagate into expanding ejecta in the shell that fuels the GRB afterglow. Both shocks can be roughly modeled as broken power-law SEDs, with the SED peak of each component expected to move towards lower frequencies with time after the explosion. 
Although the reverse shock is typically thought to be much brighter than the forward shock (the latter of which is often the focus for near-IR and higher frequencies), the reverse shock must be caught soon after the jet is launched because the peak of the reverse shock SED at a given time is at a much lower frequency than the forward shock. 
If detected, reverse shocks can play an important role in constraining the magnetization and bulk Lorentz factor of the jets driving this emission. 
To first order at FYST frequencies, the reverse shock of GRB afterglows with higher magnetic field densities will be brighter, while having their reverse shock peaks at slightly earlier times; similarly, GRB afterglows with lower initial Lorentz factors will have their reverse shocks peak at later times, while being a little less bright \citep[see figure~6 in][]{2015Urata}. 
While radio emission can also play a role in detecting reverse shocks, the best constraints come from systems that can be observed at submillimeter-to-radio frequencies.

In Fig.~\ref{fig:extragaltransients}, we summarize the millimeter/submillimeter detections of core-collapse SNe, faint blue optical transients (FBOTs), long-duration GRBs (LGRBs), low-luminosity GRBs (LLGRBs), and TDEs. While most CCSNe will be too faint for FYST, 7-mJy detection limit 350 GHz observations would track the rest of the source classes' submillimeter evolution on timescales from $\lesssim$ 10 days (GRBs) to 100 days at reasonable distances (100--300 Mpc for FBOTs and some LLGRBs, 1 Gpc for brighter LGRBs and TDEs). Fainter 3 mJy detection limits (and the sensitivity to $\sim 50\%$ farther sources) are more suitable for 220 GHz and 280 GHz (see Table \ref{tab:TD-2}).
Although the displayed CCSNe (SN 1993J, SN 2008D, SN 2011df, and SN 2020oi) are relatively faint millimeter/submillimeter sources, radio light curves of other SNe (SNe 2009bb, 2003L, 2003bg, and 2007bg) suggest that there exists a larger sample of SNe that may be similarly luminous (sub)mm transients as compared to the FBOTs AT2018cow and AT2020xnd \citep{2019Ho,2021Ho}. In addition, the radio luminosities of other FBOTs have peak radio spectral luminosities that are as bright or brighter than AT2018cow and AT2020xnd. FYST is well suited for early detection and characterization of the reverse shock (or detection limits on the reverse shock strength), provided that scheduling allows FYST to get on the source within a few hours.  Compared to GRBs, the longer delay of the rise from the initial detection makes FBOTs easier targets to schedule for FYST observations. We note, however, that the underlying rates of sufficiently bright sources for FYST are poorly known at this time, and are expected to amount to no more than a few sources per year.

The deep survey (DSS) is sensitive to transient events like those listed above; however, the DSS has a limited angular extent. Therefore, it is unlikely to catch many (and possibly any) such transients. We may instead consider that target of opportunity (ToO) observations triggered by other observatories would enable FYST to help answer significant questions in this regime. 
While there may only be a handful of such sources over a few years,
FYST could target southern sources that cannot easily be observed multiple times at 220--850\,GHz by other facilities (for instance due to telescope pressure on ALMA) or target bright sources where FYST's uniquely sensitive 850-GHz window could provide needed high-frequency detections.

Finally, FYST could play a significant role in identifying potential submillimeter counterparts of multimessenger astrophysical sources.  Although short GRBs tend to be less luminous than long GRBs, the association of short GRBs with gravitational wave events from neutron star mergers \citep{2017AbbottApJ} make targeted FYST followup of GW events another priority. When FYST comes online, the typical uncertainty region of a gravitational wave event will be well matched to the smaller areas considered in surveys like the transient YSO survey. Similarly, astrophysical neutrinos \citep[which may be connected to blazar activity or other transient events;][]{IcecubeMMA2018} also have an uncertain localization that is well matched to the capacities of FYST. Thus, FYST can efficiently cover wide patches of the sky for gravitational wave and neutrino triggers to potentially localize their electromagnetic counterparts. Targeting southern GW events from merging neutron stars within 200\,Mpc provides a relatively well-constrained program that could detect such sources down to one-tenth the brightness of GRBs 030329, 100418A, or 120326A (both for gravitational wave events with no prompt electromagnetic counterpart and those where the electromagnetic counterpart has been identified at another frequency). Similarly, targeting a subset of the tens of neutrino triggers per year provides a second relatively well-constrained program. 

The details of ToO implementation within FYST have not yet been determined.
The time-constrained nature of ToOs will naturally use relatively limited telescope resources (e.g., 1 hr per non-localized trigger, $5\rm{-}20\times1$-hr epochs per localized FBOT, GRB, TDE, or  multimessenger counterpart).
ToOs can play a role in optimizing telescope use through observing at 220\,GHz and 280\,GHz during poor weather conditions. Observing at higher frequencies in better weather conditions for high-impact targets will need to be balanced with other FYST scientific programs (see Table~\ref{tab:TD-2} for example sensitivities).
Dependent on the final implementation of a ToO program, 
there is potential for FYST to make major advancements in explosive transients, tidal disruption events, and especially multimessenger astronomy.

\subsubsection{The Fast and the Furious}

Magnetic reconnection events in stellar flares have been detected at mm and submillimeter wavelengths from young objects such as M dwarf and T Tauri stars, as well as chromospherically active binaries whose dwarf star may be about to arrive, or recently has arrived, on the main sequence \citep{2003Bower,2019Mairs,2020MacGregor,2021Guns,2021Naess}.
While commensal observations (see Section \ref{sec:TD-commensal}) will capture a significant number of stellar flares, some nearby sources that undergo multiple flares per day reaching 100\,mJy \citep[e.g., Proxima Cen;][]{2020MacGregor} in 30 seconds could be targeted by FYST. Other stellar flaring sources have flares that reach 900 mJy and decay over 15-minute timescales; these may be captured commensally in the protostellar survey. 
\citep[e.g., the T Tauri Binary System JW 566;][]{2019Mairs}.
These interesting bright events have usually gone undetected, as few data sets have been analysed at the necessarily rapid timescales.
Nevertheless, they are already revealing the presence of high-energy electrons during the flares, while constraining the density and magnetic field
\citep{2020MacGregor}.

During transient outbursts of stellar-mass black holes accreting from nearby stars (i.e., black hole X-ray binaries), a relativistic jet may be launched.  This is an important outflow event that is connected to accretion inflow through poorly understood physics. While astronomers have largely used radio observations to probe the jet properties of this connection, the submillimeter regime provides a much-needed tracer of the jet properties closer to the black hole. Radio and submillimeter observations of black hole X-ray binaries have demonstrated that the jets from X-ray binaries can vary significantly on day-to-day timescales \citep[e.g.,][]{RussellD2013}. By measuring the evolution of an approximately broken power-law SED whose break frequency travels through the submillimeter to the radio, sophisticated models can constrain the evolving power injected into the jet base, the radius of the jet base, the location where particle acceleration in the jets starts, and the slope of the injected  non-thermal particle distribution \citep{Lucchini2021}. These properties can, in turn, be connected to the accretion inflow properties measured by the X-ray emission. Moreover, recent observations have shown that the most extreme sources (e.g., V404 Cyg and MAXI J1820$+$070) demonstrate submillimeter variability on timescales as fast as tens of seconds
\citep{2017Tetarenko,2021Tetarenko}. Combined with simultaneous multi-wavelength observations, this rapid variability can reveal poorly understood jet properties, like how the jet shape changes with distance from where particles are first accelerated.

FYST can provide the critical submillimeter data through short campaigns where an object is observed nightly for a short period of time over multiple frequencies (a handful of X-ray binaries per year reach $\sim 5$\,mJy flux-density levels, where FYST sensitivity in a 6-hr multiple camera observation is sufficient for detecting the sources in the submillimeter at 220--410 GHz; see  Table~\ref{tab:TD-2}).
FYST could also observe for long periods during a single night to detect the most rapid variability (e.g., 30 s) in a single camera for sources with bright rapid flares (e.g., brighter flares from young stars at 100--800\,mJy or X-ray binaries at 100--8000\,mJy).
As with the case for GRBs, TDEs, and multimessenger sources, even a small number of campaigns can make a large impact to the burgeoning field of rapid submillimeter variability.

\section{Summary}

\label{sec:summary}
The CCAT-prime Consortium is building the 6-m aperture  Fred Young Submillimeter Telescope to be sited at more than 5600 m elevation on Cerro Chajnantor in northern Chile.  Our science cases center on large scale mapping at very high sensitivity in the submillimeter-to-millimeter wave telluric windows and are enabled by the wide-field of view and excellent surface accuracy of FYST, together with the excellent site and dedicated instrumentation. Our primary science instruments are the CCAT-prime Heterodyne Array Instrument (CHAI) and the \pcam direct detection instrument. \pcam consists of seven instrument modules, two EoR-Spec imaging spectrometer modules and five imaging polarimeters, one each at 220, 280, 350, 410, and 850\,GHz.  The focal planes of each module are kinetic inductance detector arrays filling the available 1.3$^\circ$ diameter image plane with diffraction-limited pixels.  The \pcam instrument on the FYST telescope sited on Cerro Chajnantor is a combination that promises unsurpassed mapping speeds, enabling our science goals.

Our science programs are survey-centered.  The primary surveys are the Wide Field Survey (WFS) and the Deep Spectroscopic Survey (DSS) which enable much of our science program.  Each of these will take 4000 hours of the top three weather quartiles each to complete. In addition, we have narrower or shallower 
programs to address first-light or demonstration science.  

We have discussed at length our planned science program in Sections 4 through 10 above.  Here we summarize our main goals and expectations.

\begin{itemize}
  \item \textbf{Reionization and galaxy evolution traced through line intensity mapping:} We will use EoR-Spec to map the large-scale spatial distribution of the 158 $\mu$m [\ion{C}{2}] and the 88 $\mu$m [\ion{O}{3}] lines to
 
  outline large scale structures from the epoch of reionization to cosmic noon. This line intensity mapping thereby yields constraints on structure formation and growth, the reionization process and the formation and growth of galaxies.
  \item \textbf{Dusty galaxy evolution traced in far-IR surveys:} The multi-band, wide-area and deep continuum surveys enabled by \pcam will: constrain FIR luminosities and therefore cosmic starformation histories of galaxies from z $\sim$ 1 to 5; advance studies of the evolution of high luminosity DSFGs at high z into present day elliptical galaxies; and reveal "exotic" sources such as extreme luminosity protoclusters or the rare, very strongly-lensed galaxies at high z.
  \item \textbf{Improving constraints on primordial gravitational waves through removal of Galactic foregrounds:} By combining sub-millimeter \pcam measurements with longer wavelength data from Simons Observatory we will improve our understanding of polarized dust foregrounds to help searches for primordial gravitational waves. In particular, \pcam measurements have the potential to reduce bias in constraints on the tensor-to-scalar ratio, $r$, from SO and thereby aid in constraining models of the early Universe such as cosmic inflation. 
  \item \textbf{Constraints on the role of magnetic fields in star formation, MHD turbulence, and dust grain composition as revealed through polarimetric imaging:}
\pcam will map hundreds of molecular clouds in polarized dust emission at high fidelity on scales ranging from entire clouds down to the scales where individual stellar systems form. This will enable detailed studies of the role of magnetic fields in the formation of molecular clouds and support of dense substructures against gravitational collapse.  These observations will also allow the investigation of magnetized turbulence in the ISM, while the high-frequency coverage will enable tests of models of dust grain composition as a function of density and environment.
\item \textbf{Galaxy and cluster evolution revealed through SZ effects:} \pcam will provide broad coverage of the tSZ increment at 5 \textendash~
20 times better resolution than currently available. Combining this spectral coverage with Simons Observatory observations, \pcam will enable the clean separation of SZ signals from correlated thermal dust emission. This separation provides unbiased observational windows into the thermodynamic properties of galaxies and clusters, the dust-to-gas mass ratio of the ICM, cosmic ray populations, and magnetic field strengths in the ICM.
\item \textbf{The recombination epoch traced through Rayleigh scattering:} 
\pcam observations, together with data from other observatories such as Simons Observatory and {\it Planck} will place limits on CMB Rayleigh scattering. 
Atmospheric sky noise and foreground contamination are serious challenges to signal detection, but data from the multi-frequency large-scale surveys with \pcam on FYST promise to help significantly in both regards. The additional cosmological information provided by Rayleigh scattering motivates continued efforts in this direction.
\item \textbf{Astrophysical phenomena tracked in the time domain}: Prime-Cam will dramatically improve our understanding of the submillimeter time-domain Universe. This will be achieved through both commensal monitoring of the primary WFS and DSS survey regions and specially crafted targeted campaigns.  Together, these approaches will allow for the systematic study of submillimeter variability, on timescales from seconds to many years, monitoring sources with both thermal and non-thermal origins, from young stellar objects through super-massive black holes. 
\end{itemize}

The FYST telescope is under construction by Vertex Antennentechnik-GMBH in Germany and will begin to be assembled at the Cerro Chajnantor site in the third quarter of 2023.  \pcam with four instrument modules: the 280, 350 and 850\,GHz polarization sensitive cameras and the first of the two EoR-Spec modules are funded and under construction at Cornell and partner institutions. First light is anticipated in the second quarter of 2024.

\section{Acknowledgements}
\label{sec:Acknowledgements}
We thank an anonymous referee for their very complete comments, criticisms and requests for clarifications of earlier drafts of this manuscript.

The CCAT-prime project, FYST telescope and \pcam instrument have been supported by generous contributions from Fred Young, Cornell University, and the Canada Foundation for Innovation and the Provinces of Ontario, Alberta, and British Columbia.  The construction of the FYST telescope was supported by the Gro\ss ger\"ate-Programm of the German Science Foundation (Deutsche Forschungsgemeinschaft, DFG) under 
grant INST 216/733-1 FUGG. 
The construction of the EoR-Spec instrument module for \pcam is supported by NSF grant AST-2009767. The construction of the 350 GHz instrument module for \pcam is supported by NSF grant AST-2117631.

MA acknowledges partial support from Fondecyt through grant 1211951. MA has been supported by the grant ``CONICYT + PCI + INSTITUTO MAX PLANCK DE ASTRONOMIA MPG190030'' and ``CONICYT+PCI+REDES 190194.'' SKC acknowledges support from NSF award AST-2001866. SEC acknowledges support by the Friends of the Institute for Advanced Study Membership, and by the National Science Foundation under Grant No. 2106607. FB acknowledges funding from the European Research Council (ERC) under the European Union's Horizon 2020 research and innovation programme (grant agreement No.726384/Empire). LMF acknowledges support from an NSERC Discovery Grant (RGPIN-2020-06266). SF is supported in part by the Government of Canada through the Department of Innovation, Science and Industry Canada and by the Province of Ontario through the Ministry of Colleges and Universities. PG is sponsored (in part) by the Chinese Academy of Sciences (CAS), through a grant to the CAS South America Center for Astronomy (CASSACA) in Santiago, Chile, by the ``Comisi\'on Nacional de Ciencia y Tecnolog\'ia (CONICYT)'' via Project FONDECYT de Iniciaci\'on 11170551. DJ is supported by the National Research Council of Canada and by an NSERC Discovery Grant. EK is supported in part by the Excellence Cluster ORIGINS, which is funded by the Deutsche Forschungsgemeinschaft (DFG, German Research Foundation) under Germany's Excellence Strategy $-$EXC-2094$-$390783311. PDM acknowledges support from the Netherlands organization for scientific research (NWO) VIDI grant (dossier 639.042.730). JM is supported by the US Department of Energy under grant no.~DE-SC0010129. DR acknowledges support from the National Science Foundation under grant numbers AST-1614213 and AST-1910107.  DR also acknowledges support from the Alexander von Humboldt Foundation through a Humboldt Research Fellowship for Experienced Researchers. MDN acknowledges support from NSF grant AST-2117631. ER acknowledges the support of the Natural Sciences and Engineering Research Council of Canada (NSERC), funding reference number RGPIN-2017-03987. University of Cologne and University Bonn CCAT-prime members acknowledge support by the Collaborative Research Centre 956 funded by the Deutsche Forschungsgemeinschaft (DFG) - project ID 184018867. GRS acknowledges support from NSERC Discovery Grants (RGPIN-2016-06569 and  RGPIN-2021-04001). GJS acknowledges support from the National Science Foundation under grant numbers AST-1716229 and AST-2009767, and NASA grant NNX16AC72G. AS gratefully acknowledges funding support through Fondecyt Regular (project code 1180350) and from the Chilean Centro de Excelencia en Astrof\'isica y Tecnolog\'ias Afines (CATA) BASAL grant AFB-170002. DTC is supported by a CITA/Dunlap Institute postdoctoral fellowship. The Dunlap Institute is funded through an endowment established by the David Dunlap family and the University of Toronto.
https://www.overleaf.com/project/60f8597c20bff1785f12c72e

\bibliographystyle{apj}
\bibliography{references}

\end{document}